\documentclass{article}

\usepackage{cite}
\usepackage[utf8]{inputenc}
\usepackage{geometry}
 \usepackage{lmodern}     
\usepackage[T1]{fontenc}

\usepackage{authblk}

 \usepackage{scalerel}
\usepackage{mathtools}
\usepackage[utf8]{inputenc}
\usepackage{tikz}
\usepackage[justification=centering]{caption}

\usepackage{pgfplots}
\pgfplotsset{width=15 cm,compat=1.9}
\pgfplotsset{compat=newest}

\usepackage{booktabs}
\usepackage{graphicx}

\usepackage{hyperref}

\usepackage{amssymb} 
\usepackage{amsmath, bm}
\allowdisplaybreaks
\usepackage{amsthm}
\mathchardef\mhyphen="2D 

\usepackage{notations}

\newtheorem{theorem}{Theorem}
\newtheorem{statement}[theorem]{\textbf{Statement}}

\newtheorem{lemma}{Lemma}[section]
\newtheorem{proposition}{Proposition}

\usepackage{nicefrac}

\theoremstyle{definition}
\newtheorem{remark}{Remark}

\newtheorem{assumption}{Assumption}

\usepackage{subcaption}
\usepackage{caption}
\newcommand{\blam}{\bm{\lambda}}
\newcommand{\bzeta}{\bm{\zeta}}

\newcommand{\sH}{\mathsf{H}}
\newcommand{\cH}{\mathcal{H}}
\newcommand{\cS}{\mathcal{S}}
\newcommand{\cR}{\mathcal{R}}

\newcommand{\bzt}{\bm{\zeta}}

\newcommand{\bE}{\mathbb{E}}

\newcommand{\bR}{\mathbb{R}}
\newcommand{\bP}{\mathbb{P}}

\title{\bf{Matrix Inference in Growing Rank Regimes}}

\author{Farzad Pourkamali$^{\dagger,*}$, Jean Barbier$^{\ddagger}$,
Nicolas Macris$^{\dagger}$}
\affil{$\dagger$ \emph{School of Computer and Communication Science, Ecole Polytechnique F{\'e}d{\'e}rale de Lausanne}\\ 
$\ddagger$ \emph{The Abdus Salam International Center for Theoretical Physics, Trieste}}
\date{}

\begin{document}

\maketitle

{
\let\thefootnote\relax\footnotetext{ Authors contributed equally to this work. \\
\indent $*$ Corresponding author: \url{farzad.pourkamali@epfl.ch}}
}

\begin{abstract}
The inference of a large symmetric signal-matrix $\bS\in\mathbb{R}^{N\times N}$ corrupted by additive Gaussian noise, is considered for two regimes of growth of the rank $M$ as a function of $N$. For {\it sub-linear} ranks $M=\Theta(N^\alpha)$ with $\alpha\in(0,1)$ the mutual information and minimum mean-square error (MMSE) are derived for two classes of signal-matrices: (a) $\bS=\bX\bX^\intercal$ with entries of $\bX\in\mathbb{R}^{N\times M}$ independent identically distributed; (b) $\bS$ sampled from a rotationally invariant distribution. Surprisingly, the formulas match the rank-one case. Two efficient algorithms are explored and conjectured to saturate the MMSE when no statistical-to-computational gap is present: (1) Decimation Approximate Message Passing; (2) a spectral algorithm based on a Rotation Invariant Estimator. For {\it linear} ranks $M=\Theta(N)$ the mutual information is rigorously derived for signal-matrices from a rotationally invariant distribution.
Close connections with scalar inference in free probability are uncovered, which allow to deduce a simple formula for the MMSE as an integral involving the limiting spectral measure of the data matrix only. 
An interesting issue is whether the known information theoretic phase transitions for rank-one, and hence also sub-linear-rank, still persist in linear-rank. Our analysis suggests that only a smoothed-out trace of the transitions persists. Furthermore, the change of behavior between low and truly high-rank regimes only happens at the linear scale $\alpha=1$. 
\end{abstract}

\section{Introduction}
We consider the estimation of a large matrix from noisy observations. A real symmetric signal matrix $\bS \in \bR^{N \times N}$  is observed through an additive white Gaussian noise channel with output 
\begin{align}\label{observ-model}
    \bY = \sqrt{\gamma} \bS + \bZ
\end{align}
with $\bZ$ a symmetric Gaussian matrix and $\gamma > 0$ proportional to the signal-to-noise ratio (SNR). We adopt the point of view of Bayesian inference where the prior law of $\bS$ and the channel are known, and the task is to infer the signal $\bS$ from observations $\bY$.
This problem is known as matrix denoising and is part of many modern data analysis tasks, for example cleaning covariance matrices in statistics or finance \cite{bun2017cleaning}, signal and image analysis \cite{bouwmans2016handbook}, matrix recovery or completion \cite{donoho2013phase,wang2011denoising}. Matrix denoising is intimately connected to the more involved problem of matrix factorization, where one must estimate two matrices $\bX \in \bR^{N \times M}$ and $\bT \in \bR^{K \times M}$ (up to certain symmetries) from noisy observation of $\bX \bT^\intercal$. Many problems in signal processing and learning can be formulated as matrix factorization, for example sparse coding \cite{olshausen1996emergence,olshausen1997sparse}, blind source separation \cite{belouchrani1997blind}, robust principal component analysis \cite{candes2011robust}, interpretations of patterns in images \cite{lee1999learning} or also in genomics data \cite{kossenkov2010matrix}. Given this ubiquity both matrix denoising and factorization have gained much attention. But, their theoretical underpinnings still present great challenges.

Much progress has been made in the last decade when the matrix-signal $\bS$ is low-rank, meaning the rank is fixed as the matrix dimension grows. Fundamental information theoretical and  algorithmic limits, as well as phase transitions and computational-to-statistical gaps, are determined in various statistical settings, see \cite{korada2009exact, dia2016mutual, lelarge2019fundamental, pourkamali2021mismatched, barbier2022price, lesieur2015mmse, barbier2019adaptive, barbier2019adaptive-b, miolane2017fundamental, luneau2020high, pourkamali2022mismatched, montanari2022fundamental, guionnet2023estimating}. However,
when the matrix rank grows with the dimension, the problem remains very challenging and solid results are scarce. Here we consider regimes where the rank grows {\it sub-linearly} and {\it linearly} with matrix dimension. We first describe our main findings and then discuss how they compare with the literature. A global view of existing and new results is summarized in Table \ref{table:comparison}.



\subsection{Sub-linear rank regime} This corresponds to $M=\Theta(N^\alpha)$ with $N\to +\infty$ and $\alpha\in(0,1)$ fixed. We study two types of priors for the signal matrix: (a) $\bS=\bX\bX^\intercal$ with entries of $\bX\in\mathbb{R}^{N\times M}$ independent identically distributed; (b) $\bS$ sampled from a rotationally invariant distribution. 
Surprisingly we find that for {\it any} $\alpha\in (0,1)$ the statistical behavior matches the rank-one case (in particular this implies that there can be phase transitions). 
For model (a) a new version of the (non-rigorous) replica method \cite{barbier2021statistical} is used to derive the asymptotic mutual information between signal and observation matrices in the form of a low dimensional variational formula. The result coincides with the asymptotic mutual information of the rank-one case; see Statement \ref{statementSubLin}. For model (b) we adapt methods of \cite{bun2016rotational, bun2017cleaning} using so-called Rotation Invariant Estimators (RIE). These turn out to be Bayesian optimal estimators which can be used to compute the MMSE and hence the mutual information under Gaussian noise.
Two efficient algorithms are also discussed and conjectured to saturate the MMSE when no statistical-to-computational gap is present. For model (a) we introduce Decimation Approximate Message Passing inspired from \cite{camilli2022matrix}, while for model (b) a RIE estimator \cite{bun2016rotational, bun2017cleaning} adapted to the sub-linear regime provides a natural spectral algorithm.



\subsection{Linear rank regime} This corresponds to $M=\Theta(N)$ with $N\to +\infty$. In this regime, we restrict ourselves to rotationally invariant prior distributions for $\bS$. The mutual information (between signal and observation matrices) and MMSE are rigorously computed under natural assumptions. In this linear regime, the behavior of these quantities is qualitatively different from the low-rank case. We answer a previously open question about the continuity of MMSE as a function of SNR in the affirmative, which rules out the existence of first-order phase transitions in general.
However, it remains unclear if higher-order phase transitions might exist in specific examples. Based on a mix of numerical and analytical evidence on specific examples, we observe that only smoothed-out traces of phase transitions seem to persist. We also check that for $\bS = \frac{1}{N}\bX\bX^\intercal$ with $\bX\in \mathbb{R}^{N\times M}$ a Gaussian matrix and $M/N =\epsilon \to 0$ (taken after the infinite size limit) the MMSE curve tends to the rank-one MMSE of a Gaussian spiked model, thus recovering a second-order phase transition (MMSE continuous and first derivative discontinuous). This limiting behavior is ``compatible'' with the findings of the sub-linear rank regime.

We first prove Theorem \ref{theoremMain} relating
the asymptotic mutual information to the limit of a log-spherical-integral, known as an Harish Chandra-Itzykson-Zuber (HCIZ) integral in the mathematical physics literature \cite{harish1957differential, itzykson1980planar}. Computing the asymptotic behavior of HCIZ integrals is a notoriously difficult problem that we address indirectly here. 
Indeed we approach the problem from a completely independent perspective making use of the class of RIEs introduced in \cite{bun2016rotational}. The best estimator in this class is in fact optimal and equal to the minimum-mean-square estimator. Thus using the mean-square error of RIE, we are able to directly compute the MMSE; see Theorem \ref{statementMMSE}. Then, integrating an I-MMSE relation \cite{guo2005mutual}, we derive an explicit expression for the asymptotic mutual information; see Theorem \ref{statementMI}. As a bonus, comparing Theorems \ref{theoremMain} and \ref{statementMI} we directly find the asymptotic expression of the {\it particular} log-spherical integral occurring in the present setting.

\subsection*{Comparison with literature and discussion} Table \ref{table:comparison} summarizes existing results and our main new contributions. The relationship between mutual information and log-spherical integral appearing in Theorem \ref{theoremMain} was already found in \cite{barbier2021statistical} and \cite{maillard2021perturbative} via non rigorous derivations. In general the analysis of the asymptotic log-spherical integrals requires Matytsin's hydrodynamical formalism, and one must deal with an optimal transportation problem which is usually not exactly solvable, see \cite{schmidt2018statistical}, \cite{maillard2021perturbative}. However, in the present setting with additive Gaussian noise, it turns out an exact solution can also be computed by other means as shown in \cite{maillard2021perturbative}. Our derivation of the MMSE and mutual information in Theorems \ref{statementMMSE} and \ref{statementMI} proceeds by a different route, using the RIE, and in particular it entirely avoids the use of Matytsin's formalism. This can be viewed as a new calculation of the asymptotic log-spherical integral and it remains to be seen if the method can be extended beyond the additive Gaussian noise in order to deal with more general spherical integrals (as Matytsin's formalism does). The denoising problem of rectangular matrices was investigated in \cite{barbier2021statistical} and \cite{troiani2022optimal, pourkamali2023rectangular} for rotationally invariant priors, and an optimal denoiser is also proposed in \cite{pourkamali2023rectangular} for rotational invariant distributed noise matrices. 

For non-rotational invariant priors and linear rank regimes, the matrix denoising problem is wide open. There have been several attempts to derive the mutual information and MMSE for general i.i.d priors using the replica method from statistical mechanics \cite{kabashima2016phase}, \cite{barbier2021statistical}. However, this method involves largely heuristic ansatzes that have been very difficult to justify for large rank, and it is suspected that the proposed solutions to date are doubtful.

Given the difficulties encountered by the replica method for linear rank, we believe it is of great interest that the sub-linear rank case with i.i.d. priors seems to be amenable to this method. We note that this regime is also particularly relevant practically, as it captures the ``not-so-low rank'' regime of matrix inference which was recently understood to capture numerous applications \cite{lee2023statistical}.
Although we cannot prove Statement \ref{statementSubLin} there are indications that it is correct. The rank-one formula derived here for separable priors can be shown to be an upper bound to the true mutual information \cite{private-communication}. Also, for $\bX$ i.i.d Gaussian the signal matrix $\bS$ has a rotation invariant Wishart prior and we can check consistency with the MMSE obtained from the RIE and the formula for the mutual information recently proven in rotationally invariant sub-linear rank settings \cite{husson2022spherical}. 

Addressing whether these findings hold for whole regime $M = o(N)$ (such as $M = N/\log N$ or $M=N^{(1-\delta)}$ with 'small' $\delta$) is an open intriguing question. For rotationally invariant priors all sublinear regimes  $M = o(N)$ are rigorously treated in \cite{husson2022spherical} and the rank-one formula holds. Concerning the Wigner spiked model where $\bS = \bX \bX^\intercal$ with a separable prior our replica analysis remains inconclusive. Nonetheless, recent rigorous advancements have validated the replica prediction for $M = o(N^{\nicefrac{1}{10}})$ \cite{barbier2024multiscale}.

\begin{table*}[t]
\centering
\caption{Summary of known and new results on matrix inference, on the information-theoretic limits (IT) and optimal algorithms (Alg)}
\scriptsize
\begin{tabular}{lll}
            &       {\bf Separable priors}    &      {\bf Rotationally invariant priors} \\
\midrule
{\bf Finite rank}  &       (IT) Rigorous MI, MMSE formulas \cite{korada2009exact, dia2016mutual, lelarge2019fundamental, lesieur2015mmse, barbier2019adaptive, barbier2019adaptive-b, miolane2017fundamental}  &      (IT) Rigorous MI, MMSE formulas \cite{ luneau2020high,pourkamali2021mismatched, barbier2022price, pourkamali2022mismatched}\\
$M=\Theta(1)$ & (Alg) Approximate message-passing (AMP) \cite{deshpande2014information,rangan2012iterative,lesieur2017constrained,barbier2022bayes} & (Alg) Spectral PCA \cite{Baik2005,benaych2011eigenvalues}\\
\midrule
{\bf Sub-linear rank} & (IT) \underline{New:} Stat. \ref{statementSubLin}: rank-one MI and MMSE formulas        & (IT) Rigorous rank-one MI formula \cite{husson2022spherical}     \\
$1\ll M\ll N$ &Rigorous rank-one MI and MMSE formulas  & (Alg) \underline{New:} Sub-linear RIE, conjectured optimal\\
& for $M = o(N^{\nicefrac{1}{10}})$ \cite{barbier2024multiscale} & \\
& (Alg) \underline{New:} Decimation-AMP, conjectured optimal  &\\
\midrule
{\bf Linear rank}     & Open problem & (IT) \underline{New:} Thm. \ref{theoremMain}: relation between MI and log-spherical                                       integrals \\
$M=\Theta(N)$ &              & \underline{New:} Thm. \ref{statementMMSE} and \ref{statementMI}: explicit expressions of MMSE and MI via free probability \\ 
                &           &In \cite{maillard2021perturbative} expressions are derived via other means (see discussion) and in \cite{barbier2021statistical} \\ 
                && perturbative expansions are calculated\\
                & & (Alg) Rotational Invariant Estimator (RIE) \cite{bun2016rotational}\\
\midrule
\end{tabular}
\label{table:comparison}
\end{table*}
\normalsize
\section{Inference of matrices with growing rank: settings}

\subsection{Sub-linear rank matrix factorization and denoising} We start with the setting for matrix {\it factorization}. 
Let $\bX \in \bR^{N \times M}$, with $M=\lfloor N^\alpha\rfloor$ and $\alpha\in(0,1)$ independent of $N$, be drawn with i.i.d. entries according to a symmetric distribution $p_X(x) = p_X(-x)$ and variance $\mathbb{E}_X(x^2) = \rho$. The rank $M$ of the matrix $\bS(\bX)=\bX \bX^\intercal/N$ (to be recovered) is sub-linear compared to $N$. The statistician has access to a symmetric data matrix $\bY = {\sqrt\gamma} \bS(\bX) + \bZ$
where $\bZ\sim \exp(-\frac N{4}\Tr \bZ^2)$ is a standard Wigner matrix and $\gamma\in\mathbb{R}_+$ controls SNR. The Bayesian task here is to infer $\bS$ by exploiting the knowledge that it admits a decomposition $\bX\bX^\intercal$, with a separable prior for $\bX$.

We also consider symmetric signal matrices $\bS\in \mathbb{R}^{N\times N}$ distributed according to a rotation invariant prior supported on matrices with sublinear rank $M=\lfloor N^\alpha\rfloor$. Such signal instances can be realized 
by taking $\bS = \bV \bm{\Lambda} \bV^\intercal$ with $\bV\in \mathbb{R}^{N\times M}$ a matrix uniformly sampled from the Stieffel manifold of full column-rank matrices such that $\bV^\intercal \bV = {\bf I}_M$ and $\bm{\Lambda} = {\rm diag}(\lambda_1^S, \dots, \lambda^S_M)$ a set of random i.i.d. $M$ real eigenvalues. 
Again, the statistician has access to the observation matrix $\bY = \sqrt\gamma \bS + \bZ$. This setting corresponds to a \emph{denoising} problem, namely, the task of recovering a certain matrix $\bS$ which has no specific factorized structure, but with a prior on the law of $\bS$ itself. Note that if all of the eigenvalues are set to $1$, this model and the factorized one with Gaussian i.i.d. elements for $\bX$, are asymptotically equivalent \cite{jonsson1982some}.

For both models the main objects of interest are the mutual information $I_N(\bS;\bY)/(MN)$ between signal and data and the MMSE
\begin{equation*}
    {\rm MMSE}_N(\gamma)=\frac{1}{M} \bE \| \bS - \bE [ \bS \mid \bY ]\|_{\rm F}^2
\end{equation*}
when $N\to \infty$, both suitably scaled in order to have non-trivial limits. 

\subsection{Linear rank matrix denoising}

We consider $\bS = \bS^\intercal \in \bR^{N \times N}$ distributed according to a rotation invariant prior, i.e., $d P_{S,N}(\bO \bS \bO^\intercal) = d P_{S,N} (\bS)$ for any $N \times N$ orthogonal matrix $\bO$. The matrix $\bS$ is again corrupted by Gaussian Wigner noise $\bY = {\sqrt\gamma} \bS + \bZ$ and we are interested in the same information-theoretic quantities as in the sub-linear rank regime.
%

The empirical spectral distribution of $\bS$ is denoted as 
$\rho_S^{(N)}(x)dx = \frac{1}{N} \sum_{i\le N} \delta(x-\lambda_i^S)dx$ 
where $(\lambda^S_i)$ are the eigenvalues of $\bS$. For the rigorous analysis throughout the linear rank case we shall assume the following:

\begin{assumption}\label{weak-conv-emp} 
The empirical spectral distribution $\rho_S^{(N)}$ converges almost surely weakly to a well-defined probability measure $\rho_S$ with support in  $[-C, C]$ for some finite $C>0$ independent of $N$. Moreover, the second moment of $\rho_S^{(N)}$ is almost surely bounded.
\end{assumption}

Signal instances satisfying this assumption can be constructed as $\bS= \bO \bm{\Lambda} \bO^\intercal$ with $\bO$ uniformly sampled over the manifold of orthogonal matrices (or Haar distributed) and $\bm{\Lambda} = {\rm diag}(\lambda_1^S, \dots, \lambda_N^S)$ i.i.d. eigenvalues distributed according to $\rho_S$ with compact support. Note that for measures $\rho_S$ containing a weight $\epsilon\in [0, 1]$ at $0$ the random matrices $\bS$ have rank $M=(1-\epsilon)N$. When $M=N$ there is another popular way to construct rotation invariant matrix ensembles, namely by setting $dP_{S,N}( \bS ) \propto \exp(-\frac{N}{2} \Tr V (\bS ))d\bS$ where $V (\bS )$ is a rotation invariant ``matrix potential''. For such priors almost sure weak convergence of the empirical spectral distribution is proved in \cite{anderson2010introduction} whenever $\liminf_{|x| \to \infty} V(x)/(\beta \ln |x| ) > 1$ for some $\beta > 1$. Moreover, under some additional conditions, the largest eigenvalue of such a random matrix satisfies a large deviation principle, which implies the almost sure boundedness of the top eigenvalue \cite{anderson2010introduction}. Therefore, our assumption  holds for a large class of ensembles described by rotation invariant potentials. 




\section{Linear rank matrix denoising}
We slightly abuse notation and use $\mu(x)\, dx$ for $d\mu(x)$ even if the measure is not absolutely continuous. For a sequence of matrices $\bA_N\in \mathbb{R}^{N\times N}$, we denote the limiting empirical spectral measure by $\rho_A$. The free additive convolution \cite{voiculescu1991limit} of two probability distributions is $\rho_A\boxplus \rho_B$. 

For real symmetric matrices $\bA, \bB \in \bR^{N \times N}$, the spherical integral is defined as 
\begin{equation*}
    \mathcal{I}_N(\bA, \bB) := \int D \bU \exp {\frac{N}{2} \Tr \bA \bU  \bB \bU^\intercal }
\end{equation*}
where $D \bU$ is the Haar measure over the group of orthogonal matrices. The asymptotic behavior of HCIZ integrals in the symmetric case has been extensively studied in the literature, both in mathematics \cite{guionnet2002large} and in physics communities \cite{matytsin1994large}. The log-spherical integral is defined as $\mathcal{J}_N(\bA, \bB) := \frac{1}{N^2}\ln \mathcal{I}_N(\bA, \bB)$.




Let $\mathcal{J}[\rho_{\sqrt{\gamma} S}, \rho_{\sqrt{\gamma} S}\boxplus \rho_{\rm sc}] = \lim_{N\to +\infty} \mathcal{J}_N( \sqrt\gamma \bS, \bY)$
where $\rho_{\sqrt{\gamma} S}$ is the limiting spectral distribution of $\sqrt{\gamma} \bS$, and $\rho_{\rm sc}$ is the usual Wigner semi-circle distribution. Ref. \cite{guionnet2002large} proves that under assumption \ref{weak-conv-emp} the limit exists. Our first result for matrix denoising in the linear rank regime is a rigorous formula for the mutual information. 

\begin{theorem}[\textbf{Mutual Information for linear rank matrix denoising}]\label{theoremMain}
Under assumption \ref{weak-conv-emp},
\begin{equation}
    \frac{{I}_N(\bS; \bY)}{N^2}   \xrightarrow{N\to\infty} \frac{\gamma}{2} \int \! x^2 \rho_S(x) \, dx  - \mathcal{J}[\rho_{\sqrt{\gamma} S}, \rho_{\sqrt{\gamma} S}\boxplus \rho_{\rm sc}].
    \label{asymp-mI-th}
\end{equation}
\end{theorem}
\begin{proof}
    Proof steps are outlined in section \ref{Proof-steps-thm1}.
\end{proof}

Although we know that the limiting log-spherical integral is given by a variational problem, see \cite{guionnet2002large}, its computation requires going through Matytsin's formalism \cite{matytsin1994large} and is, in general, highly non-trivial. Here we will provide another formula (Theorem \ref{statementMI}) for the mutual information, which is much simpler and explicit and which, in turn, also provides an expression for the log-spherical integral. 

The route to this program goes first through the class of RIE. An estimator $\hat{\Xi}(\bY)$ is called \textit{rotation invariant} if for any orthogonal matrix $\bO$,  $\bO \hat{\Xi}(\bY) \bO^\intercal = \hat{\Xi}(\bO \bY \bO^\intercal)$. We may define the best possible reconstruction error {\it within the RIE class} as 
\begin{align}
{\rm MMSE}_{\rm RIE, N}(\gamma) = \min_{ \hat{\Xi}\in {\rm RIE} } \frac{1}{N} \bE  \| \bS -  \hat{\Xi}(\bY) \|_{\rm F}^2 .
\label{MMSERIE}
\end{align}
Obviously ${\rm MMSE}_N(\gamma) \leq {\rm MMSE}_{\rm RIE, N}(\gamma)$. However, it is easy to check explicitly that the MMSE estimator $\mathbb{E}[\bS\mid \bY]$ belongs to the RIE class. Thus we also have ${\rm MMSE}_{\rm RIE, N}(\gamma)\leq {\rm MMSE}_N(\gamma)$ and hence ${\rm MMSE}_N(\gamma) = {\rm MMSE}_{\rm RIE, N}(\gamma)$. 

Because of rotation invariance, $\bY$ and $\hat{\Xi}\in {\rm RIE}$ can be diagonalized in the same basis $(\by_i)$ (see appendix A in \cite{pourkamali2024bayesian}, or appendix B in \cite{semerjian2024matrix}). Thus any RIE is expressed as
$\hat{\Xi}(\bY) = \sum \hat{\xi}_i \by_i \by_i^\intercal$
where $(\hat{\xi}_i)$ are the eigenvalues of the estimator. Therefore \eqref{MMSERIE} requires minimizing over $(\hat{\xi}_i)$ only.
In \cite{bun2016rotational} the heuristic replica method is used to show that {\it in the large $N$ limit}, the optimal $(\hat{\xi}_i)$ can be expressed only in terms of the limiting spectral measure $\rho_Y$ of the data $\bY$ and its eigenvalues $( \lambda_i^Y )$ (see in particular Eq. IV.3 in \cite{bun2016rotational}):
\begin{equation}
    \hat{\Xi}^*(\bY) =  \sum_{i\le N} \xi^*_i \by_i \by_i^\intercal, \quad
    \xi^*_i = \frac{1}{\sqrt{\gamma}} \big( \lambda^Y_i - 2 \pi \sH [\rho_Y](\lambda_i^Y) \big),
    \label{RIE-est}
\end{equation}
where $\sH [\rho_Y](z):= {\rm P.V.} \frac{1}{\pi} \int \rho_Y(x)/(z-x)  dx$ is the \textit{Hilbert} transform of $\rho_Y$. Note that $\rho_Y$ is given by the free convolution $\rho_Y = \rho_{\sqrt\gamma S}\boxplus \rho_{\rm sc}$ which is a continuous density due to the smoothing effect of the semi-circle law \cite{biane1997free}. Based on this result we make the following assumption here:

\begin{assumption}\label{assumptionRIE}
The estimator \eqref{RIE-est} is asymptotically optimal in the RIE class, i.e., 
\begin{align}
{\rm MMSE}_{\rm RIE, N}(\gamma) = \frac{1}{N}\mathbb{E} \Vert \bS - \hat{\Xi}^*(\bY)\Vert_{\rm F}^2 + o_N(1).
\label{RIEasymp}
\end{align}
\end{assumption}

Using \eqref{RIE-est} and \eqref{RIEasymp} we prove the following explicit formula for the MMSE in linear rank matrix denoising:
\begin{theorem}[\textbf{MMSE for linear rank matrix denoising}]\label{statementMMSE}
 Under assumptions \ref{weak-conv-emp}, \ref{assumptionRIE} we have
\begin{equation}
     {\rm MMSE}_N(\gamma) \xrightarrow{N\to\infty}  \frac{1}{\gamma} \Big( 1 - \frac{4 \pi^2}{3} \int \rho_Y^3(x) \, dx \Big)   
    \label{asymp-MMSE-th}
\end{equation}
where the data spectral density $\rho_Y = \rho_{\sqrt{\gamma} S}\boxplus \rho_{\rm sc}$.
Moreover ${\rm MMSE}(\gamma) := \lim_{N\to\infty} {\rm MMSE}_N(\gamma)$ is continuous in $\gamma >0$.
\end{theorem}
\begin{proof}
    Section \ref{RIE-app}.
\end{proof}
This is an explicit formula that can be used to concretely compute the ${\rm MMSE}(\gamma)$ curves for various models of rotation invariant signal ensembles, and in particular, allows to investigate the existence and nature of phase transitions\footnote{Phase transitions are non-analyticity points in the asymptotic mutual information as a function of SNR. This is a concave and continuous function, and $k$-th order phase transitions correspond to discontinuities in the $k$-th derivative. In particular for a first order transition the MMSE is discontinuous because of the I-MMSE relation \eqref{I-MMSE-rel}.}. The continuity of ${\rm MMSE}(\gamma)$ guarantees that there is no first order phase transition. In low-rank matrix denoising (as well as other inference problems) when there is no first order phase transition (but possibly higher order continuous transitions) the model does not display an algorithmically ``hard phase'' for low complexity algorithms (e.g., message passing algorithms are optimal). The present linear-rank rotation invariant case is no exception to this picture.
Indeed equation \eqref{RIE-est} suggests an optimal spectral algorithm to estimate the signal: given an observation $\bY$ one computes its eigenvalues and an estimate of the Hilbert transform replacing the integral by an empirical sum to use in \eqref{RIE-est}. This algorithm is optimal since as remarked above ${\rm MMSE}_N(\gamma) = {\rm MMSE}_{\rm RIE, N}(\gamma)$. Note however that in the (sub-linear rank) \emph{factorized} model it is  not the spectral algorithms that are optimal (apart from the special case of Gaussian $\bX$ and $\bX\bX^\intercal$ Wishart), but rather the decimation-AMP we provide in this paper.
Finally, we also mention that the optimality of RIE was also discussed in \cite{maillard2021perturbative} in a different manner where the authors show heuristically that the posterior mean $\mathbb{E}[\bS\mid \bY]$ equals $\hat\Xi^*(\bY)$ as $N\to +\infty$, and also before in \cite{bun2017cleaning}.

We now proceed to deduce a simpler formula for the mutual information (than in Thm. \ref{theoremMain}) using the I-MMSE relation \cite{guo2005mutual} 
\begin{equation}
    {\rm MMSE}_N(\gamma) = 4 \frac{d}{d \gamma}  \frac{{I}_N(\bS;\bY)}{N^2} 
    \label{I-MMSE-rel}
\end{equation}
and free probability.
Using the concavity of the mutual information w.r.t. the SNR, \eqref{I-MMSE-rel} also holds as $N \to +\infty$. One can thus permute the limit $N\to +\infty$ with the derivative w.r.t. $\gamma$. Therefore, it suffices to compute the integral over the {\it asymptotic} MMSE to find the {\it asymptotic} mutual information; this is done using basic results from free probability leading to:

\begin{theorem}[\textbf{Explicit Mutual Information for linear rank matrix denoising}]\label{statementMI}
Let $\rho_Y= \rho_{\sqrt\gamma S} \boxplus \rho_{\rm sc}$. Under assumptions \ref{weak-conv-emp}, \ref{assumptionRIE} we have
\begin{equation}\label{simple-formula}
     \frac{{I}_N(\bS;\bY)}{N^2}  \xrightarrow{N\to\infty} \frac{1}{2} \iint \ln |s-t| \rho_Y(s) \rho_Y(t) \, ds \, dt + \frac{1}{8}.
\end{equation}
\end{theorem}
\begin{proof}
    Section \ref{Explicit-MI-MMSE}.
\end{proof}

In appendix \ref{proof-theorem-4}, we extend Theorem \ref{statementMI} to the case where the noise matrix is a realization of a non-Gaussian rotation invariant ensemble. While we are not quite able to treat this case, we can generalize Theorem 3 to the setting $\bY_\epsilon= \sqrt{\gamma}\bS + \bZ_\epsilon$ where $\bZ_{\epsilon}= \bZ + \sqrt\epsilon\bzeta$ with $\bzeta$ from the Gaussian Wigner ensemble, and $\epsilon>0$ (so the noise is non-Gaussian rotation invariant).


\section{Sub-linear rank matrix factorization and denoising}
Our first result for the sub-linear rank case says that, from the information-theoretic perspective, for matrix factorization with separable priors, sub-linear rank is equivalent to having a rank-one signal. 

\begin{statement}[\textbf{Replica prediction}]
\label{statementSubLin}
For the setting of sub-linear rank matrix factorization described above with an entry-wise prior for $\bX$ the mutual information matches the rank-one prediction, i.e., 
\begin{equation}
        \frac{I_N(\bS; \bY)}{M N}  \xrightarrow{N \to \infty} \inf_{\sigma\ge 0}\Big\{\frac\gamma 4(\sigma-\rho)^2+I(X;\sqrt{\gamma \sigma}X+Z)\Big\}
    \label{asymp-mI-subL}
\end{equation}
where $X\sim p_X$, $Z\sim\mathcal{N}(0,1)$ and $\rho$ is the variance of $p_X$. If the minimizer $\sigma_*$ in 
\eqref{asymp-mI-subL} is unique, the MMSE reads
\begin{equation}
    {\rm MMSE}_N(\gamma) \xrightarrow{N \to \infty}\rho^2-\sigma_*^2.
    \label{subL-MI}
\end{equation}
\end{statement}

\begin{proof}
    The derivation entails a new replica calculation and is presented in appendix \ref{replica-subL}.
\end{proof}

In the limit $M=\lfloor N^\alpha\rfloor$, $N\to +\infty$, this statement holds for any $\alpha\in[0,1)$. 
The collapse of asymptotic mutual information for all sub-linear rank regimes (including finite rank) to a single rank-one formula might seem implausible. However, similar behavior is known in random matrix theory problems \cite{huang2018mesoscopic,husson2022spherical} and the surprise here is that such a reduction is {\it also} valid in the framework of Bayesian inference. 


A rigorous validation of this reduction in a special case comes from the recent work \cite{husson2022spherical} which derived the asymptotic mutual information for sub-linear matrix denoising with rotationally invariant priors. For Wishart signals with $\bX$ Gaussian the two formulas should therefore coincide. This consistency check indeed works  and is presented in the section \ref{subL-RI-Gaussian}. 
We also provide in section \ref{subL-RIE-app} an independent check using a sub-linear RIE to rederive (along lines similar to the linear case) the information theoretic quantities for a few specific priors, and among them the Wishart signals.

We now discuss novel {\it algorithmic} approaches for the sub-linear regime: a decimation-AMP for the factorized model with separable priors and a spectral sub-linear RIE algorithm for denoising with rotational invariant priors {\it and} noise. 


\paragraph{Decimation-AMP algorithm.} We use the standard AMP algorithm for \emph{rank-one} matrix inference \cite{deshpande2014information,rangan2012iterative,lesieur2017constrained} applied together with a {\it decimation} procedure. The idea is to iteratively subtract from the data at each ``inference round'' the estimated rank-one spike. This is inspired from \cite{camilli2022matrix} where the authors look at linear ranks and instead employ an exponentially costly inference algorithm at each step. The surprise is that in the sub-linear case, using AMP ``one spike at a time'' works and yields an algorithm of complexity $\Theta(MN^2)$. Its pseudo-code is as follows: set $t=1$ and $\bY^{(1)} = \bY$, then

\begin{itemize}
\item[(i)] {\it Rank-one AMP}: Run standard AMP for rank-one matrices (with an optimal denoiser associated to $p_X$ if known, or else any other denoiser) until convergence, whose output is a vector $\hat \bx^{(t)}={\rm AMP}_{M=1}(\bY^{(t)})\in \mathbb{R}^N$.
\item[(ii)]{\it Decimation}: If $t< M$ set  $\bY^{(t+1)}=\bY^{(t)}-\sqrt{\gamma}\hat\bx^{(t)}(\hat \bx^{(t)})^\intercal$, $t=t+1$ and go to step (i); else go to step (iii).
\item[(iii)]{\it Estimator}: Output $\sum_{t\leq M}\hat\bx^{(t)}(\hat \bx^{(t)})^\intercal$ as estimator of $\bS(\bX)$. 
\end{itemize}



\paragraph{ Sub-linear RIE algorithm.} We construct a RIE applicable to rotation invariant signals {\it and} noises. This estimator is derived based on the results on sub-linear perturbations of random matrices \cite{huang2018mesoscopic},  details of which can be found in section \ref{subL-RIE-app}. This yields a spectral algorithm in the spirit of PCA which requires the data $\bY$ and knowledge of the noise distribution. Let $G_{\rho_Z}$ the Cauchy transform of the limiting spectral measure $\rho_Z$ and $G'_{\rho_Z}$ its derivative. Define the following \emph{thresholding function} (which is \emph{not} the same as the one employed in the linear rank regime \eqref{RIE-est})
$$f_Z(x) = - \frac1{\sqrt{\gamma}}\mathbb{I}(x\notin [a,b])\frac{G_{\rho_Z}(x)}{G'_{\rho_Z}(x)},$$ where 
$[a,b]$ is the support of $\rho_Z$. For example with Wigner noise 
$f_Z(x) = \gamma^{-1/2}\mathbb{I}(|x| > 2){\rm sign}(x)\sqrt{x^2 - 4}$.
Algorithmic steps are:

\begin{enumerate}
\item[(i)] 
{\it Spectral step:} Compute eigen-data $(\lambda_i^Y, \by_i)_{i\leq N}$ from $\bY$.
\item[(ii)]
{\it Thresholded estimator:} Output $\hat{\bS} = \sum_{i \leq N} f_Z(\lambda_i^Y) \by_i \by_i^\intercal$.
\end{enumerate}





We conjecture that these two algorithms are Bayes-optimal in the sense that their mean square error matches the MMSE, for any regime $M=\lfloor N^\alpha\rfloor$ with $\alpha\in(0,1)$ when $N$ is large. This is supported by the theoretical checks for the Wishart priors and also, at least for some range of exponents, by the numerical results presented in Sec. \ref{numerics}. Our numerics do not yet allow to pinpoint whether the conjecture holds 
for any $\alpha<1$ or only up to some maximal exponent.




\section{Numerical examples and discussion}\label{numerics}

\subsection{Performance of algorithms in sub-linear rank regimes}
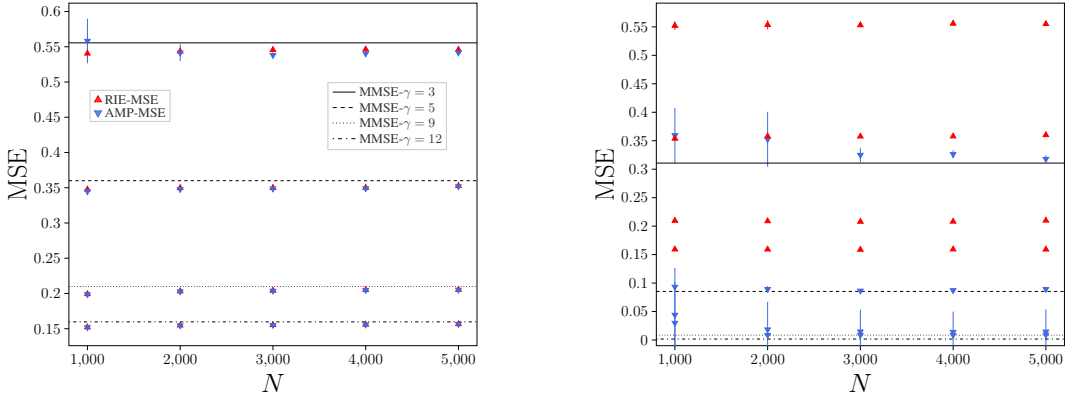
\begin{figure*}
   \captionsetup{singlelinecheck = false, justification=justified}
    \centering
    \begin{subfigure}[t]{.4\textwidth}
    \centering
\begin{tikzpicture}[scale=0.4]

\definecolor{darkgray176}{RGB}{176,176,176}
\definecolor{royalblue}{RGB}{65,105,225}

\begin{axis}[
tick align=outside,
tick pos=left,
x grid style={darkgray176},
xlabel={$N$},
xmin=800, xmax=5200,
xtick={1000,2000,3000,4000,5000},
xtick style={color=black},
y grid style={darkgray176},
ylabel={${\rm MSE}$},
ymin=0.125925824573152, ymax=0.611899810407825,
ytick style={color=black},
label style={font=\Huge},
tick label style={font=\Large}
]
\path [draw=red, semithick]
(axis cs:1000,0.531942807082658)
--(axis cs:1000,0.54907749538017);

\path [draw=red, semithick]
(axis cs:2000,0.538893234267244)
--(axis cs:2000,0.547866040798572);

\path [draw=red, semithick]
(axis cs:3000,0.541887529864766)
--(axis cs:3000,0.549414784985427);

\path [draw=red, semithick]
(axis cs:4000,0.542854541984648)
--(axis cs:4000,0.549737074703659);

\path [draw=red, semithick]
(axis cs:5000,0.543732041929107)
--(axis cs:5000,0.548034264620726);

\path [draw=royalblue, semithick]
(axis cs:1000,0.526608940395673)
--(axis cs:1000,0.589810083778976);

\path [draw=royalblue, semithick]
(axis cs:2000,0.52950649999482)
--(axis cs:2000,0.553488868694465);

\path [draw=royalblue, semithick]
(axis cs:3000,0.534530890760836)
--(axis cs:3000,0.54158618909279);

\path [draw=royalblue, semithick]
(axis cs:4000,0.537236120417101)
--(axis cs:4000,0.543346081505517);

\path [draw=royalblue, semithick]
(axis cs:5000,0.539703772681938)
--(axis cs:5000,0.543661320731805);

\addplot [semithick, black]
table {%
800 0.555555555555556
5200 0.555555555555556
};
\label{M3}
\path [draw=red, semithick]
(axis cs:1000,0.342797842612343)
--(axis cs:1000,0.352542025510306);

\path [draw=red, semithick]
(axis cs:2000,0.345224315014111)
--(axis cs:2000,0.354188181194475);

\path [draw=red, semithick]
(axis cs:3000,0.347698718304599)
--(axis cs:3000,0.352146745744484);

\path [draw=red, semithick]
(axis cs:4000,0.348850498459466)
--(axis cs:4000,0.351320782167958);

\path [draw=red, semithick]
(axis cs:5000,0.35142414663706)
--(axis cs:5000,0.35398738041229);

\path [draw=royalblue, semithick]
(axis cs:1000,0.339808948620054)
--(axis cs:1000,0.349811520638276);

\path [draw=royalblue, semithick]
(axis cs:2000,0.343527965300066)
--(axis cs:2000,0.352565240295425);

\path [draw=royalblue, semithick]
(axis cs:3000,0.346444469845065)
--(axis cs:3000,0.350820872087022);

\path [draw=royalblue, semithick]
(axis cs:4000,0.347765147299647)
--(axis cs:4000,0.350312267140333);

\path [draw=royalblue, semithick]
(axis cs:5000,0.350539596181103)
--(axis cs:5000,0.353058145810644);

\addplot [semithick, black, dashed]
table {%
800 0.36
5200 0.36
};
\label{M5}
\path [draw=red, semithick]
(axis cs:1000,0.197463578921604)
--(axis cs:1000,0.200959411718094);

\path [draw=red, semithick]
(axis cs:2000,0.201388626109714)
--(axis cs:2000,0.205007137259114);

\path [draw=red, semithick]
(axis cs:3000,0.2022752829274)
--(axis cs:3000,0.205894667348016);

\path [draw=red, semithick]
(axis cs:4000,0.203960442718274)
--(axis cs:4000,0.20602632868468);

\path [draw=red, semithick]
(axis cs:5000,0.204594147402821)
--(axis cs:5000,0.206200259710155);

\path [draw=royalblue, semithick]
(axis cs:1000,0.197013620715156)
--(axis cs:1000,0.200350799775291);

\path [draw=royalblue, semithick]
(axis cs:2000,0.200970165719881)
--(axis cs:2000,0.204583538765765);

\path [draw=royalblue, semithick]
(axis cs:3000,0.20199649174468)
--(axis cs:3000,0.205577178720098);

\path [draw=royalblue, semithick]
(axis cs:4000,0.203712170395378)
--(axis cs:4000,0.205784166083983);

\path [draw=royalblue, semithick]
(axis cs:5000,0.204356850424381)
--(axis cs:5000,0.205970927133416);

\addplot [semithick, black, dotted]
table {%
800 0.209876543209877
5200 0.209876543209877
};
\label{M9}
\path [draw=red, semithick]
(axis cs:1000,0.148302900292878)
--(axis cs:1000,0.156033602469023);

\path [draw=red, semithick]
(axis cs:2000,0.153037811268185)
--(axis cs:2000,0.156453297235568);

\path [draw=red, semithick]
(axis cs:3000,0.154438769108457)
--(axis cs:3000,0.156212262216674);

\path [draw=red, semithick]
(axis cs:4000,0.155276884231215)
--(axis cs:4000,0.156645185864022);

\path [draw=red, semithick]
(axis cs:5000,0.155642663788255)
--(axis cs:5000,0.157633608089433);

\path [draw=royalblue, semithick]
(axis cs:1000,0.148015551202001)
--(axis cs:1000,0.15564672011488);

\path [draw=royalblue, semithick]
(axis cs:2000,0.152806580385892)
--(axis cs:2000,0.15626332284114);

\path [draw=royalblue, semithick]
(axis cs:3000,0.154278995452682)
--(axis cs:3000,0.156061249449794);

\path [draw=royalblue, semithick]
(axis cs:4000,0.15512930859835)
--(axis cs:4000,0.156506284083537);

\path [draw=royalblue, semithick]
(axis cs:5000,0.155512018453909)
--(axis cs:5000,0.157512804594729);

\addplot [semithick, black, dash pattern=on 1pt off 3pt on 3pt off 3pt]
table {%
800 0.159722222222222
5200 0.159722222222222
};
\label{M12}
\addplot [semithick, red, mark=triangle*, mark size=3, mark options={solid}, only marks]
table {%
1000 0.540510151231414
2000 0.543379637532908
3000 0.545651157425096
4000 0.546295808344153
5000 0.545883153274917
};
\label{RIE}
\addplot [semithick, royalblue, mark=triangle*, mark size=3, mark options={solid,rotate=180}, only marks]
table {%
1000 0.558209512087324
2000 0.541497684344642
3000 0.538058539926813
4000 0.540291100961309
5000 0.541682546706872
};
\label{AMP}
\addplot [semithick, red, mark=triangle*, mark size=3, mark options={solid}, only marks]
table {%
1000 0.347669934061324
2000 0.349706248104293
3000 0.349922732024542
4000 0.350085640313712
5000 0.352705763524675
};
\addplot [semithick, royalblue, mark=triangle*, mark size=3, mark options={solid,rotate=180}, only marks]
table {%
1000 0.344810234629165
2000 0.348046602797746
3000 0.348632670966044
4000 0.34903870721999
5000 0.351798870995874
};
\addplot [semithick, red, mark=triangle*, mark size=3, mark options={solid}, only marks]
table {%
1000 0.199211495319849
2000 0.203197881684414
3000 0.204084975137708
4000 0.204993385701477
5000 0.205397203556488
};
\addplot [semithick, royalblue, mark=triangle*, mark size=3, mark options={solid,rotate=180}, only marks]
table {%
1000 0.198682210245224
2000 0.202776852242823
3000 0.203786835232389
4000 0.20474816823968
5000 0.205163888778899
};
\addplot [semithick, red, mark=triangle*, mark size=3, mark options={solid}, only marks]
table {%
1000 0.152168251380951
2000 0.154745554251876
3000 0.155325515662566
4000 0.155961035047618
5000 0.156638135938844
};
\addplot [semithick, royalblue, mark=triangle*, mark size=3, mark options={solid,rotate=180}, only marks]
table {%
1000 0.151831135658441
2000 0.154534951613516
3000 0.155170122451238
4000 0.155817796340944
5000 0.156512411524319
};
\end{axis}

\node [scale=0.5,fill opacity=0.8, draw opacity=1,draw=white!80!black] at (rel axis cs: 0.78,0.67) {\shortstack[l]{
\ref*{M3} MMSE-$\gamma = 3$ \\
\ref*{M5} MMSE-$\gamma = 5$ \\
\ref*{M9} MMSE-$\gamma = 9$ \\
\ref*{M12} MMSE-$\gamma = 12$}};

\node [scale=0.5,fill opacity=0.8, draw opacity=1,draw=white!80!black] at (rel axis cs: 0.15,0.7) {\shortstack[l]{
\ref*{RIE} RIE-MSE \\
\ref*{AMP} AMP-MSE}};

\end{tikzpicture}
    \label{fig:subL-Gauss}
  \end{subfigure}
  \hfil
  \begin{subfigure}[t]{.4\textwidth}
  \centering
\begin{tikzpicture}[scale=0.4]

\definecolor{darkgray176}{RGB}{176,176,176}
\definecolor{royalblue}{RGB}{65,105,225}

\begin{axis}[
legend cell align={left},
legend style={at={(0.98,0.65)},fill opacity=0.8, draw opacity=1, text opacity=1, draw=white!80!black},
tick align=outside,
tick pos=left,
x grid style={darkgray176},
xlabel={$N$},
xmin=800, xmax=5200,
xtick={1000,2000,3000,4000,5000},
xtick style={color=black},
y grid style={darkgray176},
ylabel={${\rm MSE}$},
ymin=-0.01, ymax=0.59167085155236,
ytick={0, 0.05, 0.1, 0.15, 0.2, 0.25, 0.3, 0.35, 0.4, 0.45, 0.5, 0.55, 0.6},
ytick style={color=black},
yticklabel style={
  /pgf/number format/precision=3,
  /pgf/number format/fixed},
label style={font=\Huge},
tick label style={font=\Large}
]
\path [draw=red, semithick]
(axis cs:1000,0.544767962003299)
--(axis cs:1000,0.559358342590527);

\path [draw=red, semithick]
(axis cs:2000,0.545640319299248)
--(axis cs:2000,0.561675829257291);

\path [draw=red, semithick]
(axis cs:3000,0.549165396787635)
--(axis cs:3000,0.556974267950701);

\path [draw=red, semithick]
(axis cs:4000,0.552212999029006)
--(axis cs:4000,0.559328524800965);

\path [draw=red, semithick]
(axis cs:5000,0.552235823575681)
--(axis cs:5000,0.558099168747154);

\path [draw=royalblue, semithick]
(axis cs:1000,0.311303870022614)
--(axis cs:1000,0.407710869226179);

\path [draw=royalblue, semithick]
(axis cs:2000,0.304341307648487)
--(axis cs:2000,0.400842610568218);

\path [draw=royalblue, semithick]
(axis cs:3000,0.312480340619585)
--(axis cs:3000,0.337594409878459);

\path [draw=royalblue, semithick]
(axis cs:4000,0.319562115651179)
--(axis cs:4000,0.333074847458522);

\path [draw=royalblue, semithick]
(axis cs:5000,0.311675227487298)
--(axis cs:5000,0.324211717469299);

\addplot [semithick, black]
table {%
800 0.31071482039357
5200 0.31071482039357
};

\path [draw=red, semithick]
(axis cs:1000,0.349484289535987)
--(axis cs:1000,0.358205687584852);

\path [draw=red, semithick]
(axis cs:2000,0.354356943496024)
--(axis cs:2000,0.361356591740229);

\path [draw=red, semithick]
(axis cs:3000,0.355254761607332)
--(axis cs:3000,0.359995251839879);

\path [draw=red, semithick]
(axis cs:4000,0.355189975941876)
--(axis cs:4000,0.360243837448462);

\path [draw=red, semithick]
(axis cs:5000,0.358309368143645)
--(axis cs:5000,0.361819605385709);

\path [draw=royalblue, semithick]
(axis cs:1000,0.084409764468622)
--(axis cs:1000,0.102048534916966);

\path [draw=royalblue, semithick]
(axis cs:2000,0.0837507874240893)
--(axis cs:2000,0.0948460561707227);

\path [draw=royalblue, semithick]
(axis cs:3000,0.0841575783216559)
--(axis cs:3000,0.0888188857123245);

\path [draw=royalblue, semithick]
(axis cs:4000,0.0842769245598083)
--(axis cs:4000,0.090278109660945);

\path [draw=royalblue, semithick]
(axis cs:5000,0.0877527273211987)
--(axis cs:5000,0.0908590509527508);

\addplot [semithick, black, dashed]
table {%
800 0.0852682960908877
5200 0.0852682960908877
};

\path [draw=red, semithick]
(axis cs:1000,0.206980498175587)
--(axis cs:1000,0.212071496009404);

\path [draw=red, semithick]
(axis cs:2000,0.207913431086993)
--(axis cs:2000,0.209798950161093);

\path [draw=red, semithick]
(axis cs:3000,0.207342494346791)
--(axis cs:3000,0.208334785729066);

\path [draw=red, semithick]
(axis cs:4000,0.206750863237333)
--(axis cs:4000,0.209349340108008);

\path [draw=red, semithick]
(axis cs:5000,0.208378905200612)
--(axis cs:5000,0.211371984152008);

\path [draw=royalblue, semithick]
(axis cs:1000,-0.0301586167057009)
--(axis cs:1000,0.0900473262033087);

\path [draw=royalblue, semithick]
(axis cs:2000,0.00780894192727244)
--(axis cs:2000,0.00949288124708628);

\path [draw=royalblue, semithick]
(axis cs:3000,0.00770406631510352)
--(axis cs:3000,0.00973609838732606);

\path [draw=royalblue, semithick]
(axis cs:4000,0.00788471638507698)
--(axis cs:4000,0.00947615506039042);

\path [draw=royalblue, semithick]
(axis cs:5000,0.00816331230445205)
--(axis cs:5000,0.0100400425066161);

\addplot [semithick, black, dotted]
table {%
800 0.00845867665246336
5200 0.00845867665246336
};

\path [draw=red, semithick]
(axis cs:1000,0.156761395425748)
--(axis cs:1000,0.161511062842144);

\path [draw=red, semithick]
(axis cs:2000,0.156868263377187)
--(axis cs:2000,0.161404023768954);

\path [draw=red, semithick]
(axis cs:3000,0.157080448386528)
--(axis cs:3000,0.160299183681149);

\path [draw=red, semithick]
(axis cs:4000,0.158433498782986)
--(axis cs:4000,0.160086012505633);

\path [draw=red, semithick]
(axis cs:5000,0.158548688662926)
--(axis cs:5000,0.159525947081513);

\path [draw=royalblue, semithick]
(axis cs:1000,-0.0382246166440896)
--(axis cs:1000,0.126436500749027);

\path [draw=royalblue, semithick]
(axis cs:2000,-0.0303075280280264)
--(axis cs:2000,0.0670279769073656);

\path [draw=royalblue, semithick]
(axis cs:3000,-0.0237940281427627)
--(axis cs:3000,0.0533164012935136);

\path [draw=royalblue, semithick]
(axis cs:4000,-0.0217670399506276)
--(axis cs:4000,0.0497665455078402);

\path [draw=royalblue, semithick]
(axis cs:5000,-0.0243405272630915)
--(axis cs:5000,0.0536212969160579);

\addplot [semithick, black, dash pattern=on 1pt off 3pt on 3pt off 3pt]
table {%
800 0.00165173240094618
5200 0.00165173240094618
};
\addplot [semithick, red, mark=triangle*, mark size=3, mark options={solid}, only marks]
table {%
1000 0.552063152296913
2000 0.55365807427827
3000 0.553069832369168
4000 0.555770761914985
5000 0.555167496161418
};
\addplot [semithick, royalblue, mark=triangle*, mark size=3, mark options={solid,rotate=180}, only marks]
table {%
1000 0.359507369624397
2000 0.352591959108353
3000 0.325037375249022
4000 0.32631848155485
5000 0.317943472478298
};
\addplot [semithick, red, mark=triangle*, mark size=3, mark options={solid}, only marks]
table {%
1000 0.353844988560419
2000 0.357856767618126
3000 0.357625006723606
4000 0.357716906695169
5000 0.360064486764677
};
\addplot [semithick, royalblue, mark=triangle*, mark size=3, mark options={solid,rotate=180}, only marks]
table {%
1000 0.0932291496927938
2000 0.089298421797406
3000 0.0864882320169902
4000 0.0872775171103767
5000 0.0893058891369748
};
\addplot [semithick, red, mark=triangle*, mark size=3, mark options={solid}, only marks]
table {%
1000 0.209525997092496
2000 0.208856190624043
3000 0.207838640037929
4000 0.208050101672671
5000 0.20987544467631
};
\addplot [semithick, royalblue, mark=triangle*, mark size=3, mark options={solid,rotate=180}, only marks]
table {%
1000 0.0299443547488039
2000 0.00865091158717936
3000 0.00872008235121479
4000 0.0086804357227337
5000 0.00910167740553407
};
\addplot [semithick, red, mark=triangle*, mark size=3, mark options={solid}, only marks]
table {%
1000 0.159136229133946
2000 0.159136143573071
3000 0.158689816033838
4000 0.159259755644309
5000 0.15903731787222
};
\addplot [semithick, royalblue, mark=triangle*, mark size=3, mark options={solid,rotate=180}, only marks]
table {%
1000 0.0441059420524685
2000 0.0183602244396696
3000 0.0147611865753755
4000 0.0139997527786063
5000 0.0146403848264832
};

\end{axis}

\end{tikzpicture}
    \label{fig:subL-Rad}
  \end{subfigure}
   \captionsetup{singlelinecheck = false, justification=justified}
   \caption{\small Comparison of the MSE reached by our proposed Sub-linear RIE (red) and Decimation AMP (blue) algorithms for sub-linear matrix inference as a function of the size $N$ for various SNR $\gamma$, compared to the rank-one MMSE (Statement \ref{statementSubLin}). Each point represents the average over 10 experiments, with error bars indicating 1 standard deviation. (left) Signal with Gaussian spikes, $M = \lfloor N^{0.5} \rfloor$. (right) Signal with Rademacher spikes, $M = \lfloor N^{0.3} \rfloor$. }
  \label{fig:sub-Linear-MMSE}
\end{figure*}

Fig.~\ref{fig:sub-Linear-MMSE} illustrates the performance of the proposed algorithms for two examples in the sub-linear rank regime. The left plot corresponds to a Wishart matrix signal $\bX\bX^\intercal$ where $\bX \in \bR^{N\times M}$ has i.i.d. $\mathcal{N}(0,1)$ entries. Thus $\bS_N(\bX)$ has a rotationally invariant distribution, and we observe that, as expected, both sub-linear RIE and decimation-AMP with MMSE denoiser achieve the MMSE (predicted by the rank-one formula). The right plot corresponds to a Rademacher matrix: $\bX \in \bR^{N\times M}$ has i.i.d. $\pm 1$ entries with equal probability. The regime of ranks we can attain without particular optimization and for all instances is smaller for Rademacher signals ($M = \lfloor N^{0.3} \rfloor$) than for Gaussian ($M = \lfloor N^{0.5} \rfloor$); this higher level of difficulty may be inherent to the Rademacher setting. Indeed, after relating matrix factorization to the Hopfield model \cite{camilli2022matrix} this is not suprising as it is known that continuous variables allow for greater recovery properties than binary variables \cite{lucibello2023exponential}. Note also that since the Rademacher prior is not rotationally invariant, the sub-linear RIE algorithm is sub-optimal for that case and decimation-AMP is needed.

\subsection{MMSE in linear rank regimes} As first example in the linear rank regime, we consider the case where $\rho_S = \frac{1}{2} \delta_{-1} + \frac{1}{2} \delta_{+1}$. Using the technique introduced in \cite{biane1997free}, we obtain an explicit analytical expression for $\rho_Y = \rho_{\sqrt{\gamma} S}\boxplus \rho_{\rm SC}$. For $\gamma \geq 1$ the support of $\rho_Y$ consists of two disjoint intervals, and for $\gamma < 1$ we get a single interval. Therefore, we expect that, if a phase transition in the mutual information and MMSE exists, it should happen at $\gamma_c=1$. As noted before, because of Theorem~\ref{statementMMSE} we know that ${\rm MMSE}(\gamma)$ is continuous, so a phase transition can only be second or higher order. Furthermore, from the explicit formula for $\rho_Y$ we get integral representations of the first few derivatives of ${\rm MMSE}(\gamma)$. Fig. \ref{fig: MMSE-Rad-analys} displays the results of precise numerical integrations for the MMSE and its first two derivatives, while the third derivative is computed by numerically differentiating the second derivative. These plots suggest that there is $4$-th order phase transition for this example model at $\gamma_c=1$. All details and additional figures can be found in the Sec. 7 of the SI.

In a second example we consider $\bS=\bX \bX^\intercal/N$ where $\bX \in \bR^{N \times M}$ has i.i.d. standard Gaussian entries. The limiting spectral distribution of $\bS$ when the aspect ratio $N/M \to q$ (fixed) is the \textit{Marchenko-Pastur} law. For this model it is not difficult to directly compute $\rho_Y$ (see apendix \ref{Ex-3}). Its support is a single interval for $q \leq 1$, and two disjoint intervals for $q > 1$ and $\gamma > q (q^{1/3} -1)^{-3}$. However, when the intervals merge in this case there does not seem to exist a phase transition, at least on low order derivatives (investigated numerically). Fig.~\ref{fig:linear-MMSE} (left) shows the MMSE as a function of $\log \gamma$.

\begin{figure*}
  \begin{subfigure}[t]{.18\textwidth}
    \centering
    \input{Main-paper/Figures/Rademacher/MSE_Rademach}
    \caption{MMSE \& MSE of RIE}
    \label{fig:Th-MMSE & RIE-MSE}
  \end{subfigure}
  \hfill
  \begin{subfigure}[t]{.18\textwidth}
  \centering
\begin{tikzpicture}[scale=0.2]

\definecolor{color0}{rgb}{0.12156862745098,0.466666666666667,0.705882352941177}

\begin{axis}[
tick align=outside,
tick pos=left,
x grid style={white!69.0196078431373!black},
xlabel={$\gamma$},
xmin=-0.0895, xmax=2.0995,
xtick style={color=black},
y grid style={white!69.0196078431373!black},
ymin=-1.02422657234277, ymax=-0.0639362725565435,
ytick style={color=black},
label style={font=\fontsize{52}{58}},
tick label style={font=\fontsize{20}{29}}
]
\addplot [ultra thick, color0]
table {%
0.01 -0.980577013261581
0.02 -0.962224004733308
0.03 -0.944829526655763
0.04 -0.928298971625107
0.05 -0.912551102263986
0.06 -0.897515707976817
0.07 -0.883131603898002
0.08 -0.869345161520686
0.09 -0.856109091430111
0.1 -0.843381468366843
0.11 -0.831124942968393
0.12 -0.819306095502232
0.13 -0.807894895150833
0.14 -0.79686425502466
0.15 -0.786189656015376
0.16 -0.775848830293858
0.17 -0.765821492533053
0.18 -0.756089108950357
0.19 -0.746634699664751
0.2 -0.737442670766927
0.21 -0.728498664320263
0.22 -0.719789431866388
0.23 -0.711302721355168
0.24 -0.7030271796138
0.25 -0.694952265447632
0.26 -0.687068174041136
0.27 -0.679365768849003
0.28 -0.67183652197056
0.29 -0.664472460546225
0.3 -0.657266119375222
0.31 -0.650210498107525
0.32 -0.643299022645692
0.33 -0.636525511154056
0.34 -0.629884142550942
0.35 -0.62336942862471
0.36 -0.616976188446309
0.37 -0.610699525471559
0.38 -0.60453480621549
0.39 -0.598477641429888
0.4 -0.592523868424496
0.41 -0.586669535232275
0.42 -0.580910885780934
0.43 -0.575244346685516
0.44 -0.569666514664815
0.45 -0.564174145395255
0.46 -0.558764142851221
0.47 -0.553433549672268
0.48 -0.548179538209861
0.49 -0.542999402209317
0.5 -0.537890549093608
0.51 -0.532850492672337
0.52 -0.52787684663607
0.53 -0.52296731816014
0.54 -0.518119702142741
0.55 -0.513331875627184
0.56 -0.508601792810163
0.57 -0.503927480032418
0.58 -0.49930703127969
0.59 -0.494738603858243
0.6 -0.490220414312868
0.61 -0.485750734428843
0.62 -0.481327887577674
0.63 -0.476950245085949
0.64 -0.472616222820945
0.65 -0.468324277774981
0.66 -0.464072905029667
0.67 -0.459860634293041
0.68 -0.455686026967142
0.69 -0.451547673022319
0.7 -0.447444187904606
0.71 -0.443374209422255
0.72 -0.439336394630809
0.73 -0.435329416607288
0.74 -0.43135196106334
0.75 -0.427402722910533
0.76 -0.423480402817671
0.77 -0.419583702916212
0.78 -0.415711322833699
0.79 -0.411861955048356
0.8 -0.408034279797419
0.81 -0.404226959404358
0.82 -0.400438631857791
0.83 -0.396667903311579
0.84 -0.392913339679952
0.85 -0.389173456745505
0.86 -0.385446707638323
0.87 -0.381731468545628
0.88 -0.378026021031136
0.89 -0.37432852884998
0.9 -0.37063700971249
0.91 -0.366949297852176
0.92 -0.363262993609353
0.93 -0.359575393592293
0.94 -0.355883390789458
0.95 -0.352183325233784
0.96 -0.348470744594035
0.97 -0.344739987340984
0.98 -0.340983366447204
0.99 -0.337189181962098
1 -0.333333333332004
1.01 -0.329372272858879
1.02 -0.325363846032007
1.03 -0.321340404362144
1.04 -0.317319271383051
1.05 -0.313312186996761
1.06 -0.309327762177757
1.07 -0.305372569076412
1.08 -0.301451736228911
1.09 -0.297569315098506
1.1 -0.293728524425683
1.11 -0.289931922493849
1.12 -0.286181533808963
1.13 -0.282478945289661
1.14 -0.278825381129142
1.15 -0.275221762255209
1.16 -0.271668754361758
1.17 -0.268166807215216
1.18 -0.264716187184101
1.19 -0.261317004418437
1.2 -0.257969235752845
1.21 -0.254672744115954
1.22 -0.251427295101638
1.23 -0.248232571155194
1.24 -0.245088183811384
1.25 -0.241993684263658
1.26 -0.238948572504994
1.27 -0.235952305304392
1.28 -0.233004303140768
1.29 -0.230103956256073
1.3 -0.227250629960568
1.31 -0.224443669251763
1.32 -0.221682402891893
1.33 -0.218966146970432
1.34 -0.216294208014061
1.35 -0.21366588574415
1.36 -0.21108047548014
1.37 -0.20853727023958
1.38 -0.206035562589316
1.39 -0.20357464626608
1.4 -0.201153817581655
1.41 -0.198772376659565
1.42 -0.19642962851279
1.43 -0.194124883948154
1.44 -0.19185746040037
1.45 -0.189626682601437
1.46 -0.187431883175695
1.47 -0.185272403149249
1.48 -0.183147592365531
1.49 -0.181056809839477
1.5 -0.178999424052688
1.51 -0.176974813186133
1.52 -0.174982365302126
1.53 -0.173021478494129
1.54 -0.171091560982481
1.55 -0.169192031190222
1.56 -0.16732231777757
1.57 -0.165481859651248
1.58 -0.163670105953317
1.59 -0.161886516033867
1.6 -0.160130559382971
1.61 -0.158401715578728
1.62 -0.156699474198737
1.63 -0.155023334719273
1.64 -0.153372806414886
1.65 -0.151747408243307
1.66 -0.150146668722757
1.67 -0.148570125797039
1.68 -0.147017326705102
1.69 -0.145487827842044
1.7 -0.143981194605547
1.71 -0.142497001260522
1.72 -0.141034830781813
1.73 -0.139594274704106
1.74 -0.138174932973558
1.75 -0.1367764137809
1.76 -0.135398333427616
1.77 -0.134040316152536
1.78 -0.13270199399098
1.79 -0.131383006614417
1.8 -0.130083001180775
1.81 -0.128801632184384
1.82 -0.127538561310128
1.83 -0.126293457270478
1.84 -0.12506599568056
1.85 -0.123855858894505
1.86 -0.122662735873667
1.87 -0.121486322042888
1.88 -0.120326319153458
1.89 -0.119182435142466
1.9 -0.118054384003504
1.91 -0.11694188565478
1.92 -0.11584466580343
1.93 -0.114762455830035
1.94 -0.113694992651001
1.95 -0.11264201860387
1.96 -0.111603281327196
1.97 -0.110578533641026
1.98 -0.109567533429054
1.99 -0.108570043533446
2 -0.107585831637736
};
\end{axis}

\end{tikzpicture}%
    \vspace{1pt}
    \caption{{\small$\frac{d}{d \gamma} {\rm MMSE}(\gamma)$}}
  \end{subfigure}
  \hfill
  \begin{subfigure}[t]{.18\textwidth}
    \centering
    \input{Main-paper/Figures/Rademacher/ddMSE_Rademach_wCL}
    \caption{{\small$\frac{d^2}{d \gamma^2} {\rm MMSE}(\gamma)$}}
  \end{subfigure}
  \hfill
  \begin{subfigure}[t]{.18\textwidth}
    \centering
    \input{Main-paper/Figures/Rademacher/dddMSE_Rademach_wCL_numeric}
      
   \caption{{\small $\frac{d^3}{d \gamma^3} {\rm MMSE}(\gamma)$ }}
  \end{subfigure}
  \hfill
  \begin{subfigure}[t]{.18\textwidth}
    \centering
      \input{Main-paper/Figures/Rademacher/ddddMSE_Rademach_wCL_numeric}
    \caption{{\small $\frac{d^4}{d \gamma^4} {\rm MMSE}(\gamma)$ }}
  \end{subfigure}
  \captionsetup{singlelinecheck = false, justification=justified}
    \caption{\small MMSE for the Rademacher spectral distribution. From left to right: Plot (a) ${\rm MMSE}(\gamma)$ computed from \eqref{asymp-MMSE-th} and ${\rm MSE}_{\rm N, RIE}(\gamma)$ points computed from \eqref{RIE-est} for $N=1000$ averaged over 20 runs (error bars are invisible). Plots (b) and (c):  first and second derivatives of ${\rm MMSE}(\gamma)$ computed using their integral representation (integral computed numerically). Plots (d) and (e): first and second numerical differentiation of (c). These suggest that the ${\rm MMSE}^{\prime\prime}(\gamma)$ has a vertical tangent at $\gamma_c = 1$, and a possible phase transition (if present) would be $4$-th order. A numerical analysis in appendix \ref{Example} is compatible with a weak singularity at  $\gamma_c=1$ of the form $(\gamma - 1)^3\ln |\gamma - 1|$.
    }\label{fig: MMSE-Rad-analys}
\end{figure*}

\begin{figure*}
   \captionsetup{singlelinecheck = false, justification=justified}
    \centering
    \begin{subfigure}[t]{.4\textwidth}
    \centering
    \input{Main-paper/Figures/Wishart/MMSE_Wishart_sparse}
    \label{fig:Wishart-MMSE}
  \end{subfigure}
  \quad
  \begin{subfigure}[t]{.4\textwidth}
  \centering
      \input{Main-paper/Figures/Bernoulli/MSE-Ber.tex}%
  \end{subfigure}
   \captionsetup{singlelinecheck = false, justification=justified}
   \caption{\small MMSE in the linear-rank regime with sparse spectral priors. The MMSE of the rank-one problem is also plotted for comparison. (left) Signal with Marchenko-Pastur spectral distribution for large $q$'s.The vertical dashed lines corresponds to the critical value where the support of $\rho_Y$ splits. (right) Signal with rank $M = (1-p)N$ and Bernoulli spectral distribution, $\rho_S = p \delta_{0} + (1-p) \delta_{+1}$,  for $p$'s close to 1.}
   \label{fig:linear-MMSE}
\end{figure*}

\subsection{Transition between the sub-linear and linear rank regimes}

A surprising prediction of our analysis is the collapse of the formulas for the mutual information and MMSE of sub-linear matrix inference with separable priors, onto rank-one expressions; see Statement~\ref{statementSubLin}. Our replica analysis is however not able to assess for what precise regime of $M$ this prediction fails. But, we can exploit our rigorous results on linear rank inference to pinpoint when this happens, namely, when does the transition beetween the low and truly extensive rank regime occurs. For the aforementioned case of a Wishart signal, we observe in the left part of Fig. \ref{fig:linear-MMSE} that as $q \to +\infty$ the MMSE tends to the one of the rank-one version of matrix denoising with Gaussian prior. In particular, we recover the second-order phase transition of the rank-one problem at $\gamma=1$, which matches the famous Ben Arous-Baik-P\'ech\'e transition \cite{Baik2005}. This convergence towards the rank-one prediction can also be observed in a model with Bernoulli spectral distribution, see right part of Fig.\ref{fig:linear-MMSE}.

\section{Proof steps of Theorem \ref{theoremMain}}\label{Proof-steps-thm1}
We present the steps needed to prove theorem 1. It is convenient to decompose Assumption 1 in the main text in two parts.\\

\noindent{\bf Assumption 1.A}
The empirical spectral distribution of $\bS$ converges almost surely weakly to a well-defined probability measure $\rho_S$ with support included in  $[-C, C]$ for some finite positive constant $0<C <+\infty$ independent of $N$.
\vskip 0.25cm 

\noindent{\bf Assumption 1.B}
The second moment of the empirical spectral distribution of $\bS$ is almost surely bounded.

\begin{remark}\label{conv-moment}
These assumptions taken together imply that the second moment of the empirical measure $\rho_S^{(N)}$ converges almost surely to the second moment of $\rho_S$.
\end{remark}

\begin{remark}\label{W2-conv}
    By  Theorem 7.12 in ref. \cite{villani2021topics}, these assumptions are equivalent to the convergence of the empirical distribution in the Wasserstein-$2$ metric to  $\rho_S$. 
\end{remark}

We start from the posterior distribution of the model, which is proportional to
\begin{equation}
\begin{split}
        P_N (\bX | \bY) &\propto e^{-\frac{N}{4} \| \bY - \sqrt{\gamma}\bX \|_F^2 } P_{S,N}(\bX) \\
        &\propto e^{\frac{N}{2} \Tr \big[ \sqrt{\gamma}\bX \bY - \frac{\gamma}{2} \bX^2 \big] } P_{S,N}(\bX).
\end{split}
    \label{post-model1}
\end{equation}
The partition function is defined as the normalizing factor of the posterior distribution \eqref{post-model1}
\begin{equation}
    Z(\bY) = \int d \bX e^{\frac{N}{2} \Tr \big[ \sqrt{\gamma}\bX \bY - \frac{\gamma}{2} \bX^2 \big] } P_{S,N}(\bX)
    \label{partition-function-def}
\end{equation}
and the free energy is defined as
\begin{equation}
    F_N (\gamma) = -\frac{1}{N^2} \bE_{Y} \big[ \ln Z(\bY) \big].
    \label{free-energy-def}
\end{equation}
One can easily see that the free energy is linked to the average mutual information via the identity
\begin{equation}
    \frac{1}{N^2} I_N(\bS;\bY) = F_N (\gamma) + \frac{\gamma}{4 N} \bE \big[ \Tr \bS^2 \big]
\end{equation}
in which $\frac{1}{N} \bE \big[ \Tr \bS^2 \big]$ converges to the second moment of $\rho_S$ by assumption. Therefore, to prove theorem 1 it is enough to show
\begin{equation}
     \lim_{N \to \infty} F_N (\gamma)  = \frac{\gamma}{4} \int x^2 \rho_{S}(x) \, dx - \mathcal{J}[\rho_{\sqrt{\gamma}S}, \,\rho_{\sqrt{\gamma}S} \boxplus \rho_{\rm sc}].
     \label{asymp-free energy}
\end{equation}

The proof of \eqref{asymp-free energy} is done in two main steps. First we we show that such a limit holds for the free energy of an {\it independent eigenvalue model}. Second, using the \textit{pseudo-Lipschitz} continuity of the free energy w.r.t. to the prior distribution, we deduce that the same limit holds for the free energy of the original model.

We make the convention that in the eigen-decomposition of a $N \times N$ matrix $\bS = \bU \bLambda \bU^\intercal$ with $\bLambda = {\rm diag}( \blam)$, the eignevalues $\lambda_1, \cdots, \lambda_N$ are in non-decreasing order.

\subsection{An independent eigenvalue model}
Suppose $\blam^0 \in \bR^N$ is generated with i.i.d. elements from $\rho_S$ and is ordered in non-decreasing way. Fix $\blam^0$ once for all. Let $\tilde{\bS} \in \bR^{N \times N}$ the matrix constructed as $\bU \tilde{\bLambda} \bU^\intercal$ where $\bU$ is distributed according to the Haar measure, and $\tilde{\bLambda} = {\rm diag}(\tilde{\bLam})$ is a diagonal matrix. The distribution of the matrix $\tilde{\bS}$ is
\begin{equation}
\begin{split}
        d P_{\tilde{S},N}(\tilde{\bS}) &= d \mu_N(\bU) d p_{\tilde{S},N}(\tilde{\blam}) = d \mu_N(\bU) \, \prod_{i=1}^N \delta(\tilde{\lambda}_i - \lambda^0_i) \, d \tilde{\blam}.
\end{split}
\label{model 2}
\end{equation}
The independent eigenvalue model is defined as an inference model where the
matrix $\tilde{\bS}$ is observed through an AWGN channel $\tilde{\bY} = \sqrt{\gamma} \tilde{\bS} + \tilde{\bZ}$ with SNR proportional to $\gamma$ and $\tilde{\bZ}$ a symmetric Gaussian Wigner matrix. The associated partition function and the free energy are defined in the same way as in \eqref{partition-function-def},\eqref{free-energy-def}
\begin{equation}
    \tilde{Z}(\tilde{\bY}) = \int d \bX e^{\frac{N}{2} \Tr \big[ \sqrt{\gamma}\bX \tilde{\bY} - \frac{\gamma}{2} \bX^2 \big] } P_{\tilde{S},N}(\bX).
    \label{partition-function-2}
\end{equation}
\begin{equation}
    \tilde{F}_N(\gamma) = -\frac{1}{N^2} \bE_{\tilde{Y}} \big[ \ln \tilde{Z} (\tilde{\bY}) \big] .
    \label{free-energy-2}
\end{equation}
For this independent eigenvalue model, we have
\begin{proposition}
For $\rho_S$ with compact support and any $\gamma > 0$, $\rho_S$-almost surely
\begin{equation}
    \lim_{N \to \infty} \tilde{F}_N(\gamma) = \frac{\gamma}{4} \int x^2 \rho_S(x) \, dx  - \mathcal{J}[\rho_{\sqrt{\gamma}S}, \,\rho_{\sqrt{\gamma}S} \boxplus \rho_{\rm sc}] . \hspace{10 pt}
\label{asymp-free-energy-2}
\end{equation}
\label{asymp-free-energy-2-proposition}
\end{proposition}

\begin{proof}
We start from the partition function \eqref{partition-function-2},
\begin{equation}
    \begin{split}
        \tilde{Z}(\tilde{\bY}) &= \int d \bX e^{\frac{N}{2} \Tr \big[ \sqrt{\gamma}\bX \tilde{\bY} - \frac{\gamma}{2} \bX^2 \big] } P_{\tilde{S},N}(\bX) \\
        &= \int d \blam  d \mu_N(\bU)  \prod_{i=1}^N \delta(\lambda_i - \lambda^0_i) \, e^{\frac{N}{2} \Tr [ \sqrt{\gamma} \bU \bLambda \bU^\intercal \tilde{\bY} - \frac{\gamma}{2}\bLam^2 ]} \\
        &= e^{ - \frac{N}{4} \gamma \Tr {\bLambda^0}^2 } \int d \mu_N(\bU) \, e^{\frac{N}{2} \Tr [ \sqrt{\gamma} \bU \bLambda^0 \bU^\intercal \tilde{\bY} ]} \\
        &= e^{ - \frac{N}{4} \gamma \Tr {\bLambda^0}^2 } \mathcal{I}_N \big( \sqrt{\gamma} \bLambda^0, \tilde{\bY} \big)
    \end{split}
\label{Z2-comp1}
\end{equation}
Recall that $\tilde{Y} = \sqrt{\gamma} \bU \bLambda^0 \bU^\intercal + \tilde{\bZ}$, so the free energy can be written as
\begin{equation}
\begin{split}
        \tilde{F}_N(\gamma) &= \bE_{\tilde{\bY}} \Big[ \frac{\gamma}{4 N} \Tr {\bLambda^0}^2 - \mathcal{J}_N\big( \sqrt{\gamma} \bLambda^0, \tilde{\bY} \big) \Big] \\
        &=  \frac{\gamma}{4 N} \Tr {\bLambda^0}^2 - \bE_{\bU} \bE_{\tilde{\bZ}} \Big[ \mathcal{J}_N\big( \sqrt{\gamma} \bLambda^0, \tilde{\bY} \big) \Big] \\
        &= \frac{\gamma}{4 N} \Tr {\bLambda^0}^2 - \bE_{\bU} \bE_{\tilde{\bZ}} \Big[ \mathcal{J}_N \big( \sqrt{\gamma} \bLambda^0, \sqrt{\gamma} \bU \bLambda^0 \bU^\intercal + \bU \tilde{\bZ} \bU^\intercal\big) \big] \hspace{10pt} (\text{By rotational invariance of } \tilde{\bZ})\\
        &= \frac{\gamma}{4 N} \Tr {\bLambda^0}^2 - \bE_{\bU} \bE_{\tilde{\bZ}} \Big[ \mathcal{J}_N \big( \sqrt{\gamma} \bLambda^0, \sqrt{\gamma} \bLambda^0  +  \tilde{\bZ}\big) \Big] \hspace{10pt} (\text{$\mathcal{J}_N$ is invariant under rotation by $\bU$}) \\
        &= \frac{\gamma}{4 N} \Tr {\bLambda^0}^2 - \bE_{\tilde{\bZ}} \Big[ \mathcal{J}_N \big( \sqrt{\gamma} \bLambda^0, \sqrt{\gamma} \bLambda^0  +  \tilde{\bZ} \big) \Big] .
\end{split}
\label{F2-com}
\end{equation}
By the strong law of large numbers the first term in \eqref{F2-com} converges to $\frac{\gamma}{4} \int x^2 \rho_S(x) \, d x$ almost surely. Finally proposition \ref{asymp-free-energy-2-proposition} follows from the lemma
\begin{lemma}
For any $\gamma \in \bR_+$, the sequence $\bE_{\tilde{\bZ}} \Big[ \mathcal{J}_N \big( \sqrt{\gamma} \bLambda^0, \sqrt{\gamma} \bLambda^0  +  \tilde{\bZ} \big) \Big]$ converges to $\mathcal{J}[\rho_{\sqrt{\gamma}S}, \,\rho_{\sqrt{\gamma}S} \boxplus \rho_{\rm sc}]$ as $N \to \infty$, $\rho_S$-almost surely.
\label{converg-Expec}
\end{lemma}

\noindent This lemma is based on an important result on the convergence of log-spherical integrals \cite{guionnet2002large}. We refer to appendix \ref{proof of prop1} for the details.
\end{proof}

\subsection{Pseudo-Lipschitz continuity of the free energy}
Consider any two rotationally invariant matrix ensembles $P^{(1)}_{N}$, $P^{(2)}_{N}$. Let $\bS \sim P^{(1)}_{N}(\bS)$, $\tilde{\bS} \sim P^{(2)}_{N}(\tilde{\bS})$ with eigendecompositions $\bS = \bU \bLambda \bU^\intercal$, $\tilde{\bS} = \tilde{\bU} \tilde{\bLambda} \tilde{\bU}^\intercal$ and 
\begin{equation}
\begin{split}
    d P_N^{(1)} (\bS )  &= d \mu_N(\bU) \,\, P_N^{(1)} (\blam ) \, d \blam, \\
        d P_N^{(2)} ( \tilde{\bS} )  &= d \mu_N(\tilde{\bU}) \,\, P_N^{(2)} (\tilde{\blam} ) \, d \tilde{\blam}
\end{split}
\end{equation}
where  $P^{(1)}_{N}(\blam)$, $P^{(2)}_{N}(\tilde{\blam})$ are the joint probability density functions for the eigenvalues, induced by the priors. Now consider the two inference problems corresponding to reconstructing the signals from outputs of an AWGNC channel and define as before the corresponding free energies 
$F_N^{(1)}(\gamma), F_N^{(2)}(\gamma)$. Then we have

\begin{proposition}
For all $\gamma > 0$ and $N$
\begin{equation}
\begin{split}
        \big| F_N^{(1)}(\gamma) - F^{(2)}_N(\gamma) \big| \leq \frac{\gamma}{4 N} \Big( \sqrt{ \bE_{\blam} \big[ \| \blam \|_2^2 \big]} + \sqrt{ \bE_{\tilde{\blam}} \big[ \| \tilde{\blam} \|_2^2 \big]} \Big) \sqrt{\bE_{\blam, \tilde{\blam}} \big[ \| \blam - \tilde{\blam} \|_2^2 \big] } .
\end{split}
\end{equation}
\label{pseudo-lip}
\end{proposition}
\begin{proof}
The proof is based on an interpolation between the two matrix ensembles. We refer to appendix \ref{proof of prop2}.
\end{proof}

\subsection{The distance between the original and independent eigenvalue models}
Consider the original and independent eigenvalue models, in other words, the models with prior distributions
\begin{equation}
\begin{split}
        d P_{S,N}(\bS)  &= d \mu_N(\bU) \, P_{S,N}(\blam) \, d \blam^S, \\ 
        d P_{\tilde{S},N}(\tilde{\bS}) &= d \mu_N(\bU) \, \prod_{i=1}^N \delta(\tilde{\lambda}_i - \lambda^0_i) \, d \tilde{\blam}
\end{split}
\end{equation}
where $P_{S,N}(\blam)$ is the joint p.d.f. of eigenvalues of $\bS$, and $\blam^0$ is generated with i.i.d. elements from $\rho_S$. Denote $\prod_{i=1}^N \delta(\tilde{\lambda}_i - \lambda^0_i)$ by $P_{\tilde{S},N}(\tilde{\blam})$.

\begin{lemma}
Under assumption 1  for $\blam \sim P_{S,N}(\blam)$ and  $\tilde{\blam} \sim P_{\tilde{S},N}(\tilde{\blam})$ we have
\begin{equation}
    \lim_{N \to \infty} \frac{1}{N} \bE_{\blam, \tilde{\blam}} \big[ \| \blam - \tilde{\blam} \|^2 \big] =0 .
    \label{W2-dis-eq}
\end{equation}
\label{W2-dis}
\end{lemma}

\begin{proof}
Since $P_{\tilde{\bS},N}(\tilde{\blam})$ is a delta distribution, we can write
\begin{equation}
    \bE_{\blam, \tilde{\blam}} \big[ \| \blam - \tilde{\blam} \|^2 \big] =  \bE_{\blam} \big[ \| \blam - \blam^0 \|^2 \big] .
    \label{W-2-1}
\end{equation}
For a vector $\blam$, denote the empirical distribution of its components by $\hat{\mu}_{\blam}$. The Wasserstein-2 distance between two empirical distributions, $\hat{\mu}_{\blam}, \hat{\mu}_{\blam^0}$ is defined as 
\begin{equation*}
\begin{split}
    W_2(\hat{\mu}_{\blam}, \hat{\mu}_{\blam^0}) = \sqrt{\inf_{\gamma \in \Gamma(\hat{\mu}_{\blam}, \hat{\mu}_{\blam^0})} \bE_{\gamma(x, y)}\big[ ( x  - y)^2 \big]}
\end{split}
\end{equation*}
with $\Gamma(\hat{\mu}_{\blam}, \hat{\mu}_{\blam^0})$ the set of couplings of $(\hat{\mu}_{\blam}, \hat{\mu}_{\blam^0})$. By lemma \ref{W-perm} in appendix \ref{proof of lemma 2} , we have
\begin{equation*}
\begin{split}
    W_2(\hat{\mu}_{\blam}, \hat{\mu}_{\blam^0}) = \sqrt{\min_{\pi \in \mathcal{S}_N}  \frac{1}{N} \| \blam  - \blam^0_{\pi} \|^2 }
\end{split}
\end{equation*}
where $\blam^0_{\pi}$ is the permuted version of $\blam^0$, and $\mathcal{S}_N$ is the group of all permutations of $N$ elements. So, for given $\blam$ and $\blam^0$ (which have a non-decreasing order), we have (considering the identity permutation)
\begin{equation}
    \| \blam - \blam^0 \|^2 \geq N W_2(\hat{\mu}_{\blam}, \hat{\mu}_{\blam^0})^2.
        \label{W-2-2}
\end{equation}
Now we recall recall the \textit{rearrangement inequality}: for real numbers $x_1 \leq x_2 \leq \cdots \leq x_n$, $y_1 \leq y_2 \leq \cdots \leq y_n$, for every permutation $\pi \in \mathcal{S}_N$ we have
\cite{hardy1952inequalities}
$$
x_n y_1 + \hdots + x_1 y_n 
\hspace{-0.5 pt}\leq \hspace{-0.5 pt} x_{\pi(1)} y_1 + \hdots + x_{\pi(n)} y_n 
\hspace{-0.5 pt}\leq \hspace{-0.5 pt} x_1 y_1 + \hdots + x_n y_n
$$
For any permutation of $\blam^0$ (in particular the one which achieves the minimum in \eqref{W-2-2}), using the rearrangement inequality, we get (recall that $\blam^0$  is ordered in non-decreasing order)
\begin{equation*}
    \begin{split}
        \| \blam  - \blam^0_{\pi} \|^2 &= \| \blam \|^2 + \| \blam^0_{\pi} \|^2 - 2 \blam^\intercal \blam^0_{\pi} \\
        & \geq \| \blam \|^2 + \| \blam^0_{\pi} \|^2 - 2 \blam^\intercal \blam^0 \\
        &= \| \blam  - \blam^0 \|^2 
    \end{split}
\end{equation*}
and consequently
\begin{equation}
    \| \blam - \blam^0 \|^2 \leq N W_2(\hat{\mu}_{\blam}, \hat{\mu}_{\blam^0})^2.
        \label{W-2-3}
\end{equation}
Finally
from \eqref{W-2-1}, \eqref{W-2-2},\eqref{W-2-3}, we obtain
\begin{equation}
  \frac{1}{N}\bE_{\blam, \tilde{\blam}} \big[ \| \blam - \tilde{\blam} \|^2 \big] = \bE_{\blam} \big[ W_2(\hat{\mu}_{\blam}, \hat{\mu}_{\blam^0})^2 \big].
    \label{W-2-5}
\end{equation}
Lemma \ref{E-W-0-l} in appendix \ref{proof of lemma 2} allows to conclude the proof.
\end{proof}

\subsection{Concluding the Proof}
By proposition \ref{pseudo-lip}, the free energies $F_N(\gamma)$ (defined in \eqref{free-energy-def}) and $\tilde{F}_N(\gamma)$  (defined in \eqref{free-energy-2}), satisfy
\begin{equation}
\begin{split}
        \big| F_N^{(1)}(\gamma) - F^{(2)}_N(\gamma) \big| \leq \frac{\gamma}{4 N} \Big( \sqrt{ \bE_{\blam} \big[ \| \blam \|^2 \big]} + \sqrt{ \bE_{\tilde{\blam}} \big[ \| \tilde{\blam} \|^2 \big]} \Big) \sqrt{\bE_{\blam, \tilde{\blam}} \big[ \| \blam - \tilde{\blam} \|^2 \big] } .
\end{split}
\label{proof-1}
\end{equation}
The term $\frac{1}{N} \| \blam \|^2 = \frac{1}{N} \sum \lambda_i^2 $ is the second moment of the empirical spectral distribution of $\bS$, which is almost surely bounded by assumption 1.B. So,  $\frac{1}{N} \bE_{\blam} \big[ \| \blam \|^2 \big]$ is bounded uniformly in $N$.
 Moreover, $\frac{1}{N} \bE \big[ \| \tilde{\blam} \|^2 \big] = \frac{1}{N}  \sum {\lambda^0}_i^2$ is also bounded uniformly in $N$. By lemma  \ref{W2-dis}, $\lim_{N \to \infty} \frac{1}{N} \bE_{\blam, \tilde{\blam}} \big[ \| \blam - \tilde{\blam} \|^2 \big] = 0$. Therefore
\begin{equation}
    \lim_{N \to \infty} |F_N(\gamma) - \tilde{F}_N(\gamma) | = 0 .
\label{proof-2}
\end{equation}
Proposition \ref{asymp-free-energy-2-proposition} together with \eqref{proof-2} conclude the proof. $\hfill \square$

\section{Proof of Theorem 2 and rotation invariance of the MMSE estimator}\label{RIE-app}
In this section we show how to use the RIE class in order to prove Theorem 2 under the extra assumption 1.B. 
Recall that an estimator $\hat{\Xi}(\bY)$ is called \textit{rotation invariant} if for any orthogonal matrix $\bO \hat{\Xi}(\bY) \bO^\intercal = \hat{\Xi}(\bO \bY \bO^\intercal)$. Such an  estimator has the same eigenvectors as the matrix $\bY$ and denoting the eigenvectors of $\bY$ by $\by_1, \hdots, \by_N$, it can be expressed as $\hat{\Xi}(\bY) = \sum_{i=1}^N \hat{\xi}_i \by_i \by_i^\intercal$
where $\hat{\xi}_1, \hdots, \hat{\xi}_N$ are the eigenvalues of the estimator. The best RIE estimator corresponds to eigenvalues chosen to minimize the mean-square-error in the RIE class and a heuristic calculation using the replica method leads to 
\begin{equation}
    \begin{cases}
    \hat{\Xi}^*(\bY) =  \sum_{i=1}^N \xi^*_i \by_i \by_i^\intercal \\
    \xi^*_i = \sum_{j=1}^N \lambda^S_j \big( \bs_j^\intercal \by_i \big)^2   =\frac{1}{\sqrt{\gamma}} \big( \lambda^Y_i - 2 \pi \sH [\rho_Y](\lambda_i^Y) \big)
    \end{cases}
    \label{RIE-est-app}
\end{equation}
where $( \lambda_i^Y )_{1 \leq i \leq N}$ are the eigenvalues of $\bY$, and $\sH [\rho_Y]$ is the \textit{Hilbert} transform of the limiting spectral distribution of $\bY$ defined as: 
\begin{align}
\sH [\rho_Y](z):= {\rm P.V.} \frac{1}{\pi} \int \frac{\rho_Y(x)}{z-x} \, dx .
\end{align}
We will need the following properties of the Hilbert transform. A proof can be found in lemma 3.1 of \cite{shlyakhtenko2020fractional}. 
\begin{lemma}\label{properties of Hilbert}
If $f : \bR \to \bR$ is compactly supported and
sufficiently regular (continuous is sufficient), then one has the identities
\begin{equation}
   \int_\bR f(x)\big( \sH [f] (x) \big)^2 \, dx = \frac{1}{3} \int_\bR f^3(x) \, dx ,
   \label{Hilbert-1}
\end{equation}
\begin{equation}
   \int_\bR \sH [f] (x) x f(x) \, dx = \frac{1}{2 \pi} \Big( \int_\bR f(x) \, dx \Big)^2 .
   \label{Hilbert-2}
\end{equation}
\label{Hilbert-iden}
\end{lemma}

\begin{proof}[{\it Proof of Theorem 2}]
As explained in the main text the best RIE estimator is optimal in the sense ${\rm MMSE}_N(\gamma) = {\rm MMSE}_{\rm RIE, N}(\gamma)$ so our proof proceeds by a computation of the limit of the r.h.s. The optimality follows from rotation invariance of the MMSE estimator $\mathbb{E}(\bS\vert \bY)$ and this rotation invariance is checked later for completeness.

From assumption 2, it suffices to compute the limit of $\frac{1}{N}\mathbb{E}\Vert \bS - \hat{\Xi}^*(\bY)\Vert^2$. Denoting the eigenvectors of $\bS$ by $\bs_1, \hdots, \bs_N$. Expanding the MSE, we find  
\begin{equation}
\begin{split}
        \big \| \bS - \hat{\Xi}^*(\bY) \big \|_F^2 &= \sum_{i=1}^N \bigg[ {\lambda^S_i}^2 + {\xi_i^*}^2 - 2 \xi^*_i \sum_{j=1}^N \lambda^S_j \big( \bs_j^\intercal \by_i \big)^2 \bigg] \\
    &= \sum_{i=1}^N ({\lambda^S_i}^2 - {\xi_i^*}^2)
\end{split}
\end{equation}
where we used \eqref{RIE-est-app} to get the last equality. 
Using again \eqref{RIE-est-app} we have
\begin{equation}
    \begin{split}
        \big \| \bS - \hat{\Xi}^*(\bY) \big \|_F^2 & = \sum_{i=1}^N ({\lambda_i^{S}}^2 - {\xi_i^*}^2)
        \\ &
        = \sum_{i=1}^N {\lambda^S_i}^2 - \frac{1}{\gamma} \big( \lambda^Y_i - 2 \pi \sH [\rho_Y](\lambda_i^Y) \big)^2 
        \\ &= \sum_{i=1}^N {\lambda^S_i}^2 - \frac{1}{\gamma} 
        \sum_{i=1}^N {\lambda^Y_i}^2 - \frac{4 \pi^2}{\gamma} \sum_{i=1}^N \big(\sH [\rho_Y](\lambda_i^Y)\big)^2 + \frac{4 \pi}{\gamma} \sum_{i=1}^N  \sH [\rho_Y](\lambda_i^Y) \lambda^Y_i
    \end{split}
\end{equation}
From linearity of expectation and, as the Hilbert transform $ \sH [\rho_Y]$ is continuous on the support of $\rho_Y$ \cite{biane1997free}, we find
\begin{equation}
\begin{split}
    \lim_{N \to \infty} \frac{1}{N} \mathbb{E}\big \| \bS - \hat{\Xi}^*(\bY) \big \|_F^2 &= \int x^2 \rho_S(x) \, dx - \frac{1}{\gamma} \int x^2 \rho_Y(x) \, dx - \frac{4 \pi^2}{\gamma} \int \rho_Y(x) \big( \sH [\rho_Y] (x) \big)^2 \, dx \\
    &\hspace{4cm}+ \frac{4 \pi}{\gamma} \int \sH [\rho_Y] (x) x \rho_Y(x) \, dx
\end{split}
\end{equation}
By linearity of expectation and the independence of $\bS$ and $\bZ$,  we have:
\begin{equation*}
    \bE \Tr \bS \bZ = \bE \sum_{i,j}^N S_{i,j} Z_{j,i} = \bE \sum_{i,j} \bE S_{i,j}  \bE Z_{j,i} = 0
\end{equation*}
Therefore, we find:
\begin{align}
     \int x^2  \rho_Y(x) \, dx  & = \lim_{N \to \infty} \frac{1}{N} \mathbb{E}\Tr \bY^2 
     \nonumber \\ &
     = \lim_{N \to \infty} \frac{1}{N} \gamma \mathbb{E}\Tr \bS^2 + \lim_{N \to \infty} \frac{1}{N} \Tr \bZ^2 + \lim_{N \to \infty} \frac{2}{N} \sqrt{\gamma} \mathbb{E}\Tr \bS \bZ 
     \nonumber \\ & 
     = \gamma \int x^2  \rho_S(x) + 1
\end{align}
where in the last line we used that the second moment of the semicircle law is equal to $1$. Finally, using the identities in lemma \ref{Hilbert-iden}, we get
\begin{equation}
\begin{split}
    \lim_{N \to \infty} \frac{1}{N} \mathbb{E}\big \| \bS - \hat{\Xi}^*(\bY) \big \|_F^2 &= - \frac{1}{\gamma}  - \frac{4 \pi^2}{3 \gamma} \int \rho_Y^3(x)  \, dx + \frac{2}{\gamma} =  \frac{1}{\gamma} \Big( 1 - \frac{4 \pi^2}{3} \int \rho_Y^3(x) \, dx \Big) .
\end{split}
\label{RIE-MSE}
\end{equation}
\end{proof}


For completeness we provide a check that the MMSE estimator belongs to the RIE class. The posterior mean given $\bY$ is
\begin{equation}
    \bE [ \bS | \bY ]= \frac{\int d\bX \, P_{S,N}(\bX) \bX e^{-\frac{N}{4} \big\| \bY - \sqrt{\gamma}\bX \big\|_F^2}}{\int d\bX \, P_{S,N}(\bX) e^{-\frac{N}{4} \big\| \bY - \sqrt{\gamma}\bX \big\|_F^2}} .
\end{equation}
By rotation invariance of $P_{S,N}(\bX)$ under any orthogonal transformation $\bX \to \bO \bX \bO^\intercal$ with Jacobian $\vert{\rm det} \bO\vert =1$ we have 
\begin{equation}
    \begin{split}
        \bE [ \bS | \bO \bY \bO^\intercal ] &= \frac{\int d\bX \, P_{S,N}(\bX) \bX e^{-\frac{N}{4} \big\| \bO \bY \bO^\intercal - \sqrt{\gamma}\bX \big\|_F^2}}{\int d\bX \, P_{S,N}(\bX) e^{-\frac{N}{4} \big\| \bO \bY \bO^\intercal - \sqrt{\gamma}\bX \big\|_F^2}} \\
        &= \frac{\int d\bX \, P_{S,N}(\bX) \bO \bX \bO^\intercal e^{-\frac{N}{4} \big\| \bO \bY \bO^\intercal - \sqrt{\gamma}\bO \bX \bO^\intercal\big\|_F^2}}{\int d\bX \, P_{S,N}(\bX) e^{-\frac{N}{4} \big\| \bO \bY \bO^\intercal - \sqrt{\gamma}\bO \bX \bO^\intercal \big\|_F^2}} \\
        &= \bO \Big\{ \frac{\int d\bX \, P_{S,N}(\bX) \bX e^{-\frac{N}{4} \big\| \bY - \sqrt{\gamma}\bX \big\|_F^2}}{\int d\bX \, P_{S,N}(\bX) e^{-\frac{N}{4} \big\| \bY - \sqrt{\gamma}\bX \big\|_F^2}}  \Big\} \bO^\intercal \\
        &= \bO \bE [ \bS | \bY ] \bO^\intercal .
    \end{split}
\end{equation}
Therefore the posterior mean estimator is an RIE.

 

\section{Proof of theorem 3: explicit expression of the mutual information}\label{Explicit-MI-MMSE}
We derive an explicit expression for the asymptotic mutual information using the I-MMSE relation \cite{guo2005mutual} and basic results in free probability. This derivation is completely independent of Theorem 1
in which the asymptotic mutual information is expressed using the asymptotic spherical integral, $\mathcal{J}[\rho_{\sqrt{\gamma} S}, \rho_{\sqrt{\gamma} S}\boxplus \rho_{\rm sc}]$. 

Free probability (see \cite{nica2006lectures, mingo2017free}) was initially introduced to study operator algebras, but has gained considerable importance in other realms due to its connection with the asymptotic behavior of random matrices. While free probability has been exploited in linear estimation and wireless communication problems \cite{muller2004random, verdu2010mismatched, nica2008free, reeves2018mutual}, the connection with matrix inference setting studied here is to the best of our knowledge new. 

Let $X$ be a self-adjoint \textit{non-commutative} random variable $X$ associated to a probability measure  $\mu_X$ with compact support on the real line. According to \cite{voiculescu1993analogues, voiculescu1994analogues} the \textit{free entropy} $\chi(X)$ and the \textit{free Fisher information} $\Phi(X)$  are given as
\begin{align}
    \chi(X) &= \iint \ln |s-t| \mu_X(s) \mu_X(t) \, ds \,  dt + \frac{3}{4} + \frac{1}{2} \ln 2 \pi,
    \label{Free Entropy}\\
    \Phi(X) &= \frac{4 \pi^2}{3} \int \mu_X^3(s) \, ds.
    \label{Free Fisher}
\end{align}
Moreover, these two quantities are related through the relation:
\begin{equation}
    \chi(X) = \frac{1}{2} \int_0^{\infty} \Big( \frac{1}{1+t} - \Phi( X + \sqrt{t}Z ) \Big) \, dt + \frac{1}{2} \ln 2\pi + \frac{1}{2}
    \label{entropy-fisher}
\end{equation}
where $Z$ is a semicircular non-commutative random variable, and $X$ and $Z$ are {\it free}. We apply these relations to $X =S$ and $Z$, two free non-commutative random variables (not to confused with the $N\times N$ matrices $\bS$ and $\bZ$) associated to the probability measures $\rho_S$ and $\rho_{\rm sc}$. Since $S$ and $Z$ are free, the sum $\sqrt\gamma S + Z$ is a non-commutative random variable associated to the measure $\rho_{\sqrt\gamma S} \boxplus \rho_{\rm sc} = \rho_Y$. 
Clearly then
\eqref{asymp-MMSE-th} can be written in the free probability language as
\begin{equation}
    {\rm MMSE}(\gamma) = \frac{1}{\gamma} \Big( 1 - \Phi(\sqrt{\gamma}S + Z) \Big).
    \label{MMSE-freeFisher}
\end{equation}

It is shown in \cite{voiculescu1997derivative} that $\Phi(\sqrt\gamma S + Z)$ is a continuous function of $\gamma$. Thus, \eqref{MMSE-freeFisher} implies that the ${\rm MMSE}(\gamma)$ is a continuous function of $\gamma$, concluding the proof of Theorem \ref{statementMMSE}.

\begin{proof}[\it Proof of Theorem 3]
An important property of the Gaussian channel is the I-MMSE relation relating the MMSE to the derivative of the mutual information w.r.t the SNR. 
The concavity of the mutual information w.r.t. SNR, implies that this relation also holds in the limit $N \to \infty$. Integrating this relation we have
\begin{equation}
    I(\bS;\bY) = \frac{1}{4} \int_0^{\gamma} {\rm MMSE}(t) \, dt + {\rm constant}
    \label{I-MMSE-integral}
\end{equation}
where $I(\bS;\bY):= \lim_{N \to \infty} \frac{1}{N^2} I_N(\bS;\bY)$ and ${\rm MMSE}(t) = \lim_{N\to +\infty}{\rm MMSE}_N(t)$.

Since for $\gamma = 0$, $I(\bS;\bY) = 0$ the integration constant vanishes. Therefore, we just need to compute the integral over the asymptotic MMSE given by Theorem 2. Using \eqref{MMSE-freeFisher}, we have
\begin{equation}
\begin{split}
    I(\bS;\bY)
    &= \frac{1}{4} \int_0^{\gamma} \Big( \frac{1}{t}  - \frac{1}{t} \Phi(\sqrt{t}S + Z) \Big) \, dt \\
    &= \frac{1}{4} \int_0^{\gamma} \Big( \frac{1}{t}  - \frac{1}{t^2} \Phi(S + \sqrt{\frac{1}{t}} Z) \Big) \, dt \\
    &= \frac{1}{4} \int_{\frac{1}{\gamma}}^{\infty} \Big( \frac{1}{x}  - \Phi(S + \sqrt{x} Z) \Big) \, dx \hspace{10pt}(\frac{1}{t} \to x) \\
    &= \frac{1}{4} \int_{\frac{1}{\gamma}}^{\infty} \Big( \frac{1}{x}  - \gamma \Phi(\sqrt{\gamma} S + \sqrt{\gamma x} Z) \Big) \, dx \hspace{10 pt} ( \Phi(a X) = \frac{1}{a^2} \Phi(X) \text{ for } a>0) \\
    &= \frac{1}{4} \int_{1}^{\infty} \Big( \frac{1}{y}  - \Phi(\sqrt{\gamma} S + \sqrt{y} Z) \Big) \, dy \hspace{10pt}(\gamma x \to y) \\
    &= \frac{1}{4} \int_{0}^{\infty} \Big( \frac{1}{t+1}  - \Phi(\sqrt{\gamma} S + \sqrt{t+1} Z) \Big) \, dt \hspace{10pt}(y \to t+1) \\
    &= \frac{1}{4} \int_{0}^{\infty} \Big( \frac{1}{t+1}  - \Phi(\sqrt{\gamma} S + Z_0 + \sqrt{t} Z) \Big) \, dt \hspace{10pt} (\text{$Z$ and $Z_0$ free semi-circulars}) \\
    &= \frac{1}{2} \chi(\sqrt{\gamma} S + Z_0) - \frac{1}{4} \ln 2 \pi - \frac{1}{4} \hspace{10pt} (\text{from \eqref{entropy-fisher} )}  
\end{split}
\label{deriv-asymp-MI}
\end{equation}
Since $S$ and $Z$ are free, $\sqrt\gamma S + Z$ is a non-commutative random variable associated to the measure $= \rho_Y =\rho_{\sqrt\gamma S} \boxplus \rho_{\rm sc}$.  Therefore, using \eqref{Free Entropy}, we obtain
\begin{equation}
    \lim_{N \to \infty} \frac{1}{N^2} I_N(\bS;\bY) = \frac{1}{2} \iint \ln |s-t| \rho_Y(s) \rho_Y(t) \, ds \, dt + \frac{1}{8}
    \label{simp-asymp-mI}
\end{equation}
where $\rho_Y = \rho_{\sqrt{\gamma}S} \boxplus \rho_{\rm sc}$.
\end{proof}

\section{Sub-linear rank RIE and MSE}\label{subL-RIE-app}

The signal matrix $\bS \in \bR^{N \times N}$ is taken from a rotationally invariant prior and observed through an additive channel,
\begin{equation*}
    \bY = \sqrt{\gamma} \bS + \bZ
\end{equation*}
where the  noise $\bZ \in \bR^{N \times N}$ is also distributed according to a rotationally invariant ensemble. Here we allow non-Gaussian noise. It is assumed that $\bS$ has $M = \lfloor{N^\alpha}\rfloor$ non-zero i.i.d eigenvalues sampled from 
a distribution $\rho_S(x)$ with finite second moment and bounded support. By construction the empirical distribution of non-zeros eigenvalues $\frac{1}{M} \sum_{i =1}^M \delta(x - \lambda_i^S)$ tends weakly to $\rho_S(x)$. Moreover we assume that $\bZ$ has a limiting spectral distribution $\rho_Z$. Since we are in the sub-linear regime, the limiting spectral measure of $\bY$ (normalized by $1/N$) is the same as the one of $\bZ$. But note that $\bY$ may have sub-linear number of eigenvalues outside the support.

We propose a sub-linear rank RIE and an associated algorithm to estimate $\bS$ from $\bY$, which we conjecture to be optimal. 
Arguing as in the linear case it is not difficult to see that the {\it optimal} RIE (i.e., the one minimizing the MSE in the general RIE class)must have a MSE which equals the MMSE. We conjecture that the {\it proposed} RIE, and associated algorithm, are indeed the optimal. Evidence for this conjecture comes from numerics discussed below (at least for some range of $\alpha$), but also from the particular case of Gaussian noise. Indeed for Gaussian noise we have the I-MMSE relation  so by integrating the MMSE we find the mutual information. Thus our proposed sub-linear rank RIE predicts an expression for the mutual information. On the other hand the mutual information has recently been rigorously computed from the asymptotics of sub-linear rank spherical integrals \cite{husson2022spherical}. By comparing the expressions obtained by these two independent approaches we validate the conjecture analytically, at least for Gaussian noise.

\subsection{ Sub-linear rank RIE}
Let the eigen-decomposition of $\bS$ and $\bY$ be $\bS = \sum_{i = 1}^M \lambda_i^S \bs_i \bs_i^\intercal$, $\bY = \sum_{i =1}^N \lambda_i^Y \by_i \by_i^\intercal$. For a RIE $\Xi(\bY) = \sum_{i =1}^N \xi_i \by_i \by_i^\intercal$, the MSE can be written as:
\begin{equation}
\begin{split}
    \frac{1}{M} \| \bS - \Xi(\bY) \|_F^2 &= \frac{1}{M} \big \| \sum_{i = 1}^M \lambda_i^S \bs_i \bs_i^\intercal - \sum_{i =1}^N \xi_i \by_i \by_i^\intercal \big\|_F^2    \\
    &= \frac{1}{M} \sum_{i = 1}^M {\lambda_i^S}^2 + \frac{1}{M} \sum_{i =1}^N \xi_i^2 - \frac{2}{M} \sum_{i=1}^N \xi_i \sum_{j =1}^M \lambda_j^S \big( \bs_j^\intercal \by_i \big)^2 .
\end{split}
\label{MSE-RIE-SubL}
\end{equation}
Minimizing \eqref{MSE-RIE-SubL} over $\xi_i$'s, the optimum is achieved at 
\begin{equation}
    \xi_i^* = \sum_{j =1}^M \lambda_j^S \big( \bs_j^\intercal \by_i \big)^2 = \by_i^\intercal \bS \by_i .
    \label{Oracle-est}
\end{equation}
This estimator is called {\it oracle estimator} since it requires the knowledge of the signal. 

To compute the optimal eigenvalues we need to know the overlap between the eigenvectors of the signal and the observation.  For finite-rank additive perturbation, $\bX$ of a rotationally invariant matrix $\bZ$, the eigenvalues and the overlap between the eigenvectors of the perturbation and the perturbed matrix $\bX + \bZ$ has been computed rigorously in \cite{benaych2011eigenvalues}. In \cite{huang2018mesoscopic} this is extended to sub-linear rank perturbations. Essentially the same formulas (as for finite-rank perturbations)  give the eigenvalues of the perturbed matrix and overlap in the large size limit. 

These results could presumably be used to get a rigorous computation of the asymptotic MSE predicted by the oracle estimator but this is left for future work. Here we use these results in an heuristic way to propose a specific RIE and associated algorithm. We use the notation $a\to b$ to mean that $|a -b|$ tends to zero with high probability as $N\to +\infty$.
Theorem 2.7 in \cite{huang2018mesoscopic} suggests 
\begin{equation}
    \big( \bs_i^\intercal \by_i \big)^2 \to \begin{cases}
        \frac{-1}{\gamma {\lambda_i^S}^2 G'_{\rho_Z} \Big( G^{-1}_{\rho_Z} \big( \frac{1}{\sqrt{\gamma} \lambda_i^S} \big) \Big)  } \hspace{10 pt} &\text{ if } \frac{1}{\sqrt{\gamma} \lambda_i^S} \in \big( G_{\rho_Z}(a^-), G_{\rho_Z}(b^+) \big), \\
        0 &\text{ else. }
    \end{cases}
    \label{Overlap-subL}
\end{equation}
where $G_{\rho_Z}(z)$ is the \textit{Cauchy transform} of $\rho_Z$, constrained on $\bR \backslash {\rm supp } \, \rho_Z$, and $a, b$ are  the infimum and supremum of the support of $\rho_Z$, and $G_{\rho_Z}(a^-) \equiv \lim_{z \to a^{-}} G_{\rho_Z}(z)$ and similarly for $b^+$.  
The overlap in \eqref{Overlap-subL} is expressed in terms of eigenvalues of $\bS$, but since the corresponding eigenvalue of $\bY$ is affected by $\lambda_i^S$, we can express the overlap in terms of eigenvalues of $\bY$. Theorem 2.1 in \cite{huang2018mesoscopic} suggests 
\begin{equation}
    \lambda_i^Y \to \begin{cases}
        G^{-1}_{\rho_Z} \big( \frac{1}{\sqrt{\gamma} \lambda_i^S} \big)  \hspace{10 pt} &\text{ if } \frac{1}{\sqrt{\gamma} \lambda_i^S} \in \big( G_{\rho_Z}(a^-), G_{\rho_Z}(b^+) \big), \\
        b &\text{if  }  \frac{1}{\sqrt{\gamma} \lambda_i^S} > G_{\rho_Z}(b^+), \\
        a &\text{if  }   \frac{1}{\sqrt{\gamma} \lambda_i^S} > G_{\rho_Z}(a^-).
    \end{cases}
    \label{Top-ev-subL}
\end{equation}
From \eqref{Top-ev-subL}, we can see that if an eigenvalue of $\bY$ is outside the support of $\rho_Z$, then the corresponding eigenvalue of the signal can computed as $\lambda_i^S \approx \frac{1}{\sqrt{\gamma} G_{\rho_Z}(\lambda_i^Y) }$.
From \eqref{Overlap-subL}, \eqref{Top-ev-subL} we deduce that for an eigenvalue $\lambda_i^Y \in (a,b)$ there is no spike aligned with $\by_i$, because otherwise $\lambda_i^Y$ would be outside of the support of $\rho_Z$. So, the corresponding $\xi_i^*$'s are zero for the eigenvalues in $(a,b)$. On the other hand, since there are $M$ spikes, at most $M$ $\xi_i^*$'s are non-zero which makes the whole expression in \eqref{MSE-RIE-SubL} for the optimum $\xi_i^*$'s $O(1)$. 

Finally the proposed RIE estimator is naturally constructed as follows
\begin{equation}
    \begin{cases}
    \hat{\Xi}^*(\bY) =  \sum_{i=1}^N \xi^*_i \by_i \by_i^\intercal, \\
    \xi^*_i = - \frac{1}{\sqrt{\gamma}}   \mathbb{I}\big( \lambda^Y_i \notin [a,b] \big) \,  \frac{G_{\rho_Z}\big(\lambda^Y_i\big)}{G'_{\rho_Z}\big(\lambda^Y_i\big)}.
    \end{cases}
    \label{subL-RIE-est-gen}
\end{equation}
This provides an algorithm with the following steps to reconstruct the signal 
\begin{itemize}
\item[(i)] 
Compute spectral data $(\lambda_i^Y,  \by_i)$ from the matrix $\bY$.
\item[(ii)] 
Apply the function $f_Z$ to the eigenvalues
\begin{equation*}
f_Z(x) = \begin{cases}
      - \frac{1}{\sqrt{\gamma}} \frac{G_{\rho_Z}(x)}{G'_{\rho_Z}(x)} \hspace{10 pt} &\text{if } x \notin [a,b] \\
       0 & \text{else.}
    \end{cases}
\end{equation*}
\item[(iii)]
Construct the estimate as $\hat{\bS} = \sum_{i =1}^N f_Z(\lambda_i^Y) \by_i \by_i^\intercal$.
\end{itemize}
The second algorithmic step requires knowledge of the limiting distribution of the noise which can in principle be computed from its distribution.

\subsection{Gaussian Noise}\label{subL-RI-Gaussian}
For $\rho_Z = \frac{1}{2 \pi} \sqrt{4 - x^2}$, we have that $G_{\rho_{\rm sc}}(z) = \frac{z - {\rm sign}(z)\sqrt{z^2 - 4}}{2}$. Thus, given matrix $\bY$ the estimator reads:
\begin{equation*}
    \begin{cases}
    \hat{\Xi}^*(\bY) =  \sum_{i=1}^N \xi^*_i \by_i \by_i^\intercal, \\
    \xi^*_i = \frac{1}{\sqrt{\gamma}}   \mathbb{I}\big( |\lambda^Y_i| > 2 \big) \, {\rm sign}(\lambda^Y_i) \,\sqrt{{\lambda^Y_i}^2 - 4}.
    \end{cases}
\end{equation*}

\subsubsection{MSE}
To compute the MSE, we use \eqref{Overlap-subL} which is in terms of the eigenvalues of $\bS$. We have
\begin{equation}
    \big( \bs_i^\intercal \by_i \big)^2 \to \begin{cases}
        1 - \frac{1}{\gamma {\lambda_i^S}^2} \hspace{10 pt} &\text{ if } \gamma {\lambda_i^S}^2 \geq 1, \\
        0 &\text{ else, }
    \end{cases}
\end{equation}
so the optimal eigenvalues of the estimator are
\begin{equation}
    \xi_i^* \to \begin{cases}
        \lambda_i^S - \frac{1}{\gamma \lambda_i^S} \hspace{10 pt} &\text{ if } \gamma {\lambda_i^S}^2 \geq 1, \\
        0 &\text{ else. }
    \end{cases}
\end{equation}
The MSE can be written as,
\begin{equation}
\begin{split}
    \frac{1}{M} \| \bS - \Xi^*(\bY) \|_F^2 &= \frac{1}{M} \sum_{i = 1}^M {\lambda_i^S}^2 - \frac{1}{M} \sum_{i = 1}^N {\xi^*_i}^2 
\end{split}
\label{MSE-expansion}
\end{equation}
where,  we used that, in the limit $N \to \infty$, at most $M$ number of $\xi_i^*$'s are non-zero. Taking the limit $N \to \infty$ (or $M \to \infty$), we find
\begin{equation}
\begin{split}
    \lim_{N \to \infty}  \frac{1}{M} \| \bS - \Xi^*(\bY) \|_F^2 & = \int x^2 \rho_S(x) \, dx - \int_{|x| \geq \frac{1}{\sqrt{\gamma}}} \big( x - \frac{1}{\gamma x} \big)^2 \rho_S(x) \, dx .
\end{split}
\label{RIE-MSE-SubL}
\end{equation}

\subsubsection{Mutual information in Gaussian noise}
As already explained the MSE computed above should be equal to the MMSE, thus by integrating over $\gamma$ we should recover the mutual information. Moreover, the mutual information can be computed using the limit of the spherical integrals of sub-linear rank \cite{husson2022spherical}. In this section we explicitly check for a few priors that the two expressions indeed coincide.

We first present the main formula of \cite{husson2022spherical}.
The asymptotic mutual information between $\bS$ and the observation $\bY = \sqrt{\gamma} \bS + \bZ$ is
\begin{equation}
\begin{split}
        \lim_{N \to \infty} \frac{1}{M N} I_N(\bS; \bY) &= \frac{\gamma}{2} \int x^2 \rho_S(x) \, dx - \lim_{N \to \infty} \frac{1}{M N} \ln \int D \bU e^{\frac{N}{2} \Tr \sqrt{\gamma}\bS \bU  \bY \bU^\intercal }.
\end{split}
\label{MI-subL-Spherical-integral}
\end{equation}
The integral $\int D \bU e^{\frac{N}{2} \Tr \sqrt{\gamma}\bS \bU  \bY \bU^\intercal }$ is the spherical integral of sub-linear rank, and its asymptotic limit has been studied in \cite{husson2022spherical}. For matrix $\bA$ with $M$ positive eigenvalues, by Theorem 2.5 in \cite{husson2022spherical}, we have that
\begin{equation}
    \lim_{N \to \infty} \frac{1}{M N} \ln \int D \bU e^{\frac{N}{2} \Tr \bA \bU  \bB \bU^\intercal } = \frac{1}{2} \lim_{N \to \infty} \frac{1}{M} \sum_{i =1}^M K( \theta_i, \lambda_i, \mu_B)
    \label{subL-spherical}
\end{equation}
where $\theta_i$'s are non-zero eigenvalues of $\bA$, $\lambda_i$'s are the top eigenvalues of $\bB$, and $\mu_B$ is the limiting spectral distribution of $\bB$. Note that the result of \cite{husson2022spherical} also covers the case of negative eigenvalues, however for simplicity we only consider matrices with positive eigenvalues. The function $K$ is defined as
\begin{equation*}
K( \theta, \lambda, \mu ) = \theta \lambda' + (v - \lambda') G_{\mu} (v) - \ln |\theta| - \int \ln |v -x| \, d \mu(x) - 1
\end{equation*}
where $\lambda' = \max( \lambda, r(\mu) )$ ($r(\mu)$ is the rightmost point of the support of $\mu$),
\begin{equation*}
    v := v(\lambda, \theta) = \begin{cases}
    \lambda' &\hspace{5pt} \text{ when } 0 \leq G_{\mu}(\lambda') \leq \theta \text{ or } \theta \leq G_{\mu}(\lambda') \leq 0 \\
    G^{-1}_{\mu}(\theta) &\hspace{5pt} \text{else}
    \end{cases}
\end{equation*}
and $G_{\mu}$ is the Stieltjes transform of $\mu$.

\subsubsection{Example 1: Sub-linear Wishart signal}\label{Wishart-subL-GNoise}
Consider the signal matrix $\bS$ to be $\frac{1}{N}\bX \bX^\intercal$, with   $\bX \in \bR^{N \times M}$ has i.i.d. standard Gaussian entries, and $M=\lfloor{N^\alpha}\rfloor$. In the limit $N \to \infty$, one can show that the limiting distribution of non-zero eigenvalues of $\bS$ is $\delta(x-1)$.
For this example, the MSE given by \eqref{RIE-MSE-SubL} reads
\begin{equation}
    {\rm MSE}(\gamma) = \begin{cases}
1 \hspace{10 pt} & \text{if } \gamma \leq 1, \\
\frac{1}{\gamma} \big( 2 - \frac{1}{\gamma} \big) & \text{if } \gamma \geq 1.
    \end{cases}
    \label{MSE-Wishart-SubL-MSE}
\end{equation}
Integrating over $\gamma$, we find the mutual information to be:
\begin{equation}
\begin{split}
        \lim_{N \to \infty} \frac{1}{M N} I_N(\bS; \bY) = \begin{cases}
      \frac{\gamma}{4} &\hspace{5pt} \text{if } \gamma \leq 1,\\
     \frac{1}{4} \frac{1}{\gamma} + \frac{1}{2} \ln \gamma  &\hspace{5pt} \text{if } \gamma \geq 1.
    \end{cases}
\end{split}
\label{MI-Wishart-SubL-MSE}
\end{equation}
which is the mutual information in the rank-one case when the prior for the spike is a Gaussian vector. 

Now, we compute the mutual information using spherical integrals.  All the non-zeros eigenvalues of sub-linear rank matrix $\sqrt{\gamma} \bS$, $\theta_i$'s, converge to a single number, $\sqrt{\gamma}$. By \cite{benaych2011eigenvalues}, the limiting  top eigenvalues of $\bY$, $\lambda_i$'s, can also be computed. So, all the summands in r.h.s. of \eqref{subL-spherical} are equal in the limit, and we have:
\begin{equation}
    \lim_{N \to \infty} \frac{1}{M N} \ln \int D \bU e^{\frac{N}{2} \Tr \sqrt{\gamma}\bS \bU  \bY \bU^\intercal } =  \begin{cases}
    \frac{1}{2} K( \sqrt{\gamma}, 2, \rho_{\rm sc}) &\hspace{5pt} \text{if } \gamma < 1,\\
    \frac{1}{2} K( \sqrt{\gamma}, \sqrt{\gamma} + \frac{1}{\sqrt{\gamma}}, \rho_{\rm sc})  &\hspace{5pt} \text{if } \gamma \geq 1.
    \end{cases}
    \label{limit-spherical-subL-Wish}
\end{equation}

\noindent\underline{Case 
$\gamma < 1$}:
we have $\lambda' = 2$ and $v = G^{-1}_{\rho_{\rm sc}}(\sqrt{\gamma}) = \sqrt{\gamma} + \frac{1}{\sqrt{\gamma}}$. Thus
\begin{equation}
    \begin{split}
        K( \sqrt{\gamma}, 2, \rho_{\rm sc}) &= 2 \sqrt{\gamma} + \big( \sqrt{\gamma} + \frac{1}{\sqrt{\gamma}} - 2 \big) G_{\rho_{\rm sc}} \big( \sqrt{\gamma} + \frac{1}{\sqrt{\gamma}} \big) - \ln \sqrt{\gamma} \\
        &\hspace{4cm} - \int_{-2}^2  \frac{\sqrt{4 -x^2}}{2 \pi}   \ln \big| \sqrt{\gamma} + \frac{1}{\sqrt{\gamma}} - x|\, dx - 1 \\
        &= 2 \sqrt{\gamma} + \big( \sqrt{\gamma} + \frac{1}{\sqrt{\gamma}} - 2 \big) \sqrt{\gamma} - \frac{1}{2} \ln \gamma - \big( \frac{\gamma}{2} - \frac{1}{2} \ln \gamma \big) - 1 \\
        &= \frac{\gamma}{2} 
    \end{split}
    \label{gamma <1}
\end{equation}
where we used the integral formula
\begin{equation*}
    \frac{1}{2 \pi}\int_{-2}^2 \ln(A - Bx)\sqrt{4- x^2} dx = \frac{A}{A + \sqrt{A^2 - 4B^2}} + \ln\big(A + \sqrt{A^2 - 4B^2}\big) - \frac{1}{2} - \ln 2 .
\end{equation*}

\noindent\underline{Case 
$\gamma \geq 1$}:
we have  $\lambda' = \sqrt{\gamma} + \frac{1}{\sqrt{\gamma}}$ and $v = \lambda' = \sqrt{\gamma} + \frac{1}{\sqrt{\gamma}}$. Thus
\begin{equation}
    \begin{split}
        K( \sqrt{\gamma}, \sqrt{\gamma} + \frac{1}{\sqrt{\gamma}}, \rho_{\rm sc}) &= \sqrt{\gamma} \big( \sqrt{\gamma} + \frac{1}{\sqrt{\gamma}} \big) - \ln \sqrt{\gamma} - \int_{-2}^2  \frac{\sqrt{4 -x^2}}{2 \pi}   \ln \big| \sqrt{\gamma} + \frac{1}{\sqrt{\gamma}} - x|\, dx - 1 \\
        &= \gamma + 1 - \frac{1}{2} \ln \gamma - \big( \frac{1}{2} \frac{1}{\gamma} + \frac{1}{2} \ln \gamma \big) - 1 \\
        &= \gamma - \frac{1}{2} \frac{1}{\gamma} - \ln \gamma .
    \end{split}
    \label{gamma >1}
\end{equation}
From \eqref{limit-spherical-subL-Wish}, \eqref{gamma <1}, and \eqref{gamma >1}, we obtain
\begin{equation}
    \lim_{N \to \infty} \frac{1}{M N} \ln \int D \bU e^{\frac{N}{2} \Tr \sqrt{\gamma}\bS \bU  \bY \bU^\intercal } = \begin{cases}
    \frac{\gamma}{4} &\hspace{5pt} \text{if } \gamma < 1,\\
    \frac{\gamma}{2} - \frac{1}{4} \frac{1}{\gamma} - \frac{1}{2} \ln \gamma  &\hspace{5pt} \text{if } \gamma \geq 1.
    \end{cases}
    \label{limit-spherical-subL}
\end{equation}
Replacing this result in \eqref{MI-subL-Spherical-integral} we get
\begin{equation*}
\begin{split}
        \lim_{N \to \infty} \frac{1}{M N} I_N(\bS; \bY) &= \begin{cases}
      \frac{\gamma}{4} &\hspace{5pt} \text{if } \gamma < 1,\\
     \frac{1}{4} \frac{1}{\gamma} + \frac{1}{2} \ln \gamma  &\hspace{5pt} \text{if } \gamma \geq 1.
    \end{cases}
\end{split}
\end{equation*}
which is the same as the mutual information computed from the MSE, \eqref{MI-Wishart-SubL-MSE}.

\begin{figure} 
    \centering
\begin{tikzpicture}[scale=0.5]

\definecolor{color0}{rgb}{0.254901960784314,0.411764705882353,0.882352941176471}

\begin{axis}[
legend cell align={left},
legend style={
  fill opacity=0.8,
  draw opacity=1,
  text opacity=1,
  at={(1,0.7)},
  anchor=east,
  draw=white!80!black,
  font = \large
},
tick align=outside,
tick pos=left,
x grid style={white!69.0196078431373!black},
xlabel={N},
xmin=800, xmax=5200,
xtick={1000,2000,3000,4000,5000},
xtick style={color=black},
y grid style={white!69.0196078431373!black},
ylabel={MSE},
ymin=0.326012356385822, ymax=0.778832815714406,
ytick style={color=black},
label style={font=\Huge},
tick label style={font=\Large}
]
\path [draw=red, semithick]
(axis cs:1000,0.733666668921809)
--(axis cs:1000,0.751854718520555);

\path [draw=red, semithick]
(axis cs:2000,0.745338848726101)
--(axis cs:2000,0.755781654716344);

\path [draw=red, semithick]
(axis cs:3000,0.747172586549511)
--(axis cs:3000,0.758250067563107);

\path [draw=red, semithick]
(axis cs:4000,0.752802323457094)
--(axis cs:4000,0.757448795640642);

\path [draw=red, semithick]
(axis cs:5000,0.752941689847099)
--(axis cs:5000,0.757002634199216);

\path [draw=color0, semithick]
(axis cs:1000,0.684526382814958)
--(axis cs:1000,0.7007695543899);

\path [draw=color0, semithick]
(axis cs:2000,0.705889258479596)
--(axis cs:2000,0.714555096143906);

\path [draw=color0, semithick]
(axis cs:3000,0.712645891384258)
--(axis cs:3000,0.722797743438596);

\path [draw=color0, semithick]
(axis cs:4000,0.721610139547382)
--(axis cs:4000,0.72552692609473);

\path [draw=color0, semithick]
(axis cs:5000,0.724000114647255)
--(axis cs:5000,0.728178810694598);

\path [draw=red, semithick]
(axis cs:1000,0.552938179065565)
--(axis cs:1000,0.565091370890975);

\path [draw=red, semithick]
(axis cs:2000,0.551468705169198)
--(axis cs:2000,0.560296143688402);

\path [draw=red, semithick]
(axis cs:3000,0.551308789669176)
--(axis cs:3000,0.559582107653629);

\path [draw=red, semithick]
(axis cs:4000,0.551827981162748)
--(axis cs:4000,0.555802843748155);

\path [draw=red, semithick]
(axis cs:5000,0.551243937700404)
--(axis cs:5000,0.554570685826205);

\path [draw=color0, semithick]
(axis cs:1000,0.527907767718043)
--(axis cs:1000,0.536868850575763);

\path [draw=color0, semithick]
(axis cs:2000,0.534694144882552)
--(axis cs:2000,0.542287978322589);

\path [draw=color0, semithick]
(axis cs:3000,0.540185001805502)
--(axis cs:3000,0.547321864433004);

\path [draw=color0, semithick]
(axis cs:4000,0.542906640474852)
--(axis cs:4000,0.546255729240202);

\path [draw=color0, semithick]
(axis cs:5000,0.544166695292419)
--(axis cs:5000,0.547029086814202);

\path [draw=red, semithick]
(axis cs:1000,0.429842654783681)
--(axis cs:1000,0.441020908586985);

\path [draw=red, semithick]
(axis cs:2000,0.432263519361399)
--(axis cs:2000,0.436138538733685);

\path [draw=red, semithick]
(axis cs:3000,0.434049559305501)
--(axis cs:3000,0.436703520743343);

\path [draw=red, semithick]
(axis cs:4000,0.433418277660082)
--(axis cs:4000,0.435489831021784);

\path [draw=red, semithick]
(axis cs:5000,0.433675740990736)
--(axis cs:5000,0.436312843333693);

\path [draw=color0, semithick]
(axis cs:1000,0.419433739694536)
--(axis cs:1000,0.430301728617743);

\path [draw=color0, semithick]
(axis cs:2000,0.42649600168471)
--(axis cs:2000,0.430363159297028);

\path [draw=color0, semithick]
(axis cs:3000,0.429809785757337)
--(axis cs:3000,0.432122042744788);

\path [draw=color0, semithick]
(axis cs:4000,0.42971138495746)
--(axis cs:4000,0.4318257750007);

\path [draw=color0, semithick]
(axis cs:5000,0.430555668862521)
--(axis cs:5000,0.433129171891342);

\path [draw=red, semithick]
(axis cs:1000,0.35166520962314)
--(axis cs:1000,0.359371004948253);

\path [draw=red, semithick]
(axis cs:2000,0.354619642817062)
--(axis cs:2000,0.357182825072237);

\path [draw=red, semithick]
(axis cs:3000,0.356100504248849)
--(axis cs:3000,0.358104674422276);

\path [draw=red, semithick]
(axis cs:4000,0.356131453213213)
--(axis cs:4000,0.357429218086829);

\path [draw=red, semithick]
(axis cs:5000,0.356785176110266)
--(axis cs:5000,0.358408361342378);

\path [draw=color0, semithick]
(axis cs:1000,0.346595104537121)
--(axis cs:1000,0.353834412765741);

\path [draw=color0, semithick]
(axis cs:2000,0.351368216510539)
--(axis cs:2000,0.353913719423411);

\path [draw=color0, semithick]
(axis cs:3000,0.35359485133456)
--(axis cs:3000,0.355502936712242);

\path [draw=color0, semithick]
(axis cs:4000,0.354144466767107)
--(axis cs:4000,0.35535860556719);

\path [draw=color0, semithick]
(axis cs:5000,0.354988504643638)
--(axis cs:5000,0.356606160957392);

\addplot [semithick, black]
table {%
800 0.75
5200 0.75
};
\addlegendentry{$\gamma = 2$}
\addplot [semithick, black, dashed]
table {%
800 0.555555555555555
5200 0.555555555555555
};
\addlegendentry{$\gamma = 3$}
\addplot [semithick, black, dotted]
table {%
800 0.4375
5200 0.4375
};
\addlegendentry{$\gamma = 4$}
\addplot [semithick, black, dash pattern=on 1pt off 3pt on 3pt off 3pt]
table {%
800 0.36
5200 0.36
};
\addlegendentry{$\gamma = 5$}
\addplot [semithick, red, mark=triangle*, mark size=3, mark options={solid}, only marks]
table {%
1000 0.742760693721182
2000 0.750560251721222
3000 0.752711327056309
4000 0.755125559548868
5000 0.754972162023157
};
\addlegendentry{RIE-MSE}
\addplot [semithick, color0, mark=triangle*, mark size=3, mark options={solid,rotate=180}, only marks]
table {%
1000 0.692647968602429
2000 0.710222177311751
3000 0.717721817411427
4000 0.723568532821056
5000 0.726089462670926
};
\addlegendentry{Oracle-MSE}
\addplot [semithick, red, mark=triangle*, mark size=3, mark options={solid}, only marks, forget plot]
table {%
1000 0.55901477497827
2000 0.5558824244288
3000 0.555445448661403
4000 0.553815412455452
5000 0.552907311763305
};
\addplot [semithick, color0, mark=triangle*, mark size=3, mark options={solid,rotate=180}, only marks, forget plot]
table {%
1000 0.532388309146903
2000 0.538491061602571
3000 0.543753433119253
4000 0.544581184857527
5000 0.54559789105331
};
\addplot [semithick, red, mark=triangle*, mark size=3, mark options={solid}, only marks, forget plot]
table {%
1000 0.435431781685333
2000 0.434201029047542
3000 0.435376540024422
4000 0.434454054340933
5000 0.434994292162215
};
\addplot [semithick, color0, mark=triangle*, mark size=3, mark options={solid,rotate=180}, only marks, forget plot]
table {%
1000 0.424867734156139
2000 0.428429580490869
3000 0.430965914251063
4000 0.43076857997908
5000 0.431842420376931
};
\addplot [semithick, red, mark=triangle*, mark size=3, mark options={solid}, only marks, forget plot]
table {%
1000 0.355518107285696
2000 0.355901233944649
3000 0.357102589335562
4000 0.356780335650021
5000 0.357596768726322
};
\addplot [semithick, color0, mark=triangle*, mark size=3, mark options={solid,rotate=180}, only marks, forget plot]
table {%
1000 0.350214758651431
2000 0.352640967966975
3000 0.354548894023401
4000 0.354751536167149
5000 0.355797332800515
};
\end{axis}

\end{tikzpicture}
    \captionsetup{singlelinecheck = false, justification=justified}
    \caption{{\small Comparison of the sub-linear RIE and the oracle estimator \eqref{Oracle-est} for Gaussian noise with sub-linear Wishart signal with $M = \lfloor \sqrt{N} \rfloor$. Horizontal lines are MMSE computed from \eqref{MSE-Wishart-SubL-MSE}. Points are averaged over 10 experiments (error bars might be invisible).}}
    \label{fig:Wishart-GNoise-Subl}
\end{figure}
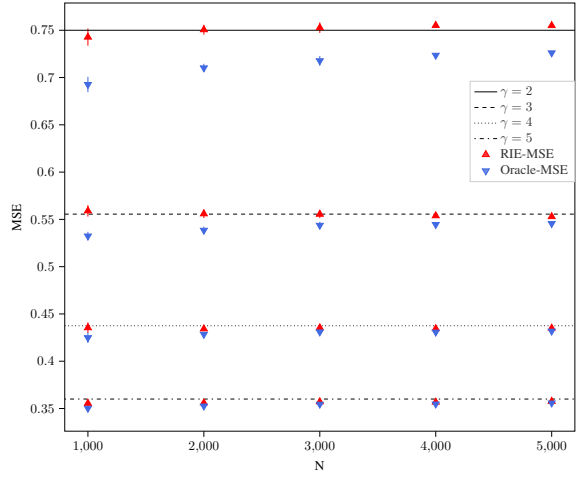

In Fig. \ref{fig:Wishart-GNoise-Subl}, we compare the performance of the sub-linear RIE and the oracle estimator \eqref{Oracle-est} with $M = \lfloor \sqrt{N} \rfloor$ for various values of $N$. We observe that the performance of the RIE is very close to the one of oracle estimator (which requires the knowledge of the signal). Moreover, the MSE is close to the theoretical predictions. 

\subsubsection{Example 2: Uniform distribution}
Let $\rho_S$ be the uniform distribution on $[1,2]$. From \eqref{RIE-MSE-SubL}, the MSE can be computed to be

\begin{equation}
    {\rm MSE}(\gamma) = \begin{cases}
        \frac{7}{3} \hspace{10 pt} &\text{for } 0 \leq \gamma \leq \frac{1}{4}, \\
         -\frac{1}{3} + \frac{4}{\gamma} - \frac{8}{3 \gamma^{3/2}} + \frac{1}{2 \gamma^2} &\text{for } \frac{1}{4} \leq \gamma \leq 1, \\
         \frac{2}{\gamma} - \frac{1}{2 \gamma^2} &\text{for } 1 \leq \gamma.
    \end{cases}
    \label{MSE-Uniform-GNoise-Subl}
\end{equation}
Integrating over $\gamma$, we find the mutual information to be:
\begin{equation}
    \lim_{N \to \infty} \frac{1}{M N} I_N(\bS; \bY) = \begin{cases}
        \frac{7}{12} \gamma \hspace{10 pt} &\text{for } 0 \leq \gamma \leq \frac{1}{4}, \\
        - \frac{\gamma}{12} + \ln \gamma +  2(\ln 2 -1)   + \frac{4}{3 \sqrt{\gamma}} - \frac{1}{8 \gamma}
         &\text{for } \frac{1}{4} \leq \gamma \leq 1, \\
         \frac{1}{2} \ln \gamma + 2 \ln 2 -1 + \frac{1}{8 \gamma} &\text{for } 1 \leq \gamma.
    \end{cases}
    \label{MI-Uniform-SubL}
\end{equation}

Now, we proceed with the computation of the mutual information using the asymptotic of spherical integrals. In the limit $N \to \infty$, r.h.s. of \eqref{subL-spherical} becomes an integral (expectation w.r.t. $\rho_S$). For $\rho_S = \mathcal{U}\big([1,2]\big)$ we have

\begin{equation}
    \lim_{N \to \infty} \frac{1}{M N} \ln \int D \bU e^{\frac{N}{2} \Tr \sqrt{\gamma}\bS \bU  \bY \bU^\intercal } = \frac{1}{2} \int_1^2 K \Big(\sqrt{\gamma} x, h \big(\sqrt{\gamma}x \big), \rho_{\rm sc} \Big) \, dx
    \label{limit-spherical-subL-Uniform}
\end{equation}
with
\begin{equation*}
    h \big(\sqrt{\gamma}x \big) = \begin{cases}
        2 \hspace{10 pt} &\text{if } \sqrt{\gamma}x \leq 1, \\
        \sqrt{\gamma}x + \frac{1}{\sqrt{\gamma}x} &\text{if } \sqrt{\gamma}x \geq 1.
    \end{cases}
\end{equation*}
\begin{equation*}
    K \Big(\sqrt{\gamma} x, h \big(\sqrt{\gamma}x \big), \rho_{\rm sc} \Big) = \begin{cases}
        \frac{1}{2} \gamma x^2 \hspace{10 pt} &\text{for } 0 < \gamma \leq \frac{1}{4}, \\
        \frac{1}{2} \gamma x^2 \hspace{10 pt} &\text{for } \frac{1}{4} \leq \gamma \leq 1 \text{ and } x \leq \frac{1}{\sqrt{\gamma}}, \\
        \gamma x^2 - \frac{1}{2} \frac{1}{\gamma x^2} - \ln \gamma - 2 \ln x \hspace{10 pt} &\text{for } \frac{1}{4} \leq \gamma \leq 1 \text{ and } x \geq \frac{1}{\sqrt{\gamma}}, \\
        \gamma x^2 - \frac{1}{2} \frac{1}{\gamma x^2} - \ln \gamma - 2 \ln x \hspace{10 pt} &\text{for } 1 \leq \gamma .
    \end{cases}
\end{equation*}
Thus we find:
\begin{equation}
    \lim_{N \to \infty} \frac{1}{M N} \ln \int D \bU e^{\frac{N}{2} \Tr \sqrt{\gamma}\bS \bU  \bY \bU^\intercal }  = \begin{cases}
        \frac{7}{12} \gamma \hspace{10 pt} &\text{for } 0 \leq \gamma \leq \frac{1}{4}, \\
        \frac{5}{4} \gamma - \ln \gamma + 2(1- \ln 2) - \frac{4}{3 \sqrt{\gamma}} + \frac{1}{8 \gamma}
         &\text{for } \frac{1}{4} \leq \gamma \leq 1, \\
        \frac{7}{6} \gamma - \frac{1}{2} \gamma + 1 - 2 \ln 2 - \frac{1}{8 \gamma} &\text{for } 1 \leq \gamma.
    \end{cases}
\end{equation}
Replacing in \eqref{MI-subL-Spherical-integral}, we find the same mutual information as in \eqref{MI-Uniform-SubL}.

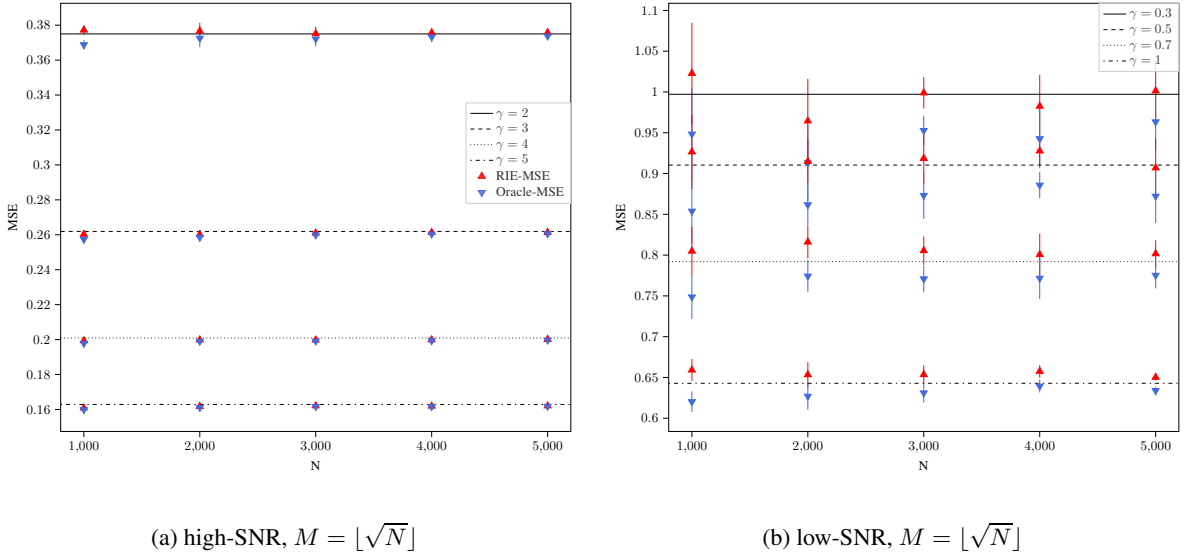
\begin{figure} 
    \centering
    \begin{subfigure}{.45\textwidth}
        \centering
\begin{tikzpicture}[scale=0.5]

\definecolor{color0}{rgb}{0.254901960784314,0.411764705882353,0.882352941176471}

\begin{axis}[
legend cell align={left},
legend style={
  fill opacity=0.8,
  draw opacity=1,
  text opacity=1,
  at={(1,0.65)},
  anchor=east,
  draw=white!80!black,
  font = \large
},
tick align=outside,
tick pos=left,
x grid style={white!69.0196078431373!black},
xlabel={N},
xmin=800, xmax=5200,
xtick={1000,2000,3000,4000,5000},
xtick style={color=black},
y grid style={white!69.0196078431373!black},
ylabel={MSE},
ymin=0.14739744714752, ymax=0.3925411044188,
ytick style={color=black},
label style={font=\Huge},
tick label style={font=\Large}
]
\path [draw=red, semithick]
(axis cs:1000,0.374555270732554)
--(axis cs:1000,0.379846705075193);

\path [draw=red, semithick]
(axis cs:2000,0.372041766906431)
--(axis cs:2000,0.381398210906469);

\path [draw=red, semithick]
(axis cs:3000,0.371248367406882)
--(axis cs:3000,0.379075161592976);

\path [draw=red, semithick]
(axis cs:4000,0.372968413311387)
--(axis cs:4000,0.378416082459333);

\path [draw=red, semithick]
(axis cs:5000,0.373613274911665)
--(axis cs:5000,0.37791149845147);

\path [draw=color0, semithick]
(axis cs:1000,0.366070245894261)
--(axis cs:1000,0.371535537293165);

\path [draw=color0, semithick]
(axis cs:2000,0.367411055415574)
--(axis cs:2000,0.377634582507151);

\path [draw=color0, semithick]
(axis cs:3000,0.367821973129673)
--(axis cs:3000,0.376316166016987);

\path [draw=color0, semithick]
(axis cs:4000,0.370260544474799)
--(axis cs:4000,0.37624443162448);

\path [draw=color0, semithick]
(axis cs:5000,0.371452997760447)
--(axis cs:5000,0.376049962413308);

\path [draw=red, semithick]
(axis cs:1000,0.258336339241446)
--(axis cs:1000,0.26230489076482);

\path [draw=red, semithick]
(axis cs:2000,0.257524532415374)
--(axis cs:2000,0.262782309627513);

\path [draw=red, semithick]
(axis cs:3000,0.259517617919271)
--(axis cs:3000,0.262462103217073);

\path [draw=red, semithick]
(axis cs:4000,0.259815099775714)
--(axis cs:4000,0.262866381931502);

\path [draw=red, semithick]
(axis cs:5000,0.260005161374835)
--(axis cs:5000,0.262738454626832);

\path [draw=color0, semithick]
(axis cs:1000,0.255419044832956)
--(axis cs:1000,0.259337473364742);

\path [draw=color0, semithick]
(axis cs:2000,0.255937844274957)
--(axis cs:2000,0.261277211422935);

\path [draw=color0, semithick]
(axis cs:3000,0.258319024646355)
--(axis cs:3000,0.261408904077346);

\path [draw=color0, semithick]
(axis cs:4000,0.258791829481276)
--(axis cs:4000,0.26197767679185);

\path [draw=color0, semithick]
(axis cs:5000,0.259182626404731)
--(axis cs:5000,0.262027061889653);

\path [draw=red, semithick]
(axis cs:1000,0.197169787430234)
--(axis cs:1000,0.201556156236994);

\path [draw=red, semithick]
(axis cs:2000,0.1988499934747)
--(axis cs:2000,0.200838126544153);

\path [draw=red, semithick]
(axis cs:3000,0.198500732457028)
--(axis cs:3000,0.201056948879234);

\path [draw=red, semithick]
(axis cs:4000,0.19906780054325)
--(axis cs:4000,0.200641228712817);

\path [draw=red, semithick]
(axis cs:5000,0.199283249318483)
--(axis cs:5000,0.20097769529827);

\path [draw=color0, semithick]
(axis cs:1000,0.195650253520794)
--(axis cs:1000,0.200103432051304);

\path [draw=color0, semithick]
(axis cs:2000,0.19797139299193)
--(axis cs:2000,0.200034374523537);

\path [draw=color0, semithick]
(axis cs:3000,0.19785662083972)
--(axis cs:3000,0.200464931111375);

\path [draw=color0, semithick]
(axis cs:4000,0.198553970603962)
--(axis cs:4000,0.20016695183687);

\path [draw=color0, semithick]
(axis cs:5000,0.198843469816714)
--(axis cs:5000,0.200568173547589);

\path [draw=red, semithick]
(axis cs:1000,0.159528426370739)
--(axis cs:1000,0.162165311606256);

\path [draw=red, semithick]
(axis cs:2000,0.160924303633166)
--(axis cs:2000,0.162488593362583);

\path [draw=red, semithick]
(axis cs:3000,0.160902391389901)
--(axis cs:3000,0.163260702297851);

\path [draw=red, semithick]
(axis cs:4000,0.161106823011692)
--(axis cs:4000,0.1627217497245);

\path [draw=red, semithick]
(axis cs:5000,0.161651406401781)
--(axis cs:5000,0.162433511020133);

\path [draw=color0, semithick]
(axis cs:1000,0.158540340659851)
--(axis cs:1000,0.161214117655572);

\path [draw=color0, semithick]
(axis cs:2000,0.16030979467122)
--(axis cs:2000,0.161945339072199);

\path [draw=color0, semithick]
(axis cs:3000,0.160497727078962)
--(axis cs:3000,0.162884380106536);

\path [draw=color0, semithick]
(axis cs:4000,0.160770458832752)
--(axis cs:4000,0.16238692644237);

\path [draw=color0, semithick]
(axis cs:5000,0.161390206078731)
--(axis cs:5000,0.162161303939132);

\addplot [semithick, black]
table {%
800 0.375
5200 0.375
};
\addlegendentry{$\gamma = 2$}
\addplot [semithick, black, dashed]
table {%
800 0.261904761904762
5200 0.261904761904762
};
\addlegendentry{$\gamma = 3$}
\addplot [semithick, black, dotted]
table {%
800 0.200892857142857
5200 0.200892857142857
};
\addlegendentry{$\gamma = 4$}
\addplot [semithick, black, dash pattern=on 1pt off 3pt on 3pt off 3pt]
table {%
800 0.162857142857143
5200 0.162857142857143
};
\addlegendentry{$\gamma = 5$}
\addplot [semithick, red, mark=triangle*, mark size=3, mark options={solid}, only marks]
table {%
1000 0.377200987903873
2000 0.37671998890645
3000 0.375161764499929
4000 0.37569224788536
5000 0.375762386681568
};
\addlegendentry{RIE-MSE}
\addplot [semithick, color0, mark=triangle*, mark size=3, mark options={solid,rotate=180}, only marks]
table {%
1000 0.368802891593713
2000 0.372522818961362
3000 0.37206906957333
4000 0.37325248804964
5000 0.373751480086877
};
\addlegendentry{Oracle-MSE}
\addplot [semithick, red, mark=triangle*, mark size=3, mark options={solid}, only marks, forget plot]
table {%
1000 0.260320615003133
2000 0.260153421021443
3000 0.260989860568172
4000 0.261340740853608
5000 0.261371808000833
};
\addplot [semithick, color0, mark=triangle*, mark size=3, mark options={solid,rotate=180}, only marks, forget plot]
table {%
1000 0.257378259098849
2000 0.258607527848946
3000 0.25986396436185
4000 0.260384753136563
5000 0.260604844147192
};
\addplot [semithick, red, mark=triangle*, mark size=3, mark options={solid}, only marks, forget plot]
table {%
1000 0.199362971833614
2000 0.199844060009427
3000 0.199778840668131
4000 0.199854514628034
5000 0.200130472308376
};
\addplot [semithick, color0, mark=triangle*, mark size=3, mark options={solid,rotate=180}, only marks, forget plot]
table {%
1000 0.197876842786049
2000 0.199002883757733
3000 0.199160775975547
4000 0.199360461220416
5000 0.199705821682151
};
\addplot [semithick, red, mark=triangle*, mark size=3, mark options={solid}, only marks, forget plot]
table {%
1000 0.160846868988498
2000 0.161706448497875
3000 0.162081546843876
4000 0.161914286368096
5000 0.162042458710957
};
\addplot [semithick, color0, mark=triangle*, mark size=3, mark options={solid,rotate=180}, only marks, forget plot]
table {%
1000 0.159877229157712
2000 0.16112756687171
3000 0.161691053592749
4000 0.161578692637561
5000 0.161775755008931
};
\end{axis}

\end{tikzpicture}
        \caption{high-SNR, $M = \lfloor \sqrt{N} \rfloor $}
    \end{subfigure}
    \hfil
    \begin{subfigure}{.45\textwidth}
        \centering
\begin{tikzpicture}[scale=0.5]

\definecolor{color0}{rgb}{0.254901960784314,0.411764705882353,0.882352941176471}

\begin{axis}[
legend cell align={left},
legend style={
  fill opacity=0.8,
  draw opacity=1,
  text opacity=1,
  at={(1,0.92)},
  anchor=east,
  draw=white!80!black,
  font =\large
},
tick align=outside,
tick pos=left,
x grid style={white!69.0196078431373!black},
xlabel={N},
xmin=800, xmax=5200,
xtick={1000,2000,3000,4000,5000},
xtick style={color=black},
y grid style={white!69.0196078431373!black},
ylabel={MSE},
ymin=0.583791186121482, ymax=1.10875654598248,
ytick style={color=black},
label style={font=\Huge},
tick label style={font=\Large}
]
\path [draw=red, semithick]
(axis cs:1000,0.960815575493745)
--(axis cs:1000,1.08489448417061);

\path [draw=red, semithick]
(axis cs:2000,0.913048125401593)
--(axis cs:2000,1.01585844448528);

\path [draw=red, semithick]
(axis cs:3000,0.979722220071167)
--(axis cs:3000,1.01809333195006);

\path [draw=red, semithick]
(axis cs:4000,0.944243668698909)
--(axis cs:4000,1.02093668342377);

\path [draw=red, semithick]
(axis cs:5000,0.964452045338307)
--(axis cs:5000,1.0382876205217);

\path [draw=color0, semithick]
(axis cs:1000,0.891340188651157)
--(axis cs:1000,1.00544796574789);

\path [draw=color0, semithick]
(axis cs:2000,0.866013500514207)
--(axis cs:2000,0.960047195122499);

\path [draw=color0, semithick]
(axis cs:3000,0.93482498518056)
--(axis cs:3000,0.970756111529459);

\path [draw=color0, semithick]
(axis cs:4000,0.906444066991506)
--(axis cs:4000,0.978915760519387);

\path [draw=color0, semithick]
(axis cs:5000,0.928474029070062)
--(axis cs:5000,0.998283770041598);

\path [draw=red, semithick]
(axis cs:1000,0.88092965572445)
--(axis cs:1000,0.97212587638142);

\path [draw=red, semithick]
(axis cs:2000,0.88783039653976)
--(axis cs:2000,0.941609913393092);

\path [draw=red, semithick]
(axis cs:3000,0.887167696549056)
--(axis cs:3000,0.949966778530177);

\path [draw=red, semithick]
(axis cs:4000,0.910395417938971)
--(axis cs:4000,0.945077506795876);

\path [draw=red, semithick]
(axis cs:5000,0.870940291784652)
--(axis cs:5000,0.94299928045235);

\path [draw=color0, semithick]
(axis cs:1000,0.814200051101877)
--(axis cs:1000,0.893550706442373);

\path [draw=color0, semithick]
(axis cs:2000,0.836906877362174)
--(axis cs:2000,0.886508179701693);

\path [draw=color0, semithick]
(axis cs:3000,0.84440953765064)
--(axis cs:3000,0.901942823120588);

\path [draw=color0, semithick]
(axis cs:4000,0.869771276435411)
--(axis cs:4000,0.901820138871555);

\path [draw=color0, semithick]
(axis cs:5000,0.839476343237612)
--(axis cs:5000,0.905031925761764);

\path [draw=red, semithick]
(axis cs:1000,0.775184433024185)
--(axis cs:1000,0.834571971890358);

\path [draw=red, semithick]
(axis cs:2000,0.79588345907419)
--(axis cs:2000,0.83622285679364);

\path [draw=red, semithick]
(axis cs:3000,0.78804164268507)
--(axis cs:3000,0.823018066163216);

\path [draw=red, semithick]
(axis cs:4000,0.774991194840363)
--(axis cs:4000,0.826253036436849);

\path [draw=red, semithick]
(axis cs:5000,0.785088653606025)
--(axis cs:5000,0.818225266446697);

\path [draw=color0, semithick]
(axis cs:1000,0.721779982372887)
--(axis cs:1000,0.775570622962876);

\path [draw=color0, semithick]
(axis cs:2000,0.754647624294284)
--(axis cs:2000,0.793826912990619);

\path [draw=color0, semithick]
(axis cs:3000,0.754551105475082)
--(axis cs:3000,0.787222834531037);

\path [draw=color0, semithick]
(axis cs:4000,0.746235209055009)
--(axis cs:4000,0.797203632476472);

\path [draw=color0, semithick]
(axis cs:5000,0.759212664851135)
--(axis cs:5000,0.791285192136432);

\path [draw=red, semithick]
(axis cs:1000,0.645732131480629)
--(axis cs:1000,0.672509826816284);

\path [draw=red, semithick]
(axis cs:2000,0.638041259939301)
--(axis cs:2000,0.668694906892285);

\path [draw=red, semithick]
(axis cs:3000,0.642330057106856)
--(axis cs:3000,0.664769195658966);

\path [draw=red, semithick]
(axis cs:4000,0.649786905732629)
--(axis cs:4000,0.665010456212907);

\path [draw=red, semithick]
(axis cs:5000,0.646093907425105)
--(axis cs:5000,0.654473247872459);

\path [draw=color0, semithick]
(axis cs:1000,0.607653247933345)
--(axis cs:1000,0.632945977108595);

\path [draw=color0, semithick]
(axis cs:2000,0.610346041803899)
--(axis cs:2000,0.643581237177786);

\path [draw=color0, semithick]
(axis cs:3000,0.619262260520803)
--(axis cs:3000,0.642906861371201);

\path [draw=color0, semithick]
(axis cs:4000,0.631578665420074)
--(axis cs:4000,0.647311047498248);

\path [draw=color0, semithick]
(axis cs:5000,0.628896223059695)
--(axis cs:5000,0.638781701398638);

\addplot [semithick, black]
table {%
800 0.997173872950271
5200 0.997173872950271
};
\addlegendentry{$\gamma = 0.3$}
\addplot [semithick, black, dashed]
table {%
800 0.910369000290069
5200 0.910369000290069
};
\addlegendentry{$\gamma = 0.5$}
\addplot [semithick, black, dotted]
table {%
800 0.792046585343264
5200 0.792046585343264
};
\addlegendentry{$\gamma = 0.7$}
\addplot [semithick, black, dash pattern=on 1pt off 3pt on 3pt off 3pt]
table {%
800 0.642857142857143
5200 0.642857142857143
};
\addlegendentry{$\gamma = 1$}
\addplot [semithick, red, mark=triangle*, mark size=3, mark options={solid}, only marks, forget plot]
table {%
1000 1.02285502983218
2000 0.964453284943438
3000 0.998907776010612
4000 0.982590176061341
5000 1.00136983293
};
\addplot [semithick, color0, mark=triangle*, mark size=3, mark options={solid,rotate=180}, only marks, forget plot]
table {%
1000 0.948394077199526
2000 0.913030347818353
3000 0.952790548355009
4000 0.942679913755446
5000 0.96337889955583
};
\addplot [semithick, red, mark=triangle*, mark size=3, mark options={solid}, only marks, forget plot]
table {%
1000 0.926527766052935
2000 0.914720154966426
3000 0.918567237539617
4000 0.927736462367424
5000 0.906969786118501
};
\addplot [semithick, color0, mark=triangle*, mark size=3, mark options={solid,rotate=180}, only marks, forget plot]
table {%
1000 0.853875378772125
2000 0.861707528531933
3000 0.873176180385614
4000 0.885795707653483
5000 0.872254134499688
};
\addplot [semithick, red, mark=triangle*, mark size=3, mark options={solid}, only marks, forget plot]
table {%
1000 0.804878202457271
2000 0.816053157933915
3000 0.805529854424143
4000 0.800622115638606
5000 0.801656960026361
};
\addplot [semithick, color0, mark=triangle*, mark size=3, mark options={solid,rotate=180}, only marks, forget plot]
table {%
1000 0.748675302667881
2000 0.774237268642452
3000 0.77088697000306
4000 0.77171942076574
5000 0.775248928493784
};
\addplot [semithick, red, mark=triangle*, mark size=3, mark options={solid}, only marks, forget plot]
table {%
1000 0.659120979148456
2000 0.653368083415793
3000 0.653549626382911
4000 0.657398680972768
5000 0.650283577648782
};
\addplot [semithick, color0, mark=triangle*, mark size=3, mark options={solid,rotate=180}, only marks, forget plot]
table {%
1000 0.62029961252097
2000 0.626963639490842
3000 0.631084560946002
4000 0.639444856459161
5000 0.633838962229167
};
\end{axis}

\end{tikzpicture}
        \caption{low-SNR, $M = \lfloor \sqrt{N} \rfloor $}
    \end{subfigure}
    \captionsetup{singlelinecheck = false, justification=justified}
    \caption{{\small Comparison of the sub-linear RIE and the oracle estimator \eqref{Oracle-est} for Gaussian noise with sub-linear signal with $\rho_S = \mathcal{U}([1,2])$. Horizontal lines are MSE computed from \eqref{MSE-Uniform-GNoise-Subl}. Points are averaged over 10 experiments (error bars might be invisible).}}
    \label{fig:Uniform-GNoise-Subl}
\end{figure}

In Fig. \ref{fig:Uniform-GNoise-Subl}, we compare the performance of the sub-linear RIE and the oracle estimator \eqref{Oracle-est} with $M = \lfloor \sqrt{N} \rfloor$ for various values of $N$ as well as with theoretical predictions for a uniformly distributed signal.

\subsection{Uniform Noise}
We now consider non-Gaussian noise, a noise matrix with eigenvalues which are uniformly distributed
$\rho_Z = \mathcal{U}\big([1,2]\big)$. We have $G_{\mathcal{U}([1,2])}(z) = \ln \frac{z-1}{z-2}$. Thus the eigenvalues of the proposed estimator are:
\begin{equation*}
    \xi^*_i = \frac{1}{\sqrt{\gamma}}   \mathbb{I}\big( \lambda^Y_i \notin [1,2] \big) \, \ln \frac{\lambda^Y_i-1}{\lambda^Y_i-2} \big(\lambda^Y_i-1\big)\big(\lambda^Y_i-2\big).
\end{equation*}

\subsubsection{MSE}
Writing the overlap in terms of the eigenvalues of the signal, for $1 \leq i \leq M$ we have
\begin{equation}
    \big( \bs_i^\intercal \by_i \big)^2 \to 
        \frac{1}{4} \frac{1}{\gamma {\lambda_i^S}^2} \big({\rm csch} \frac{1}{2 \sqrt{\gamma} {\lambda_i^S}} \big)^2 .
        \label{Overlap-Uniform-Noise}
\end{equation}
Note that, $G_{\mathcal{U}([1,2])}(1^-) = -\infty$ and $G_{\mathcal{U}([1,2])}(2^+) = +\infty$, so for any $\gamma>0$ we have an outlier eigenvalue for each spike in the signal. From \eqref{Overlap-Uniform-Noise}, the eigenvalues of the estimator are
\begin{equation}
    \xi_i^* \to \frac{1}{4} \frac{1}{\gamma {\lambda_i^S}} \big({\rm csch} \frac{1}{2 \sqrt{\gamma} {\lambda_i^S}} \big)^2.
\end{equation}
Using \eqref{MSE-expansion} we find in the asymptotic limit
\begin{equation}
\begin{split}
    \lim_{N \to \infty}  \frac{1}{M} \| \bS - \Xi^*(\bY) \|_F^2 & = \int \Big[ x^2 - \frac{1}{16}\frac{1}{\gamma^2 x^2} \big({\rm csch} \frac{1}{2 \sqrt{\gamma} x} \big)^4 \Big] \rho_S(x) \, dx.
    \end{split}
\label{RIE-MSE-UniformNoise-SubL}
\end{equation}

\subsubsection{Example: Sub-linear Wishart Signal}
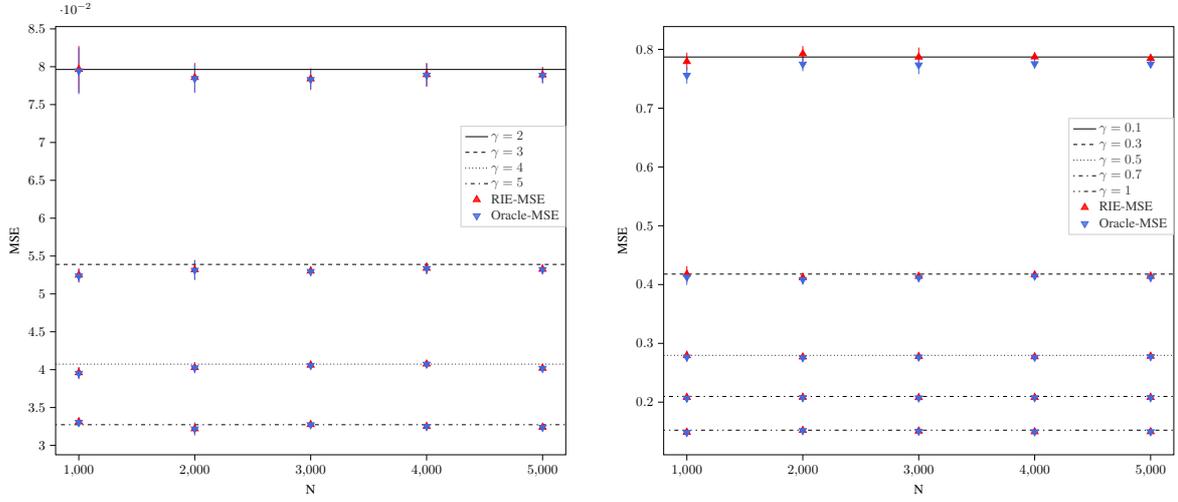
\begin{figure} 
    \centering
    \begin{subfigure}{.45\textwidth}
        \centering
\begin{tikzpicture}[scale=0.5]

\definecolor{color0}{rgb}{0.254901960784314,0.411764705882353,0.882352941176471}

\begin{axis}[
legend cell align={left},
legend style={
  fill opacity=0.8,
  draw opacity=1,
  text opacity=1,
  at={(1,0.65)},
  anchor=east,
  draw=white!80!black,
  font =\large
},
tick align=outside,
tick pos=left,
x grid style={white!69.0196078431373!black},
xlabel={N},
xmin=800, xmax=5200,
xtick={1000,2000,3000,4000,5000},
xtick style={color=black},
y grid style={white!69.0196078431373!black},
ylabel={MSE},
ymin=0.0287608168292293, ymax=0.0853024737672825,
ytick style={color=black},
label style={font=\Huge},
tick label style={font=\Large}
]
\path [draw=red, semithick]
(axis cs:1000,0.076585917378855)
--(axis cs:1000,0.0827323984519165);

\path [draw=red, semithick]
(axis cs:2000,0.0766781325512671)
--(axis cs:2000,0.0804866794128625);

\path [draw=red, semithick]
(axis cs:3000,0.0770271889969018)
--(axis cs:3000,0.0797824101079619);

\path [draw=red, semithick]
(axis cs:4000,0.0774250539008867)
--(axis cs:4000,0.0804993776863356);

\path [draw=red, semithick]
(axis cs:5000,0.0778619366866651)
--(axis cs:5000,0.0799827347105424);

\path [draw=color0, semithick]
(axis cs:1000,0.0763868023079984)
--(axis cs:1000,0.0825106367556997);

\path [draw=color0, semithick]
(axis cs:2000,0.0765485832982204)
--(axis cs:2000,0.0803194714584055);

\path [draw=color0, semithick]
(axis cs:3000,0.0769289816851065)
--(axis cs:3000,0.0796671994936632);

\path [draw=color0, semithick]
(axis cs:4000,0.0773340214728068)
--(axis cs:4000,0.0804141851102622);

\path [draw=color0, semithick]
(axis cs:5000,0.0777950812273843)
--(axis cs:5000,0.0799129352643127);

\path [draw=red, semithick]
(axis cs:1000,0.0516070264666059)
--(axis cs:1000,0.0533805545815468);

\path [draw=red, semithick]
(axis cs:2000,0.0519073789538836)
--(axis cs:2000,0.0544817002074495);

\path [draw=red, semithick]
(axis cs:3000,0.0523731207118066)
--(axis cs:3000,0.0536706147786686);

\path [draw=red, semithick]
(axis cs:4000,0.0526393205404704)
--(axis cs:4000,0.0541575010959501);

\path [draw=red, semithick]
(axis cs:5000,0.0526222053685577)
--(axis cs:5000,0.0538990549934259);

\path [draw=color0, semithick]
(axis cs:1000,0.0514753842640347)
--(axis cs:1000,0.0532400352993132);

\path [draw=color0, semithick]
(axis cs:2000,0.0518304988078132)
--(axis cs:2000,0.0544033934017677);

\path [draw=color0, semithick]
(axis cs:3000,0.0523202284278745)
--(axis cs:3000,0.0536042977127375);

\path [draw=color0, semithick]
(axis cs:4000,0.0525999079230591)
--(axis cs:4000,0.0541090621927017);

\path [draw=color0, semithick]
(axis cs:5000,0.0525776397680235)
--(axis cs:5000,0.0538615377062335);

\path [draw=red, semithick]
(axis cs:1000,0.0388433810092981)
--(axis cs:1000,0.0403560929882302);

\path [draw=red, semithick]
(axis cs:2000,0.0395682668741779)
--(axis cs:2000,0.0409972080246957);

\path [draw=red, semithick]
(axis cs:3000,0.0400105248179052)
--(axis cs:3000,0.0411853332869228);

\path [draw=red, semithick]
(axis cs:4000,0.0403214578005886)
--(axis cs:4000,0.0411748276016618);

\path [draw=red, semithick]
(axis cs:5000,0.0396858720963355)
--(axis cs:5000,0.0407020441372438);

\path [draw=color0, semithick]
(axis cs:1000,0.0387432629054724)
--(axis cs:1000,0.0402773457534399);

\path [draw=color0, semithick]
(axis cs:2000,0.0395239358781894)
--(axis cs:2000,0.04094526179145);

\path [draw=color0, semithick]
(axis cs:3000,0.0399771144863401)
--(axis cs:3000,0.0411397278948728);

\path [draw=color0, semithick]
(axis cs:4000,0.0402914747744348)
--(axis cs:4000,0.0411464310940816);

\path [draw=color0, semithick]
(axis cs:5000,0.0396616157281839)
--(axis cs:5000,0.040670724829994);

\path [draw=red, semithick]
(axis cs:1000,0.0325507694712263)
--(axis cs:1000,0.0336760448614999);

\path [draw=red, semithick]
(axis cs:2000,0.0313652792955138)
--(axis cs:2000,0.0330256140656536);

\path [draw=red, semithick]
(axis cs:3000,0.0322987197162089)
--(axis cs:3000,0.0332086562472225);

\path [draw=red, semithick]
(axis cs:4000,0.0321741578507535)
--(axis cs:4000,0.0328415665150699);

\path [draw=red, semithick]
(axis cs:5000,0.0319581980105462)
--(axis cs:5000,0.0328420078293044);

\path [draw=color0, semithick]
(axis cs:1000,0.0324474779302568)
--(axis cs:1000,0.0336146153385876);

\path [draw=color0, semithick]
(axis cs:2000,0.0313308921445953)
--(axis cs:2000,0.0329952207441336);

\path [draw=color0, semithick]
(axis cs:3000,0.0322692183533015)
--(axis cs:3000,0.0331732178592526);

\path [draw=color0, semithick]
(axis cs:4000,0.0321532588433516)
--(axis cs:4000,0.0328157266134135);

\path [draw=color0, semithick]
(axis cs:5000,0.0319383464327365)
--(axis cs:5000,0.0328210067175495);

\addplot [semithick, black]
table {%
800 0.0796385827691968
5200 0.0796385827691968
};
\addlegendentry{$\gamma = 2$}
\addplot [semithick, black, dashed]
table {%
800 0.0538953061405
5200 0.0538953061405
};
\addlegendentry{$\gamma = 3$}
\addplot [semithick, black, dotted]
table {%
800 0.0407276051993132
5200 0.0407276051993132
};
\addlegendentry{$\gamma = 4$}
\addplot [semithick, black, dash pattern=on 1pt off 3pt on 3pt off 3pt]
table {%
800 0.0327303337547009
5200 0.0327303337547009
};
\addlegendentry{$\gamma = 5$}
\addplot [semithick, red, mark=triangle*, mark size=3, mark options={solid}, only marks]
table {%
1000 0.0796591579153858
2000 0.0785824059820648
3000 0.0784047995524319
4000 0.0789622157936112
5000 0.0789223356986038
};
\addlegendentry{RIE-MSE}
\addplot [semithick, color0, mark=triangle*, mark size=3, mark options={solid,rotate=180}, only marks]
table {%
1000 0.079448719531849
2000 0.0784340273783129
3000 0.0782980905893849
4000 0.0788741032915345
5000 0.0788540082458485
};
\addlegendentry{Oracle-MSE}
\addplot [semithick, red, mark=triangle*, mark size=3, mark options={solid}, only marks, forget plot]
table {%
1000 0.0524937905240763
2000 0.0531945395806666
3000 0.0530218677452376
4000 0.0533984108182102
5000 0.0532606301809918
};
\addplot [semithick, color0, mark=triangle*, mark size=3, mark options={solid,rotate=180}, only marks, forget plot]
table {%
1000 0.0523577097816739
2000 0.0531169461047904
3000 0.052962263070306
4000 0.0533544850578804
5000 0.0532195887371285
};
\addplot [semithick, red, mark=triangle*, mark size=3, mark options={solid}, only marks, forget plot]
table {%
1000 0.0395997369987642
2000 0.0402827374494368
3000 0.040597929052414
4000 0.0407481427011252
5000 0.0401939581167896
};
\addplot [semithick, color0, mark=triangle*, mark size=3, mark options={solid,rotate=180}, only marks, forget plot]
table {%
1000 0.0395103043294562
2000 0.0402345988348197
3000 0.0405584211906064
4000 0.0407189529342582
5000 0.040166170279089
};
\addplot [semithick, red, mark=triangle*, mark size=3, mark options={solid}, only marks, forget plot]
table {%
1000 0.0331134071663631
2000 0.0321954466805837
3000 0.0327536879817157
4000 0.0325078621829117
5000 0.0324001029199253
};
\addplot [semithick, color0, mark=triangle*, mark size=3, mark options={solid,rotate=180}, only marks, forget plot]
table {%
1000 0.0330310466344222
2000 0.0321630564443645
3000 0.0327212181062771
4000 0.0324844927283825
5000 0.032379676575143
};
\end{axis}

\end{tikzpicture}
    \end{subfigure}
    \hfil
    \begin{subfigure}{.45\textwidth}
        \centering
\begin{tikzpicture}[scale=0.5]

\definecolor{color0}{rgb}{0.254901960784314,0.411764705882353,0.882352941176471}

\begin{axis}[
legend cell align={left},
legend style={
  fill opacity=0.8,
  draw opacity=1,
  text opacity=1,
  at={(1,0.65)},
  anchor=east,
  draw=white!80!black,
  font = \large
},
tick align=outside,
tick pos=left,
x grid style={white!69.0196078431373!black},
xlabel={N},
xmin=800, xmax=5200,
xtick={1000,2000,3000,4000,5000},
xtick style={color=black},
y grid style={white!69.0196078431373!black},
ylabel={MSE},
ymin=0.110484541682998, ymax=0.83870533806949,
ytick style={color=black},
label style={font=\Huge},
tick label style={font=\Large}
]
\path [draw=red, semithick]
(axis cs:1000,0.764234915329616)
--(axis cs:1000,0.79432526158904);

\path [draw=red, semithick]
(axis cs:2000,0.779592622742933)
--(axis cs:2000,0.805604392779195);

\path [draw=red, semithick]
(axis cs:3000,0.77068071934719)
--(axis cs:3000,0.803128401492857);

\path [draw=red, semithick]
(axis cs:4000,0.778793692193355)
--(axis cs:4000,0.795481412593214);

\path [draw=red, semithick]
(axis cs:5000,0.777555453438781)
--(axis cs:5000,0.791974026598852);

\path [draw=color0, semithick]
(axis cs:1000,0.741714362699429)
--(axis cs:1000,0.769908584325792);

\path [draw=color0, semithick]
(axis cs:2000,0.762866966032638)
--(axis cs:2000,0.78678192802058);

\path [draw=color0, semithick]
(axis cs:3000,0.757738251870806)
--(axis cs:3000,0.788987328723349);

\path [draw=color0, semithick]
(axis cs:4000,0.767193639290417)
--(axis cs:4000,0.783674063140833);

\path [draw=color0, semithick]
(axis cs:5000,0.768032457369806)
--(axis cs:5000,0.781889639310707);

\path [draw=red, semithick]
(axis cs:1000,0.404287874967189)
--(axis cs:1000,0.431319965187483);

\path [draw=red, semithick]
(axis cs:2000,0.403481149772281)
--(axis cs:2000,0.420071659484469);

\path [draw=red, semithick]
(axis cs:3000,0.407303828070511)
--(axis cs:3000,0.420860993435469);

\path [draw=red, semithick]
(axis cs:4000,0.412705510210193)
--(axis cs:4000,0.420107875030732);

\path [draw=red, semithick]
(axis cs:5000,0.406330000459835)
--(axis cs:5000,0.421712223478569);

\path [draw=color0, semithick]
(axis cs:1000,0.399437214854903)
--(axis cs:1000,0.425680199971258);

\path [draw=color0, semithick]
(axis cs:2000,0.399907227868053)
--(axis cs:2000,0.416421475468175);

\path [draw=color0, semithick]
(axis cs:3000,0.404816759765206)
--(axis cs:3000,0.418322217833501);

\path [draw=color0, semithick]
(axis cs:4000,0.41056943479383)
--(axis cs:4000,0.417988459800076);

\path [draw=color0, semithick]
(axis cs:5000,0.404453390252745)
--(axis cs:5000,0.419754652203414);

\path [draw=red, semithick]
(axis cs:1000,0.270755849170918)
--(axis cs:1000,0.287554015156509);

\path [draw=red, semithick]
(axis cs:2000,0.272249078446852)
--(axis cs:2000,0.281510667631655);

\path [draw=red, semithick]
(axis cs:3000,0.271613951842882)
--(axis cs:3000,0.28370884260004);

\path [draw=red, semithick]
(axis cs:4000,0.273145005940275)
--(axis cs:4000,0.281176534225448);

\path [draw=red, semithick]
(axis cs:5000,0.275479914116236)
--(axis cs:5000,0.280906131986505);

\path [draw=color0, semithick]
(axis cs:1000,0.26827203794768)
--(axis cs:1000,0.284842206158771);

\path [draw=color0, semithick]
(axis cs:2000,0.270893654148447)
--(axis cs:2000,0.280101390999888);

\path [draw=color0, semithick]
(axis cs:3000,0.270578082412814)
--(axis cs:3000,0.282557415941947);

\path [draw=color0, semithick]
(axis cs:4000,0.272275406089581)
--(axis cs:4000,0.280231892765595);

\path [draw=color0, semithick]
(axis cs:5000,0.274627682075173)
--(axis cs:5000,0.280033608605057);

\path [draw=red, semithick]
(axis cs:1000,0.20108765148086)
--(axis cs:1000,0.215035930732609);

\path [draw=red, semithick]
(axis cs:2000,0.204515534106675)
--(axis cs:2000,0.212249573480341);

\path [draw=red, semithick]
(axis cs:3000,0.205011424925197)
--(axis cs:3000,0.210638418626608);

\path [draw=red, semithick]
(axis cs:4000,0.204833219794216)
--(axis cs:4000,0.211230188264365);

\path [draw=red, semithick]
(axis cs:5000,0.205055313355013)
--(axis cs:5000,0.211109904904123);

\path [draw=color0, semithick]
(axis cs:1000,0.199844800436521)
--(axis cs:1000,0.213537600701287);

\path [draw=color0, semithick]
(axis cs:2000,0.203678741655324)
--(axis cs:2000,0.211374532850855);

\path [draw=color0, semithick]
(axis cs:3000,0.204341589935883)
--(axis cs:3000,0.209878343189385);

\path [draw=color0, semithick]
(axis cs:4000,0.204283297066169)
--(axis cs:4000,0.210709229056819);

\path [draw=color0, semithick]
(axis cs:5000,0.204643472083196)
--(axis cs:5000,0.210657936971353);

\path [draw=red, semithick]
(axis cs:1000,0.144218559962257)
--(axis cs:1000,0.153511531509454);

\path [draw=red, semithick]
(axis cs:2000,0.149047970040577)
--(axis cs:2000,0.154955416848398);

\path [draw=red, semithick]
(axis cs:3000,0.148249209375587)
--(axis cs:3000,0.153388194337925);

\path [draw=red, semithick]
(axis cs:4000,0.148530147259173)
--(axis cs:4000,0.150784324789727);

\path [draw=red, semithick]
(axis cs:5000,0.148324642438598)
--(axis cs:5000,0.151271380810841);

\path [draw=color0, semithick]
(axis cs:1000,0.143585486973293)
--(axis cs:1000,0.152851318150913);

\path [draw=color0, semithick]
(axis cs:2000,0.148605691641338)
--(axis cs:2000,0.154552541960996);

\path [draw=color0, semithick]
(axis cs:3000,0.147930543076359)
--(axis cs:3000,0.153024764054303);

\path [draw=color0, semithick]
(axis cs:4000,0.148271586796563)
--(axis cs:4000,0.150511404382376);

\path [draw=color0, semithick]
(axis cs:5000,0.148084701571445)
--(axis cs:5000,0.151039028430027);

\addplot [semithick, black]
table {%
800 0.786982356176059
5200 0.786982356176059
};
\addlegendentry{$\gamma = 0.1$}
\addplot [semithick, black, dashed]
table {%
800 0.417771414079225
5200 0.417771414079225
};
\addlegendentry{$\gamma = 0.3$}
\addplot [semithick, black, dotted]
table {%
800 0.279599367201854
5200 0.279599367201854
};
\addlegendentry{$\gamma = 0.5$}
\addplot [semithick, black, dash pattern=on 1pt off 3pt on 3pt off 3pt]
table {%
800 0.20968475690483
5200 0.20968475690483
};
\addlegendentry{$\gamma = 0.7$}
\addplot [semithick, black, dash pattern=on 1pt off 3pt on 1pt off 3pt on 3pt off 3pt on 3pt off 3pt]
table {%
800 0.152360132928505
5200 0.152360132928505
};
\addlegendentry{$\gamma = 1$}
\addplot [semithick, red, mark=triangle*, mark size=3, mark options={solid}, only marks]
table {%
1000 0.779280088459328
2000 0.792598507761064
3000 0.786904560420024
4000 0.787137552393285
5000 0.784764740018816
};
\addlegendentry{RIE-MSE}
\addplot [semithick, color0, mark=triangle*, mark size=3, mark options={solid,rotate=180}, only marks]
table {%
1000 0.755811473512611
2000 0.774824447026609
3000 0.773362790297078
4000 0.775433851215625
5000 0.774961048340257
};
\addlegendentry{Oracle-MSE}
\addplot [semithick, red, mark=triangle*, mark size=3, mark options={solid}, only marks, forget plot]
table {%
1000 0.417803920077336
2000 0.411776404628375
3000 0.41408241075299
4000 0.416406692620463
5000 0.414021111969202
};
\addplot [semithick, color0, mark=triangle*, mark size=3, mark options={solid,rotate=180}, only marks, forget plot]
table {%
1000 0.41255870741308
2000 0.408164351668114
3000 0.411569488799354
4000 0.414278947296953
5000 0.412104021228079
};
\addplot [semithick, red, mark=triangle*, mark size=3, mark options={solid}, only marks, forget plot]
table {%
1000 0.279154932163714
2000 0.276879873039254
3000 0.277661397221461
4000 0.277160770082861
5000 0.27819302305137
};
\addplot [semithick, color0, mark=triangle*, mark size=3, mark options={solid,rotate=180}, only marks, forget plot]
table {%
1000 0.276557122053225
2000 0.275497522574168
3000 0.27656774917738
4000 0.276253649427588
5000 0.277330645340115
};
\addplot [semithick, red, mark=triangle*, mark size=3, mark options={solid}, only marks, forget plot]
table {%
1000 0.208061791106734
2000 0.208382553793508
3000 0.207824921775903
4000 0.208031704029291
5000 0.208082609129568
};
\addplot [semithick, color0, mark=triangle*, mark size=3, mark options={solid,rotate=180}, only marks, forget plot]
table {%
1000 0.206691200568904
2000 0.20752663725309
3000 0.207109966562634
4000 0.207496263061494
5000 0.207650704527274
};
\addplot [semithick, red, mark=triangle*, mark size=3, mark options={solid}, only marks, forget plot]
table {%
1000 0.148865045735856
2000 0.152001693444488
3000 0.150818701856756
4000 0.14965723602445
5000 0.149798011624719
};
\addplot [semithick, color0, mark=triangle*, mark size=3, mark options={solid,rotate=180}, only marks, forget plot]
table {%
1000 0.148218402562103
2000 0.151579116801167
3000 0.150477653565331
4000 0.14939149558947
5000 0.149561865000736
};
\end{axis}

\end{tikzpicture}
    \end{subfigure}
    \captionsetup{singlelinecheck = false, justification=justified}
    \caption{{\small Comparison of the sub-linear RIE and the oracle estimator \eqref{Oracle-est} for noise with uniform spectral distribution and sub-linear Wishart signal, for $M = \lfloor \sqrt{N} \rfloor$. Horizontal lines are MSE computed from $1 - \frac{1}{16}\frac{1}{\gamma^2} \big({\rm csch} \frac{1}{2 \sqrt{\gamma}} \big)^4$. Points are averaged over 10 experiments (error bars might be invisible). Note that unlike the Gaussian noise, for any SNR the estimation is possible and MSE is less than 1.}}
    \label{fig:Wishart-UniformNoise-Subl}
\end{figure}

As mentioned in section \ref{Wishart-subL-GNoise}, the limiting measure of the signal in this case is $\rho_S = \delta_{+1}$, so the MSE is $1 - \frac{1}{16}\frac{1}{\gamma^2} \big({\rm csch} \frac{1}{2 \sqrt{\gamma}} \big)^4$. 
In Fig. \ref{fig:Wishart-UniformNoise-Subl}, we compare the performance of the sub-linear RIE and the oracle estimator \eqref{Oracle-est} for $M = \lfloor \sqrt{N} \rfloor$ for various values of $N$. We observe that the performance of the RIE is very close to the one of oracle estimator (which requires the knowledge of the signal). Moreover, the MSE is close to the theoretical predictions. 

\section*{Acknowledgments} 
We acknowledge important insights from Léo Miolane in early stages of this project. We are also thankful to Antoine Maillard for interesting discussions. The work of F. P has been supported by Swiss National Science Foundation grant no 200021-204119. J. B was funded by the European Union (ERC, CHORAL, project number 101039794). Views and opinions expressed are those of the author(s) only and do not reflect those of the European Union or the European Research Council. 
Neither the European Union nor the granting authority are responsible for them.

\bibliographystyle{unsrt}
\bibliography{References}

\clearpage
\appendix

\section{Discussion of models with rotation invariant noise}\label{proof-theorem-4}

Suppose that the noise matrix $\bZ$ in our basic model is the realization of a rotation invariant ensemble. While we are not quite able to treat this case, we can generalize Theorem 3 to the setting $\bY_\epsilon= \sqrt{\gamma}\bS + \bZ_\epsilon$ where $\bZ_{\epsilon}= \bZ + \sqrt\epsilon\bzeta$ with $\bzeta$ from the Gaussian Wigner ensemble, and $\epsilon>0$ (so the noise is non-Gaussian rotation invariant).
We call this model the {\it Additive Rotation Invariant Noise} (ARIN) model.
\begin{proposition}[\textbf{Explicit Mutual Information for the ARIN model}]
Assume that the conditions in assumption 1 hold for both $\bS$and $\bZ$. Then, we have for the ARIN model:
\begin{equation}
 \frac{{I}_N(\bS; \bY_\epsilon)}{N^2}  \xrightarrow{N\to\infty}   \frac{1}{2} \iint \ln |s-t| \rho_{Y_\epsilon}(s) \rho_{Y_\epsilon}(t) \, ds \, dt - \frac{1}{2} \iint \ln |s-t| \rho_{Z_\epsilon}(s) \rho_{Z_\epsilon} (t) \, ds \, dt.
\end{equation}
\end{proposition}
The proof leverages only on the simple formula for the  mutual information for Gaussian noise in theorem 3, and does not really hinge on assumption 2 (in main text).
Finally we would like to take the limit $\epsilon\to 0$. This however is a subtle problem which would require more specific hypothesis on the rotation invariant ensemble of $\bZ$. For example it is quite apparent that one would need the existence of a density for the limiting empirical measures $\rho_Z$ and $\rho_{Y_0}$. Moreover, this also requires an argument to permute the limits 
$N\to +\infty$ and $\epsilon\to 0$.


\begin{proof}
Set 
$\bX= \sqrt\gamma \bS + \bZ$. We have the information theoretic equalities ($\mathcal{H}_N$ are Shannon entropies)
\begin{equation}
    \begin{split}
        I_N(\bY_\epsilon; \bS) & = \mathcal{H}_N(\bY_\epsilon) - \mathcal{H}_N(\bY_\epsilon \vert \bS)  \\
& = \mathcal{H}_N(\bY_\epsilon) - \mathcal{H}_N(\bZ_\epsilon) \\ 
&=[\mathcal{H}_N(\bY_\epsilon) - \mathcal{H}_N(\bzeta)] - [\mathcal{H}_N(\bZ_\epsilon) - \mathcal{H}_N(\bzeta)]\\
&=[\mathcal{H}_N(\bY_\epsilon) - \mathcal{H}_N(\bY_\epsilon \vert \bX)] - [\mathcal{H}_N(\bZ_\epsilon) - \mathcal{H}_N(\bZ_\epsilon\vert \bZ)]\\ 
& =I_N(\bY_\epsilon; \bX) - I_N(\bZ_\epsilon; \bZ).
    \end{split}
\end{equation}
The two mutual informations in the last line correspond to the inference models for two AWGN models with strength $\epsilon$, namely 
$\bY_\epsilon = \bX + \sqrt\epsilon \bzeta$ and $\bZ_\epsilon = \bZ + \sqrt\epsilon\zeta$, and  inputs from rotation invariant ensembles. By the formula in Theorem 2 which holds for these channels we obtain 
\begin{align}
\lim_{N\to +\infty}\frac{1}{N^2} I_N(\bY_\epsilon; \bS) = \frac{1}{2}\int\int\ln|s-t| \rho_{Y_\epsilon}(s)\rho_{Y_\epsilon}(t) ds dt - 
\frac{1}{2}\int\int\ln|s-t| \rho_{Z_\epsilon}(s) \rho_{Z_\epsilon}(t) ds dt .
\end{align}
\end{proof}

\section{Replica symmetric formula for sub-linear rank matrix factorization}\label{replica-subL}

In this section, we provide the heuristic approach based on the replica symmetric method from statistical physics to derive the mutual information and minimum mean-square error of sub-linear rank matrix factorization (Statement 4). In order to be self-contained, we first recall the definition of the model.


The ground-truth matrix signal ${\bm X}=(X_{ik})\in\mathbb{R}^{N\times M}$ with $M= \lfloor N^{\alpha}\rfloor$ is generated from a prior distribution
assumed to be fully factorized 
\begin{align}
 P_{X,N}(\bX)=\prod_{i\le N} \prod_{k\le M} p_{X}(X_{ik})
\end{align}
for a law $p_X$ with finite variance, symmetric $p_X(x)=p_X(-x)$ and hence centered $\int dp_X(x)x=0$. Let $\bZ=\bZ^\intercal\in\mathbb{R}^{N\times N}$ a noise Wigner matrix with p.d.f. 
$$
P_{Z,N}(\bZ)\propto \exp\Big(-\frac{1}4{\rm Tr}\bZ^2\Big).
$$ 
Note the scaling here which corresponds to eigenvalues of order $O(\sqrt{N})$ for $\bZ$. The symmetric data matrix $\bY=(Y_{ij})\in\mathbb{R}^{N\times N}$ has entries generated through the following observation channel
\begin{align}\label{model}
\bY=\sqrt{\frac\gamma N}\,\bX\bX^\intercal + {\bm Z}.
\end{align}
Matrix $\sqrt{\gamma/N}\,\bX\bX^\intercal$ has  $O(\sqrt N)$ eigenvalues like the noise (and generically $O(1)$ entries), hence the scaling 
$\sqrt{\gamma/N}$ of the signal-to-noise ratio. After basic simplifications the Bayesian posterior reads 
\begin{align*}
&dP_{X\mid Y, N}(\bx\mid \bY) =\frac{1}{\mathcal{Z}(\bY)}dP_{X,N}(\bx)\exp\frac{1}{2}{\rm Tr}\Big(\sqrt{\frac{\gamma}{N}}\bY\bx\bx^\intercal-\frac{\gamma}{2N}(\bx\bx^\intercal)^2\Big).
\end{align*}
The mutual information $I(\bY;\bX)$, which we aim at computing, is 
\begin{align*}
I(\bY;\bX)&=-\mathbb{E}\ln \int dP_{X,N}(\bx)\exp\frac{1}{2}{\rm Tr}\Big(\sqrt{\frac{\gamma}{N}}\bY\bx\bx^\intercal-\frac{\gamma}{2N}(\bx\bx^\intercal)^2\Big)+\frac{\gamma}{4N}\EE{\rm Tr}(\bX\bX^\intercal)^2  
\end{align*}
where the first term is minus the expected free entropy defined as $$\EE f_N:=\frac{1}{MN}\mathbb{E}\ln \mathcal{Z}(\bY).$$
It is useful to note that model \eqref{model} is equivalent to 
\begin{align*}
\begin{cases}
Y_{ij}\sim\mathcal{N}\big(\sqrt{\frac{\gamma}{N}}\langle\bX_{i}, \bX_{j}\rangle,1\big) \quad &\text{for} \quad i<j \in[N]^2,\nn
Y_{ii}\sim\mathcal{N}\big(\sqrt{\frac{\gamma}{N}}\|\bX_{i}\|^2,2\big) \quad &\text{for} \quad i \in[N].
\end{cases}
\end{align*}
The average free entropy then concretely reads
\begin{align}
\EE f_N&=\frac{1}{NM}\EE \ln \int dP_{X,N}(\bx)\exp\frac12\sum_{i\le N}\Big(\sqrt{\frac{\gamma}{N}}Y_{ii}\|\bx_{i}\|^2-\frac{\gamma}{2N}\|\bx_{i}\|^4\Big)\nn
&\qquad\qquad\qquad\qquad\times \exp\sum_{i<j}^{1,N}\Big(\sqrt{\frac{\gamma}{N}}  Y_{ij}\langle\bx_{i}, \bx_{j}\rangle-\frac{\gamma}{2N}\langle\bx_{i}, \bx_{j}\rangle^2\Big).\label{eq3}
\end{align}

\subsection{Replica method: first steps}

The approach starts from the replica trick:
\begin{align}
\lim_{N\to+\infty}\EE f_N &=\lim_{N\to+\infty}\lim_{u\to0_+}\frac{1}{NMu}\ln\EE \mathcal{Z}(\bY)^u=\lim_{u\to0_+}\lim_{N\to+\infty}\frac{1}{NMu}\ln\EE \mathcal{Z}(\bY)^u.\label{replicaTrick}
\end{align}
As it is standard in the replica method, we assume that the $u$ and $N$ limits commute, and that the formulas derived for integer $u$ can be "analytically continued" to real $u\to 0_+$. We therefore evaluate the expectation $\EE \mathcal{Z}(\bY)^u$ of the replicated partition function for integer $u$. We directly integrate the quenched Gaussian observations in \eqref{eq3} using the following useful formula: 
\begin{align}
Y\sim\mathcal{N}\Big(\sqrt{\frac\gamma N}f_0,t\Big)\quad \Rightarrow\quad  \ \EE_{Y|f_0} \prod_{a\le u}\exp\frac1t\Big(\sqrt{\frac{\gamma}N}Yf_a-\frac{\gamma}{2N}f_a^2\Big)=\prod_{a< b}^{0,u}\exp\frac{\gamma}{t N}f_af_b.\label{useful_form}
\end{align}
In the Bayes-optimal setting the ground-truth $\bX$ plays the role of one additional replica, so we rename it $\bx^0:= \bX$. We set $\bx_i^a=(x_{ik}^a)_{k\le M}\in \mathbb{R}^M$ and introduce the notation $$\int dP_{X,N}(\{\bx\}_0^u)\,\cdots := \int_{\mathbb{R}^{MN(u+1)}}\prod_{a=0}^u\prod_{i\le N}\prod_{k\le M} dp_{X,N}(x^a_{ik})\,\cdots\,.$$ 
For replica indices $0\le a<b\le u$, define the $M\times M$ overlap matrices
\begin{align}
{\bQ}^{ab}:=\Big(\frac{1}{N}\sum_{i\le N}x_{ik}^a x_{i\ell}^b\Big)_{k,\ell\le M}=\frac{1}{N}(\bx^a)^\intercal  \bx^b=({\bQ}^{ba})^\intercal.\label{overlap_HDiL}
\end{align}
Then the above formula \eqref{useful_form} yields
\begin{align}
\EE \mathcal{Z}(\bY)^u&=   \int dP_{X,N}(\{\bx\}_0^u)\prod_{a< b}^{0,u}\exp\Big(\frac{\gamma}{2N}\sum_{i\le N}\|\bx_{i}^a\|^2\|\bx_{i}^b\|^2+\frac{\gamma}{N}\sum_{i<j}^{1,N}\langle \bx_{i}^a, \bx_{j}^a\rangle\langle\bx_{i}^b,  \bx_{j}^b\rangle\Big)\nn
&=  \int dP_{X,N}(\{\bx\}_0^u) \prod_{a< b}^{0,u} \exp\frac\gamma2  N\sum_{k,\ell\le M}\Big(\frac1{N}\sum_{i\le N}x_{ik}^a  x_{i\ell}^b \Big)^2 \nn
&=   \int dP_{X,N}(\{\bx\}_0^u) \prod_{a< b}^{0,u} \exp\frac\gamma2 N {\rm Tr}{\bQ}^{ab}({\bQ}^{ab})^\intercal.\label{complex_eq}
\end{align}

We introduce Dirac delta functions and their Fourier representation in order to linearize the above quadratic term in the overlaps (and thus quartic in the replicas $(\bx^a)$)
\begin{align*}
&\EE \mathcal{Z}(\bY)^u= \int dP_{X,N}(\{\bx\}_0^u)  \prod_{a< b}^{0,u}d\bQ^{ab}\delta\big((\bx^a)^\intercal \bx^b-N\bQ^{ab}\big) \exp \frac{\gamma}2 N {\rm Tr}{\bQ}^{ab}({\bQ}^{ab})^\intercal\nn
&\quad= \int dP_{X,N}(\{\bx\}_0^u)  \prod_{a< b}^{0,u} d\bQ^{ab}d\bRR^{ab}\exp {\rm Tr}\Big((\bRR^{ab})^\intercal(\bx^a)^\intercal \bx^b-N(\bRR^{ab})^\intercal\bQ^{ab} + \frac\gamma 2 N {\bQ}^{ab}({\bQ}^{ab})^\intercal\Big).
\end{align*}
In this expression we used the notations $d\bQ^{ab}=\prod_{k,\ell\le M} dQ^{ab}_{k\ell}$ and similarly for $d\bRR^{ab}$. We have $${\rm Tr} (\bRR^{ab})^\intercal(\bx^a)^\intercal \bx^b=\sum_{i\le N}\sum_{k,\ell\le M}R^{ab}_{k\ell}x_{ik}^a x_{i\ell}^b.$$
This implies that the integral can be re-written as
\begin{align}
\EE \mathcal{Z}(\bY)^u &= \int  \prod_{a< b}^{0,u} d\bQ^{ab}d\bRR^{ab}\exp N{\rm Tr}\Big( -(\bRR^{ab})^\intercal\bQ^{ab}+ \frac\gamma 2 {\bQ}^{ab}({\bQ}^{ab})^\intercal\Big)\nn
&\qquad\times \prod_{i\le N} \int \prod_{a=0}^u \prod_{k\le M} dp_X(x^a_{ik})  \exp\sum_{a<b}^{0,u}\sum_{k,\ell\le M}R^{ab}_{k\ell}x_{ik}^a x_{i\ell}^b\nn
&= \int  \prod_{a< b}^{0,u} d\bQ^{ab}d\bRR^{ab}\exp N{\rm Tr}\Big( -(\bRR^{ab})^\intercal\bQ^{ab}+ \frac\gamma 2 {\bQ}^{ab}({\bQ}^{ab})^\intercal\Big)\nn
&\qquad\times \exp N \ln \int \prod_{a=0}^u  dP_X(\by^a)\exp\sum_{a<b}^{0,u}(\by^a)^\intercal \bRR^{ab}\by^b 
\end{align}
where $\by^a=(y_k^a)_{k\le M}\in\mathbb{R}^M$ and $dP_X(\by^a)=\prod_{k\le M} dp_X(y^a_k)$. The factorization of the above integral over $i\le N$ is the first key mechanism.

We will use the singular value decompositions (SVDs)
\begin{align*}
\bQ^{ab}={\bA}^{ab}{\bsig}^{ab}({\bB}^{ab})^\intercal, \qquad \bRR^{ab}={\bC}^{ab}{\btau}^{ab}({\bD}^{ab})^\intercal,
\end{align*}
where $({\bA}^{ab},{\bB}^{ab},{\bC}^{ab},\bD^{ab})$ are orthogonal matrices of singular vectors, and $(\bsig^{ab},\btau^{ab})$ are positive semi-definite diagonal matrices of singular values. All matrices are of size $M\times M$. For diagonal matrices we simply index their non-zero entries with a single index. We also introduce the Vandermonde determinant 
$$
\Delta(\bsig):=\prod_{k< \ell}(\sigma_k-\sigma_\ell)
$$ 
and the Haar measure $\mu_M$ over the orthogonal group of orthogonal $M\times M$ matrices. Vandermonde determinants appear as Jacobians when introducing the SVDs. Thus
\begin{align}
\EE \mathcal{Z}(\bY)^u = &\int  \prod_{a< b}^{0,u} d\bsig^{ab}d\btau^{ab}|\Delta((\bsig^{ab})^2)\Delta((\btau^{ab})^2)| \exp N \frac\gamma 2{\rm Tr} (\bsig^{ab})^2\nn
&\quad\times\int \prod_{a<b}^{0,u}d\mu_M(\bC^{ab})d\mu_M(\bD^{ab}) \exp N \ln \int \prod_{a=0}^u  dP_X(\by^a)\exp\sum_{a<b}^{0,u}(\by^a)^\intercal {\bC}^{ab}{\btau}^{ab}({\bD}^{ab})^\intercal\by^b \nn
&\quad\quad\times\int \prod_{a<b}^{0,u} d\mu_M(\bA^{ab})d\mu_M(\bB^{ab}) \exp N{\rm Tr}\big( -\bD^{ab}\btau^{ab}(\bC^{ab})^\intercal {\bA}^{ab}{\bsig}^{ab}({\bB}^{ab})^\intercal\big).\label{9}
\end{align}
The singular values $(\sigma_k^{ab})_{k\le M}$ and $(\tau_k^{ab})_{k\le M}$ are considered ordered in decreasing order $$
\sigma_1^{ab}\ge \sigma_2^{ab}\ge \cdots\ge \sigma_M^{ab}, \qquad \tau_1^{ab}\ge \tau_2^{ab}\ge \cdots\ge \tau_M^{ab},
$$
and the integration over these variables takes that ordering into account (their permutations are taken into account by the integration over the singular vectors). The second key mechanism is the fact that we can absorb the singular vectors of $\bRR^{ab}$ into those of $\bQ^{ab}$ in the last line of \eqref{9} when integrating $(\bA^{ab},\bB^{ab})_{a<b}$: we change $(\bB^{ab})^\intercal\bD^{ab}\to (\bB^{ab})^\intercal$ and $(\bC^{ab})^\intercal\bA^{ab}\to \bA^{ab}$; these transformations have unit Jacobian. This decouples the integration over the orthogonal matrices in the second and last lines. 
%

\subsection{Integration over the orthogonal group by saddle point}\label{sec:saddleOrtho}

In this section we are going to evaluate certain integrals by saddle point and Laplace methods. In spin glass models, when doing so, it may happen that due to the replica limit $u\to 0_+$, instead of maximizing the exponent, one must minimize it; see, e.g., \cite{mezard2009information,mezard1987spin}. This, however, does not happen in the present setting due to the Bayes-optimality which implies that the model lies on its Nishimori line \cite{nishimori2001statistical}. Therefore, we will follow the ``normal'' recipes of saddle point and Laplace methods even in the limit $u\to 0_+$.

We argue that, in the regime $M = \lfloor N^{\alpha}\rfloor$, $\alpha\in (0, 1)$, the last two terms in \eqref{9} can be integrated out by the saddle point method. Indeed, in both terms the exponent is $\Theta(NM) = \Theta(N^{1+\alpha})$ while the number of degrees of freedom over which we integrate is much smaller $\Theta(M^2)= \Theta(N^{2\alpha})$ (as long as $\alpha<1$). Moreover, in this regime Gaussian fluctuations around the saddle points are sub-leading.

\subsubsection*{First saddle point} We start with the last term appearing in \eqref{9} corresponding to the rectangular spherical integral \cite{guionnet2021large} 
\begin{align}
 &\int d\mu_M(\bA^{ab})d\mu_M(\bB^{ab}) \exp N{\rm Tr}\big(- \btau^{ab} {\bA}^{ab}{\bsig}^{ab}({\bB}^{ab})^\intercal\big) \nn
 &\qquad=C_M\exp (o(NM))\sum_{*}\exp \big(-N{\rm Tr}\,\btau^{ab} {\bA}_*^{ab}{\bsig}^{ab}({\bB}_*^{ab})^\intercal\big)\label{10}
\end{align}
where the sum $\sum_*$ is over all saddle points $({\bA}_*^{ab},{\bB}_*^{ab})$ and $C_M$ is a volume factor coming from the normalization of the Haar measure. From now on $C_M$ or $C_N$ denotes a generic irrelevant constant that can change from place to place and that may include $\exp (o(NM))$ corrections. We focus on a given pair of replica indices $(ab)$ so we drop this notation when there is no ambiguity. To find the saddle point equation we perturb the matrices $(\bA,\bB)$ as $(\bA(I_M+\delta \bS),\bB(I_M+\delta\bT))$ for two generic anti-symmetric matrices $\bS^\intercal =-\bS$ and $\bT^\intercal =-\bT$ and a small $\delta$. Then at a saddle point the first order variation in $\delta$ of ${\rm Tr}\, \btau\bA\bsig\bB^\intercal$ under this perturbation must cancel, namely,
\begin{align}
  \Big|\Big(\frac{\partial}{\partial \delta}{\rm Tr}\,\btau\bA(I_M+\delta \bS)\bsig(I_M-\delta\bT)\bB^\intercal\Big)_{\delta=0}\Big|=\big|{\rm Tr}\bB^\intercal\btau \bA (\bS\bsig -  \bsig \bT)\big|=0.
\end{align}
This trace being zero means that, given $(\btau,\bsig)$, the matrix $\bB^\intercal\btau \bA$ lies in the orthogonal complement of $(\bS\bsig -  \bsig \bT)^\intercal$ under the Hilbert-Schmidt inner product ${\rm Tr}\bX\bY^\intercal$. 
%
As will see, the only way to find a solution is to look for symmetric saddle points belonging to the subspace of matrices verifying $\bB=\bA$; this is strongly suggested by the $\bA\leftrightarrow \bB$ symmetry of ${\rm Tr}\, \btau\bA\bsig\bB^\intercal$. 
%
%
The argument goes a follows. Because $\bsig$ is diagonal and $\bS,\bT$ have zero diagonal (by antisymmetry), both $\bS\bsig$ and $\bsig \bT$ have zero entries on their diagonal as well. The orthogonal complement of the set of matrices with vanishing diagonal is the set of diagonal matrices. Therefore ${\rm Tr}\bB^\intercal\btau \bA (\bS\bsig -  \bsig \bT)=0$ if and only if $\bB^\intercal\btau \bA$ is diagonal. This is possible only if $\bA=\bB$ is a (possibly signed) permutation matrix (because $\btau$ is already diagonal and the set of signed permutation matrices is the normalizer of the diagonal matrices in the group of orthogonal matrices). So for any perturbation $(\bS,\bT)$ and choice of $(\bsig,\btau)$, if $\bA=\bB$ belongs to the set
$$\Pi_M:=\{\boldsymbol{\pi} : \boldsymbol{\pi} \ \mbox{is a} \ M\times M \ \mbox{permutation matrix with signed non-zero entries}\},$$
then the first order variation around $\bA=\bB$ cancels. This means that the set of saddle points is $\Pi_M$. Denoting $$\pi(\bsig):=\boldsymbol{\pi}\bsig\boldsymbol{\pi}^\intercal, \qquad \sigma_{\pi_k}:=\pi(\bsig)_k,$$ where $\pi(\bsig)$ is the permuted diagonal matrix $\bsig$, \eqref{10} becomes
\begin{align}
 &C_M\sum_{\Pi_M}\exp\Big(-N\sum_{k\le M}\tau_k^{ab} \sigma^{ab}_{\pi_k}  \Big).\label{14}
\end{align}

The above sum on $\Pi_M$, whose cardinal is $2^MM!=\Theta(\exp(M\ln M))$, is over exponentially large terms of order $\exp(\Theta(MN))$. Therefore it can be estimated by Laplace approximation. In other words only the dominating terms matter, so we need to understand which choices of permutations minimize $\sum_{k}\tau_k^{ab} \sigma^{ab}_{\pi_k}$. By the {\it rearrangement inequality} \cite{hardy1952inequalities} this corresponds to the permutation which orders $(\tau_k^{ab})_{k}$ and $(\sigma_k^{ab})_k$ in opposite order. Therefore \eqref{14} becomes
\begin{align}
 &\int d\mu_M(\bA^{ab})d\mu_M(\bB^{ab}) \exp N{\rm Tr}\big(- \btau^{ab} {\bA}^{ab}{\bsig}^{ab}({\bB}^{ab})^\intercal\big) =C_M\exp\Big(-N\sum_{k\le M}\tau_k^{ab} \sigma^{ab}_{M+1-k}  \Big)\label{14_ordered}
\end{align}
where $\tau_{k}^{ab}\ge \tau_{k+1}^{ab}$ and $\sigma_{k}^{ab}\ge \sigma_{k+1}^{ab}$ for all $k\le M$.



\subsubsection*{Second saddle point}

We now consider the second term appearing in \eqref{9}. Let $\EE_{\mu}$ be the joint expectation over the Haar distributed matrices $(\bC^{ab},\bD^{ab})_{a<b}$.
We need to compute 
\begin{align}
\EE_{\mu} \exp N \phi_M\big((\bC^{ab},\bD^{ab},\btau^{ab})_{a<b}\big)  ,\label{Ephi}
\end{align}
where $\phi_M$, which plays the role of effective action for $(\bC^{ab},\bD^{ab})_{a<b}$, is
\begin{align}
\phi_M\big((\bC^{ab},\bD^{ab},\btau^{ab})_{a<b}\big):=\ln \int \prod_{a=0}^u  dP_X(\by^a)\exp\sum_{a<b}^{0,u}(\by^a)^\intercal {\bC}^{ab}{\btau}^{ab}({\bD}^{ab})^\intercal\by^b  . \label{phi} 
\end{align}
Note that it is of order $M$. A saddle point approximation to the integral over the orthogonal matrices yields
\begin{align}
\EE_{\mu} \exp N \phi_M\big((\bC^{ab},\bD^{ab},\btau^{ab})_{a<b}\big) = C_M  \sum_{*}\exp N \phi_M\big((\bC^{ab}_*,\bD^{ab}_*,\btau^{ab})_{a<b}\big)  ,\label{Ephi_saddle}
\end{align}
where $(\bC^{ab}_*,\bD^{ab}_*)_{a<b,*}$ are the saddle points. 
Symmetry suggests again that $\bC^{ab}_*=\bD^{ab}_*$.
We are going to argue that the saddle points are the elements of $\Pi_M$ as in the previous saddle point estimate. We consider a generic perturbation $(\bC^{ab}(I_M+\delta \bS^{ab}),\bD^{ab}(I_M+\delta\bT^{ab}))$ for anti-symmetric matrices $(\bS^{ab})^\intercal =-\bS^{ab}$ and $(\bT^{ab})^\intercal =-\bT^{ab}$. Then the first order variation in $\delta$ of the action reads
\begin{align}
&\Big|\Big(\frac{\partial}{\partial \delta}\phi_M\big((\bC^{ab}(I_M+\delta \bS^{ab}),\bD^{ab}(I_M+\delta\bT^{ab}),\btau^{ab})\big)\Big)_{\delta=0}\Big|\nn
&\qquad=  \Big|\sum_{a<b}{\rm Tr}(\bD^{ab})^\intercal\langle \by^b(\by^a)^\intercal \rangle \bC^{ab}({\btau}^{ab}\bT^{ab}-\bS^{ab}{\btau}^{ab})\Big| .\label{17}
\end{align}
The expectation $\langle \, \cdot\, \rangle$ acting on $(\by^a)_{a=0}^u$ is induced by the Hamiltonian defined by the exponent in \eqref{phi} with generic matrices $(\bC^{ab},\bD^{ab})_{a<b}$, under the reference measure $P_X$ and at fixed $(\btau^{ab})_{a<b}$:
\begin{align}
\langle f((\by^a)) \rangle  :=\frac{\int \prod_{a=0}^u  dP_X(\by^a) \, f((\by^a)) \, \exp\sum_{a<b}^{0,u}(\by^a)^\intercal {\bC}^{ab}{\btau}^{ab}({\bD}^{ab})^\intercal\by^b    }{\int \prod_{a=0}^u  dP_X(\by^a)  \exp\sum_{a<b}^{0,u}(\by^a)^\intercal {\bC}^{ab}{\btau}^{ab}({\bD}^{ab})^\intercal\by^b }.\label{brackety}
\end{align}
%
We assume that in order to make the sum \eqref{17} vanish, each terms in the sum must cancel. For a given pair $a<b$, the trace in \eqref{17} can thus cancel only if $(\bD^{ab})^\intercal\langle \by^b(\by^a)^\intercal \rangle \bC^{ab}$ lies in the orthogonal complement of $({\btau}^{ab}\bT^{ab}-\bS^{ab}{\btau}^{ab})^\intercal$. 
Because as before ${\btau}^{ab}\bT^{ab}-\bS^{ab}{\btau}^{ab}$ has zero diagonal, it thus requires $(\bD^{ab})^\intercal\langle \by^b(\by^a)^\intercal \rangle \bC^{ab}$ to be diagonal.
We now show that a sufficient condition for this is that $\bC^{ab}=\bD^{ab}=\boldsymbol{\pi}^{ab}\in\Pi_M$. In this case we can see from \eqref{phi} that $y^a_k$ is independent of $y^b_\ell$, 
\begin{align}
\langle y^a_ky^b_\ell\rangle= \langle y^a_k\rangle \langle y^b_\ell\rangle \  \forall \  a<b \ \mbox{and} \ k\neq \ell.\label{decop}
\end{align}
This is because the following interaction matrices are diagonal: $$\pi^{ab}(\btau^{ab}):=\boldsymbol{\pi}^{ab}\btau^{ab}(\boldsymbol{\pi}^{ab})^\intercal.$$  
Thanks to the aforementioned decoupling taking place when considering $\bC^{ab}=\bD^{ab}=\boldsymbol{\pi}^{ab}\in \Pi_M$,
\begin{align}
 \langle y_k^a\rangle\propto \int \prod_{b=0}^u  dp_X(y^b)\,y^a\,\exp\Big(y^a\sum_{b (\neq a)}^{0,u} y^b \pi^{ab}(\btau^{ab})_k \Big).\label{ferro}
\end{align}
%
This is the local mean magnetization of a spin system $(y^a)_{a=0}^{u}$. Because the prior is symmetric and centered, this local magnetization is null by the global sign symmetry $(y^a)\to (-y^a)$ of its associated Gibbs measure: $$\langle y_k^a\rangle=0  \ \forall \  k,a.$$ 
Combined with \eqref{decop} this implies that the matrix $\langle \by^b(\by^a)^\intercal  \rangle=(\langle y_k^by_\ell^a\rangle )_{k,\ell\le M}$ is diagonal and thus the matrix $(\boldsymbol{\pi}^{ab})^\intercal\langle \by^b (\by^a)^\intercal\rangle\boldsymbol{\pi}^{ab}$ is also diagonal (as $\boldsymbol{\pi}^{ab}\in\Pi_M$). We thus have ${\rm Tr}(\boldsymbol{\pi}^{ab})^\intercal\langle \by^b (\by^a)^\intercal\rangle\boldsymbol{\pi}^{ab}({\btau}^{ab}\bT^{ab}-\bS^{ab}{\btau}^{ab}) =0$ for any antisymmetric perturbations $(\bS^{ab},\bT^{ab})$ and diagonal $(\btau^{ab})$. So \eqref{17} does cancel in that case.
%
%
This ends the argument that matrices in $\Pi_M$ are again saddle points. 

Showing that \emph{only} the matrices in $\Pi_M$ are saddle points requires showing that  $\bM^{ab}:=(\bD^{ab})^\intercal\langle \by^b (\by^a)^\intercal\rangle\bC^{ab}$ cannot be diagonal if $\bC^{ab}\neq \bD^{ab}$ (but they are allowed to be permutations) or if $\bC^{ab}= \bD^{ab}$ is not a permutation. In both scenarios, $\langle \by^b (\by^a)^\intercal\rangle$ and $\bM^{ab}$ have no reasons to be diagonal. Having $\bM^{ab}$ diagonal would require finding interaction matrices $(\bC^{ab}\btau^{ab}(\bD^{ab})^\intercal)_{a<b}$ entering the spin model associated to the log-partition function \eqref{phi} such that $(\bC^{ab},\bD^{ab})$ also represent the singular vectors of $\langle \by^b (\by^a)^\intercal\rangle$. We believe that this is not possible in general and thus make the assumption that $\Pi_M$ contains all saddle points.
%
Therefore \eqref{Ephi_saddle} becomes
\begin{align}
C_M  \sum_{(\boldsymbol{\pi}^{ab})_{a<b}\in\Pi_M^{u(u+1)/2}}\exp\Big( N\sum_{k\le M} \ln \int \prod_{a=0}^u  dp_X(y^a)\exp\sum_{a<b}^{0,u}y^a y^b \pi^{ab}(\btau^{ab})_k\Big) .\label{saddle2final}
\end{align}


\subsection{Spectral replica symmetry and simplifications}

Now we assume the following ``spectral replica symmetric ansatz'' for the order parameters: 
\begin{align}
 \mbox{Spectral replica symmetry:}  \ \ \btau^{ab}=\btau \ \mbox{and} \ \bsig^{ab}=\bsig \ \forall \ a<b.\label{specRS}
\end{align}
%
The sum over the permutation matrices in \eqref{saddle2final} simply becomes
\begin{align}
C_M  \sum_{(\boldsymbol{\pi}^{ab})_{a<b}\in\Pi_M^{u(u+1)/2}}\exp\Big( N\sum_{k\le M} \ln \int \prod_{a=0}^u  dp_X(y^a)\exp\sum_{a<b}^{0,u}y^a y^b \pi^{ab}(\btau)_k\Big) .\label{saddle2final_2}
\end{align}
The sum \eqref{saddle2final_2} is over $(M!)^{u(u+1)/2}$ exponentially large terms of order $\exp(\Theta(MN))$ so a before it can be estimated by Laplace approximation. We thus need to understand which choices of permutations $(\boldsymbol{\pi}^{ab})_{a<b}$ maximize the ``free entropy''
\begin{align}
g_M:=\sum_{k\le M} \ln \int \prod_{a=0}^u  dp_X(y^a)\exp\sum_{a<b}^{0,u}y^a y^b \pi^{ab}(\btau)_k.
\end{align}
We assume here that there are only two potential candidates for the maximizers corresponding to opposite extreme cases: the fully ordered and fully disordered scenarios.

\subsubsection*{Ordered scenario} 
This corresponds to consider matrices $(\boldsymbol{\pi}^{ab})_{a<b}$ all equal, i.e., an additional layer of replica symmetric ansatz:
\begin{align}
\boldsymbol{\pi}^{ab}=\boldsymbol{\pi} \  \forall \ a<b. 
\end{align}
In this case the free entropy $g_M$ reads
\begin{align}
&g_M=\sum_{k\le M} \ln \int \prod_{a=0}^u  dp_X(y^a)\exp\Big(\tau_{\pi_k}\sum_{a<b}^{0,u}y^a y^b\Big)=\sum_{k\le M} \ln \int \prod_{a=0}^u  dp_X(y^a)\exp\Big(\tau_{k}\sum_{a<b}^{0,u}y^a y^b\Big)
\label{free_entrop_ferro}
\end{align}
as this is the same for any permutation $\boldsymbol{\pi}\in\Pi_M$. We call this the ``ordered'' scenario as this free entropy becomes, at fixed $k$, the one of a standard \emph{ferromagnetic} model (i.e., with only positive interactions) without disorder, that is with same interaction strength $\tau_{k}>0$ (which is a quenched random variable in this spin model) for all pairs of spins $(y^a,y^b)_{a<b}$. We can simplify it as follows. Let $Z$ be an i.i.d. standard normal random variable. We thus compute
\begin{align}
& \sum_{k\le M}\ln \int \prod_{a=0}^u  dp_X(y^a)\exp\Big( \frac{1}2\Big(\sqrt{\tau_{k}}\sum_{a}^{0,u}y^a\Big)^2-\frac{\tau_{k}}2\sum_{a}^{0,u}(y^a)^2\Big)\nn
&\qquad= \sum_{k\le M}\ln \EE \Big(\int    dp_X(y)\exp\Big(\sqrt{\tau_{k}}Zy-\frac{\tau_{k}}2y^2\Big)\Big)^{u+1}\nn
&\qquad=u \sum_{k\le M}\EE\int    dp_X(y_0)\exp\Big(\sqrt{\tau_{k}}Zy_0-\frac{\tau_{k}}2y_0^2\Big) \nn
&\qquad\qquad\times \ln \int    dp_X(y)\exp\Big(\sqrt{\tau_{k}}Zy-\frac{\tau_{k}}2y^2\Big)+ O(u^2)\nn
&\qquad=u \sum_{k\le M}\EE\ln \int    dp_X(y)\exp\Big(\sqrt{\tau_{k}}Zy+\tau_{k}yY_0-\frac{\tau_{k}}2y^2\Big)+ O(u^2) \label{phistar}
\end{align}
using the change of variable $Z\to Z-\sqrt{\tau_{\pi_k}}y_0$ for the last step, and where $Y_0\sim p_X$. So with this choice \eqref{saddle2final} would become
\begin{align}
C_M\exp\Big(u N\sum_{k\le M}\EE\ln \int    dp_X(y)\exp\Big(\sqrt{\tau_{k}}Zy+\tau_{k}yY_0-\frac{\tau_{k}}2y^2\Big)+ O(u^2) \Big) .\label{28}
\end{align}

\subsubsection*{Disordered scenario}
In the opposite scenario, the permutations $(\boldsymbol{\pi}^{ab})_{a<b}$ are ``typical'' (i.e., taken randomly and independently for different pairs $a<b$). In this case, in the regime $M\gg 1$ with finitely many replicas, we have
\begin{align}
 \frac{g_M}{M}\approx \EE \psi((\tau^{ab})_{a<b}):=\EE_{(\tau^{ab})}\ln \int \prod_{a=0}^u  dp_X(y^a)\exp\sum_{a<b}^{0,u}y^a y^b \tau^{ab}\label{gtau}
\end{align}
where $(\tau^{ab})_{a<b}$ are i.i.d. random ferromagnetic interactions $\tau^{ab}\sim \rho_\tau$ whose law is the weak limit  
$$\rho_\tau:=\lim_{M\to\infty}\frac1M\sum_{k\le M}\delta_{\tau_k}.$$ 
This is because picking a fixed diagonal entry $k$ of random permutations $\pi^{ab}(\btau)_k$ of a diagonal matrix $\btau$ is the same as randomly picking diagonal entries of $\btau$. Then $\EE \psi$ is the average free entropy of a \emph{disordered} mean-field ferromagnetic spin system. 

\subsubsection*{Comparing the two scenarios}
We now argue that the ordered scenario yields a greater free entropy than the disordered scenario, and therefore must be selected in the Laplace approximation to the sum \eqref{saddle2final}, which implies that \eqref{saddle2final} equals \eqref{28} at leading order. To do so we are going to compare two random free entropies (we work in the large $M$ limit directly). Let $\tau,(\tau^{ab})_{a<b}$ be random variables all i.i.d. with distribution $\rho_\tau$ defined above, and just for this section set $\btau:=(\tau^{ab})_{a<b}\in\mathbb{R}_{>0}^{u(u+1)/2}$. Let
\begin{align}
 \psi(\btau)&:=\ln \int \prod_{a=0}^u  dp_X(y^a)\exp\sum_{a<b}^{0,u}y^a y^b \tau^{ab},\label{psigene}\\
\xi(\tau)&:=\ln \int \prod_{a=0}^u  dp_X(y^a)\exp\Big(\tau\sum_{a<b}^{0,u}y^a y^b\Big).
 \label{psis}
\end{align}
We also define the Gibbs measure for the disordered scenario:
\begin{align}
\langle f((y^a)) \rangle_{\btau}  :=\frac{\int \prod_{a=0}^u  dp_X(y^a) \, f((y^a)) \, \exp\sum_{a<b}^{0,u}y^a y^b \tau^{ab}   }{\int \prod_{a=0}^u  dp_X(y^a) \exp\sum_{a<b}^{0,u}y^a y^b \tau^{ab}  }.\label{bracketpsi}
\end{align}
We denote it $\langle f((y^a)) \rangle_\tau $ for the ordered scenario where all $\tau^{ab}=\tau$.

Showing that the ordered scenario yields a greater free entropy than the disordered scenario is then equivalent (when $M$ is large) to show that
\begin{align}
\EE \xi(\tau)\ge \EE\psi(\btau)\label{orderGretDiso} .
\end{align}
Note that $\psi((\tau))=\xi(\tau)$ (where $(\tau)$ is the constant vector), so in particular
\begin{align}
\psi((\EE\tau))=\xi(\EE\tau).  
\end{align}
So the statement would follow if one could show the following equivalent inequality on the Jensen gaps: 
\begin{align}
\EE \xi(\tau)-\xi(\EE\tau)\ge \EE\psi(\btau)-\psi((\EE\tau)).\label{jensengap} 
\end{align}

An heuristic suggesting the validity of \eqref{jensengap} goes as follows. These two Jensen gaps can be approximated by evaluating the Laplacian
\begin{align}
 \Delta \xi(\tau)&= \Big\langle \Big(\sum_{a<b}^{0,u} y^a y^b\Big)^2\Big\rangle_\tau-\Big\langle\sum_{a<b}^{0,u} y^a y^b\Big\rangle_\tau^2=\sum_{a<b}^{0,u}\sum_{c<d}^{0,u}\big(\langle y^a y^b y^cy^d\rangle_\tau-\langle y^a y^b\rangle_\tau \langle y^cy^d\rangle_\tau\big)\ge 0,\label{Laplacianxi}
\end{align}
and comparing it to 
\begin{align}
  \Delta\psi(\btau)&= \sum_{a<b}^{0,u}\big(\langle (y^a y^b)^2 \rangle_{\btau}-\langle y^a y^b\rangle_{\btau}^2\big)\ge 0.
\end{align}
Because the models corresponding to the two free entropies $\psi,\xi$ are ferromagnetic and invariant under spin-flip, the Griffiths inequality implies that each term in the double sum in \eqref{Laplacianxi} is non-negative, and thus suggests that $\Delta \xi(\tau)\ge   \Delta\psi(\btau)$ may be true. This argument points towards $\xi$ being ``more convex'' than $\psi$ and thus that inequality \eqref{orderGretDiso} holds.

\subsection{Replica symmetric free entropy}

Under these assumptions we can combine everything to obtain an explicit formula for the replicated partition function. Recall that we assume that the saddle point over the singular values verifies \eqref{specRS}. Thus 
\begin{align}
\EE \mathcal{Z}^u &=  C_N\int   d\bsig d\btau\exp NM\epsilon_N u\Big(\frac{u+1}{2}\frac1{M^2}\sum_{k<\ell\le M}\ln |\sigma_k^2-\sigma_\ell^2||\tau_k^2-\tau_\ell^2|\Big) \nn
&\qquad\times\exp NMu\Big(-\frac{u+1}2\frac1M\sum_{k\le M}\tau_k {\sigma}_{M+1-k} +\frac{u+1}2\frac{\gamma} {2M}  \sum_{k\le M}\sigma_k^2\Big) \nn
&\qquad\times\exp NM u\Big(\frac1M\sum_{k\le M}\EE\ln \int    dp_X(y)\exp\Big(\sqrt{\tau_k}Zy+\tau_kyY_0-\frac{\tau_k}2y^2\Big)+ O(u^2) \Big)
\end{align}
where $C_N$ collects all irrelevant multiplicative constants and terms $\exp(o(NM))$. Note that the ordering of $(\tau_k)$ and $(\sigma_k)$ enforced by \eqref{14_ordered} finally does not matter. Saddle point estimation and the replica trick \eqref{replicaTrick} finally yields a variational formula for the average free entropy
\begin{align}
\EE f_N = \underset{\bsig,\btau\in\mathbb{R}_{>0}^M}{\rm extr} \frac1M\sum_{k\le M}\Big\{-\frac1{2} \tau_k\sigma_k+\frac\gamma 4\sigma_k^2+\EE\ln \int    dp_X(y)\exp\Big(\sqrt{\tau_k}Zy+\tau_kyY_0-\frac{\tau_k}2y^2\Big)\Big\}  +o_N(1).
\end{align}
Note that in the present regime $M/N\to 0$ the terms coming from the Vandermonde does not play a role asymptotically. At fixed $k$, the bracket $\{\cdots\}$ above is the usual formula for the rank-one case. And because different $k$ indices do not interact, the solution is clearly the same for all $k$, so this formula collapses on the rank-one formula
\begin{align}
\EE f_N \xrightarrow{N \to \infty} \underset{\sigma,\tau\in\mathbb{R}_{>0}}{\rm extr} \Big\{-\frac1{2} \tau\sigma+\frac\gamma 4\sigma^2+\EE\ln \int    dp_X(y)\exp\Big(\sqrt{\tau}Zy+\tau yY_0-\frac{\tau}2y^2\Big)\Big\}.
\end{align}
Solving for the saddle point equation for $\sigma$ and using the connection between average free entropy and mutual information we obtain the rank-one prediction:
\begin{equation}
    \frac{I_N(\bS; \bY)}{M N}  \xrightarrow{N \to \infty} \inf_{\sigma\ge 0}\Big\{\frac\gamma 4(\sigma-\rho)^2+I(X;\sqrt{\gamma \sigma}X+Z)\Big\}
    \label{asymp-mI-subL-app}
\end{equation}
where $X\sim p_X$, $Z\sim\mathcal{N}(0,1)$ and $\rho$ is the variance of $p_X$. Whenever it is unique, denoting $\sigma_*$ the minimizer in the above variational formula, the MMSE reads (see \cite{dia2016mutual,miolane2017fundamental}):
\begin{align}
    {\rm MMSE}_N(\gamma) \xrightarrow{N \to \infty}\rho^2-\sigma_*^2.
\end{align}

Finally, we remark that the present derivation used that the rank $M$ grows with $N$. But it can be verified numerically or analytically by studying the saddle point equations for the finite rank $M=O(1)$ that the replica prediction collapses onto the rank-one formulas when the prior is fully factorized as in the present setting.

\section{Examples and numerical calculations for linear ranks}\label{Example}
\subsection{Signal with Rademacher spectrum}\label{Ex-1}
\begin{figure}
\captionsetup{singlelinecheck = false, justification=justified}
    \centering
    \input{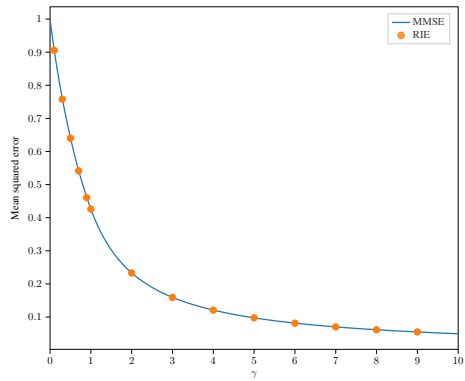}
    \caption{{\small MMSE and the MSE of RIE \eqref{RIE-est} for the signal with spectrum $\rho_S = \frac{1}{2} \delta_{-1} + \frac{1}{2} \delta_{+1}$. The MMSE is continuous w.r.t. $\gamma$. The RIE is computed for $N=1000$, and the results are averaged over 20 runs (error bars are invisible).}}
    \label{fig:Th-MMSE & RIE-MSE-app}
\end{figure}

In this example, we consider the case where $\rho_S = \frac{1}{2} \delta_{-1} + \frac{1}{2} \delta_{+1}$. Using the technique introduced in \cite{biane1997free}, we compute $\rho_Y = \rho_{\sqrt{\gamma} S}\boxplus \rho_{\rm sc}$ in appendix \ref{Deriv-Ex-1}. Fig. \ref{fig:limiting spectrum of rho_Y Ex 1} shows the support of $\rho_Y$ which consists of two disjoint intervals when $\gamma \geq 1$, and one single open set when $\gamma < 1$. Therefore, we expect that a phase transition, if it exists, should happen at a value $\gamma_c = 1$. By Theorem 2 the MMSE is a continuous function of $\gamma$, and the phase transition (if it exists) is of the second or higher order. The MMSE and the performance of the RIE for this example are plotted in figure \ref{fig:Th-MMSE & RIE-MSE-app}. 

\begin{figure}
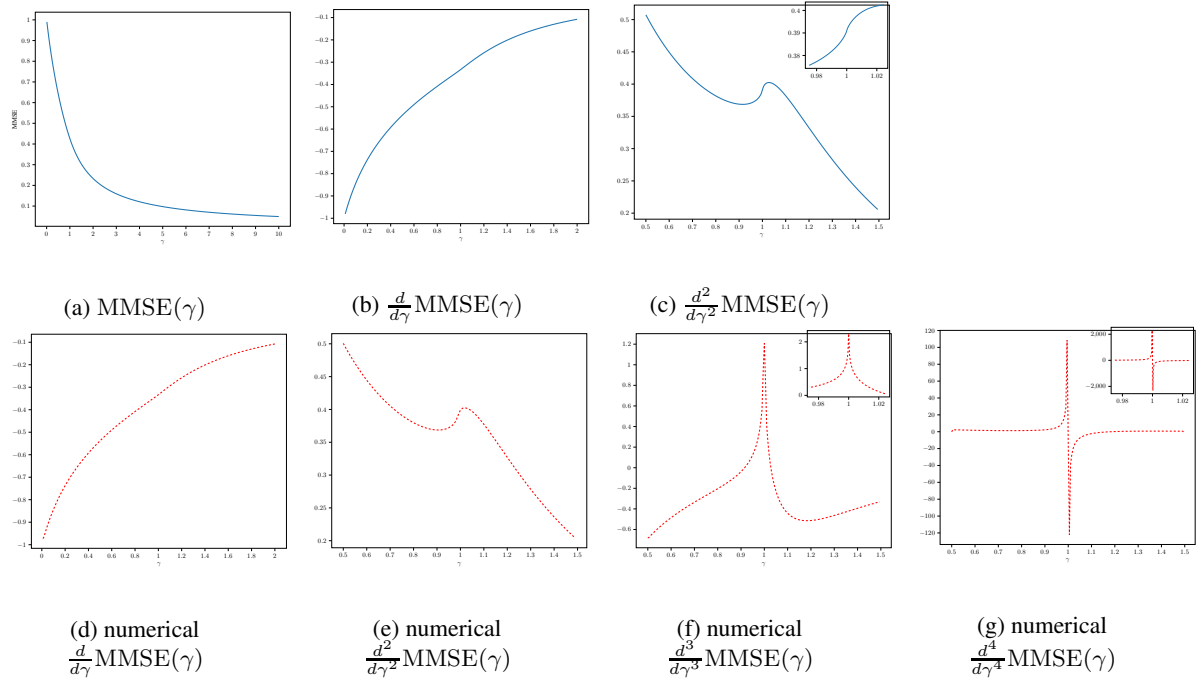

    \centering
    \begin{subfigure}{.2\textwidth}
        \centering
        \input{Supplementary/Examples/Rademacher_spectrum/MMSE_Rademach}
        \caption{{\small${\rm MMSE}(\gamma)$}}
    \end{subfigure}
    \hfil
        \begin{subfigure}{.2\textwidth}
        \centering
\begin{tikzpicture}[scale=0.25]

\definecolor{color0}{rgb}{0.12156862745098,0.466666666666667,0.705882352941177}

\begin{axis}[
tick align=outside,
tick pos=left,
x grid style={white!69.0196078431373!black},
xlabel={$\gamma$},
xmin=-0.0895, xmax=2.0995,
xtick style={color=black},
y grid style={white!69.0196078431373!black},
ymin=-1.02422657234277, ymax=-0.0639362725565435,
ytick style={color=black},
label style={font=\fontsize{40}{42}},
tick label style={font=\fontsize{20}{22}}
]
\addplot [ultra thick, color0]
table {%
0.01 -0.980577013261581
0.02 -0.962224004733308
0.03 -0.944829526655763
0.04 -0.928298971625107
0.05 -0.912551102263986
0.06 -0.897515707976817
0.07 -0.883131603898002
0.08 -0.869345161520686
0.09 -0.856109091430111
0.1 -0.843381468366843
0.11 -0.831124942968393
0.12 -0.819306095502232
0.13 -0.807894895150833
0.14 -0.79686425502466
0.15 -0.786189656015376
0.16 -0.775848830293858
0.17 -0.765821492533053
0.18 -0.756089108950357
0.19 -0.746634699664751
0.2 -0.737442670766927
0.21 -0.728498664320263
0.22 -0.719789431866388
0.23 -0.711302721355168
0.24 -0.7030271796138
0.25 -0.694952265447632
0.26 -0.687068174041136
0.27 -0.679365768849003
0.28 -0.67183652197056
0.29 -0.664472460546225
0.3 -0.657266119375222
0.31 -0.650210498107525
0.32 -0.643299022645692
0.33 -0.636525511154056
0.34 -0.629884142550942
0.35 -0.62336942862471
0.36 -0.616976188446309
0.37 -0.610699525471559
0.38 -0.60453480621549
0.39 -0.598477641429888
0.4 -0.592523868424496
0.41 -0.586669535232275
0.42 -0.580910885780934
0.43 -0.575244346685516
0.44 -0.569666514664815
0.45 -0.564174145395255
0.46 -0.558764142851221
0.47 -0.553433549672268
0.48 -0.548179538209861
0.49 -0.542999402209317
0.5 -0.537890549093608
0.51 -0.532850492672337
0.52 -0.52787684663607
0.53 -0.52296731816014
0.54 -0.518119702142741
0.55 -0.513331875627184
0.56 -0.508601792810163
0.57 -0.503927480032418
0.58 -0.49930703127969
0.59 -0.494738603858243
0.6 -0.490220414312868
0.61 -0.485750734428843
0.62 -0.481327887577674
0.63 -0.476950245085949
0.64 -0.472616222820945
0.65 -0.468324277774981
0.66 -0.464072905029667
0.67 -0.459860634293041
0.68 -0.455686026967142
0.69 -0.451547673022319
0.7 -0.447444187904606
0.71 -0.443374209422255
0.72 -0.439336394630809
0.73 -0.435329416607288
0.74 -0.43135196106334
0.75 -0.427402722910533
0.76 -0.423480402817671
0.77 -0.419583702916212
0.78 -0.415711322833699
0.79 -0.411861955048356
0.8 -0.408034279797419
0.81 -0.404226959404358
0.82 -0.400438631857791
0.83 -0.396667903311579
0.84 -0.392913339679952
0.85 -0.389173456745505
0.86 -0.385446707638323
0.87 -0.381731468545628
0.88 -0.378026021031136
0.89 -0.37432852884998
0.9 -0.37063700971249
0.91 -0.366949297852176
0.92 -0.363262993609353
0.93 -0.359575393592293
0.94 -0.355883390789458
0.95 -0.352183325233784
0.96 -0.348470744594035
0.97 -0.344739987340984
0.98 -0.340983366447204
0.99 -0.337189181962098
1 -0.333333333332004
1.01 -0.329372272858879
1.02 -0.325363846032007
1.03 -0.321340404362144
1.04 -0.317319271383051
1.05 -0.313312186996761
1.06 -0.309327762177757
1.07 -0.305372569076412
1.08 -0.301451736228911
1.09 -0.297569315098506
1.1 -0.293728524425683
1.11 -0.289931922493849
1.12 -0.286181533808963
1.13 -0.282478945289661
1.14 -0.278825381129142
1.15 -0.275221762255209
1.16 -0.271668754361758
1.17 -0.268166807215216
1.18 -0.264716187184101
1.19 -0.261317004418437
1.2 -0.257969235752845
1.21 -0.254672744115954
1.22 -0.251427295101638
1.23 -0.248232571155194
1.24 -0.245088183811384
1.25 -0.241993684263658
1.26 -0.238948572504994
1.27 -0.235952305304392
1.28 -0.233004303140768
1.29 -0.230103956256073
1.3 -0.227250629960568
1.31 -0.224443669251763
1.32 -0.221682402891893
1.33 -0.218966146970432
1.34 -0.216294208014061
1.35 -0.21366588574415
1.36 -0.21108047548014
1.37 -0.20853727023958
1.38 -0.206035562589316
1.39 -0.20357464626608
1.4 -0.201153817581655
1.41 -0.198772376659565
1.42 -0.19642962851279
1.43 -0.194124883948154
1.44 -0.19185746040037
1.45 -0.189626682601437
1.46 -0.187431883175695
1.47 -0.185272403149249
1.48 -0.183147592365531
1.49 -0.181056809839477
1.5 -0.178999424052688
1.51 -0.176974813186133
1.52 -0.174982365302126
1.53 -0.173021478494129
1.54 -0.171091560982481
1.55 -0.169192031190222
1.56 -0.16732231777757
1.57 -0.165481859651248
1.58 -0.163670105953317
1.59 -0.161886516033867
1.6 -0.160130559382971
1.61 -0.158401715578728
1.62 -0.156699474198737
1.63 -0.155023334719273
1.64 -0.153372806414886
1.65 -0.151747408243307
1.66 -0.150146668722757
1.67 -0.148570125797039
1.68 -0.147017326705102
1.69 -0.145487827842044
1.7 -0.143981194605547
1.71 -0.142497001260522
1.72 -0.141034830781813
1.73 -0.139594274704106
1.74 -0.138174932973558
1.75 -0.1367764137809
1.76 -0.135398333427616
1.77 -0.134040316152536
1.78 -0.13270199399098
1.79 -0.131383006614417
1.8 -0.130083001180775
1.81 -0.128801632184384
1.82 -0.127538561310128
1.83 -0.126293457270478
1.84 -0.12506599568056
1.85 -0.123855858894505
1.86 -0.122662735873667
1.87 -0.121486322042888
1.88 -0.120326319153458
1.89 -0.119182435142466
1.9 -0.118054384003504
1.91 -0.11694188565478
1.92 -0.11584466580343
1.93 -0.114762455830035
1.94 -0.113694992651001
1.95 -0.11264201860387
1.96 -0.111603281327196
1.97 -0.110578533641026
1.98 -0.109567533429054
1.99 -0.108570043533446
2 -0.107585831637736
};
\end{axis}

\end{tikzpicture}
        \caption{{\small$\frac{d}{d \gamma} {\rm MMSE}(\gamma)$}}
    \end{subfigure}
    \hfil
        \begin{subfigure}{.2\textwidth}
        \centering
        \input{Supplementary/Examples/Rademacher_spectrum/ddMSE_Rademach_wCL}
        \caption{{\small$\frac{d^2}{d \gamma^2} {\rm MMSE}(\gamma)$}}
    \end{subfigure}
    \hfil
    \begin{subfigure}{.2\textwidth}
\begin{tikzpicture}[scale=0.25]

\end{tikzpicture}
    \end{subfigure}\\
    \begin{subfigure}{.2\textwidth}
        \centering
\begin{tikzpicture}[scale=0.25]

\begin{axis}[
tick align=outside,
tick pos=left,
x grid style={white!69.0196078431373!black},
xlabel={$\gamma$},
xmin=-0.0895, xmax=2.0995,
xtick style={color=black},
y grid style={white!69.0196078431373!black},
ymin=-1.01606597707709, ymax=-0.0643248721436706,
ytick style={color=black},
label style={font=\fontsize{40}{42}},
tick label style={font=\fontsize{20}{22}}
]
\addplot [ultra thick, red, dashed]
table {%
0.01 -0.972805017761939
0.02 -0.963796602439484
0.03 -0.944829077960156
0.04 -0.928298602059739
0.05 -0.912550812729216
0.06 -0.897515445099277
0.07 -0.883131394725563
0.08 -0.869344985548928
0.09 -0.856108935918646
0.1 -0.84338133885856
0.11 -0.831124833798976
0.12 -0.819305996897455
0.13 -0.807894807986868
0.14 -0.796864178768975
0.15 -0.786189592565227
0.16 -0.77584877470598
0.17 -0.765821436830029
0.18 -0.756089061359893
0.19 -0.746634659403754
0.2 -0.737442632984997
0.21 -0.728498631513642
0.22 -0.719789401912098
0.23 -0.711302694413804
0.24 -0.703027155583077
0.25 -0.694952244436996
0.26 -0.68706815379362
0.27 -0.679365749163944
0.28 -0.671836504160723
0.29 -0.664472446209162
0.3 -0.657266105215308
0.31 -0.650210485000247
0.32 -0.643299011450108
0.33 -0.63652550001561
0.34 -0.629884132098148
0.35 -0.623369419316369
0.36 -0.616976179067737
0.37 -0.610699516508169
0.38 -0.604534798671778
0.39 -0.59847763479934
0.4 -0.592523862199877
0.41 -0.586669529352317
0.42 -0.580910879853546
0.43 -0.575244341373548
0.44 -0.569666510150765
0.45 -0.564174140345948
0.46 -0.558764138455501
0.47 -0.553433545770218
0.48 -0.548179534821659
0.49 -0.542999398143366
0.5 -0.537890545058457
0.51 -0.53285048971261
0.52 -0.527876843916144
0.53 -0.522967315412857
0.54 -0.518119699841176
0.55 -0.513331873597755
0.56 -0.508601790337911
0.57 -0.50392747763863
0.58 -0.499307029655994
0.59 -0.494738602261784
0.6 -0.49022041217294
0.61 -0.485750732438058
0.62 -0.481327885486461
0.63 -0.476950242694639
0.64 -0.472616222622186
0.65 -0.468324277860778
0.66 -0.464072903308361
0.67 -0.459860633333489
0.68 -0.455686026624618
0.69 -0.451547672482089
0.7 -0.447444187157032
0.71 -0.443374208715323
0.72 -0.43933639243393
0.73 -0.435329413675092
0.74 -0.431351962601636
0.75 -0.42740272538029
0.76 -0.423480402674127
0.77 -0.419583703370118
0.78 -0.415711323489764
0.79 -0.411861955605528
0.8 -0.408034279754384
0.81 -0.404226956799068
0.82 -0.400438629445922
0.83 -0.396667909081783
0.84 -0.392913346082735
0.85 -0.389173458250281
0.86 -0.385446714530843
0.87 -0.381731470807793
0.88 -0.378026016004742
0.89 -0.374328534602682
0.9 -0.370637028322726
0.91 -0.366949315079166
0.92 -0.363262998109453
0.93 -0.359575403181591
0.94 -0.355883426325363
0.95 -0.352183367176801
0.96 -0.348470805690947
0.97 -0.344740116405601
0.98 -0.340983721835821
0.99 -0.337190469446676
1 -0.333327689608505
1.01 -0.329373765750738
1.02 -0.325364286233956
1.03 -0.321340584184515
1.04 -0.317319373562946
1.05 -0.313312254370879
1.06 -0.309327810206802
1.07 -0.305372604467429
1.08 -0.30145176390844
1.09 -0.297569337331663
1.1 -0.293728541912324
1.11 -0.289931936848908
1.12 -0.286181546141903
1.13 -0.282478955478493
1.14 -0.278825389419793
1.15 -0.27522176931084
1.16 -0.271668760517035
1.17 -0.268166812441944
1.18 -0.264716191675462
1.19 -0.261317008285009
1.2 -0.257969239073139
1.21 -0.254672746771199
1.22 -0.25142729724954
1.23 -0.248232573407141
1.24 -0.24508818572537
1.25 -0.241993685661174
1.26 -0.238948573742355
1.27 -0.235952306251458
1.28 -0.233004304009414
1.29 -0.230103956837519
1.3 -0.227250630274488
1.31 -0.224443669818057
1.32 -0.221682403168778
1.33 -0.218966147045802
1.34 -0.216294208394286
1.35 -0.213665886016315
1.36 -0.211080475392875
1.37 -0.208537270099549
1.38 -0.206035562391915
1.39 -0.20357464600958
1.4 -0.201153817682314
1.41 -0.198772376473922
1.42 -0.196429628182557
1.43 -0.194124883874842
1.44 -0.191857460044283
1.45 -0.18962668232107
1.46 -0.187431882978913
1.47 -0.185272402789443
1.48 -0.183147592064166
1.49 -0.181056809633932
1.5 -0.17899942366803
1.51 -0.176974812752985
1.52 -0.174982364898789
1.53 -0.173021478135719
1.54 -0.171091560749258
1.55 -0.169192030875509
1.56 -0.167322317339635
1.57 -0.165481859403815
1.58 -0.163670105560467
1.59 -0.161886515546387
1.6 -0.1601305592609
1.61 -0.158401715348615
1.62 -0.156699473772038
1.63 -0.155023334358529
1.64 -0.153372806194945
1.65 -0.151747407927073
1.66 -0.150146668310823
1.67 -0.148570125556976
1.68 -0.147017326310865
1.69 -0.14548782740932
1.7 -0.143981194363862
1.71 -0.14249700098438
1.72 -0.14103483063348
1.73 -0.139594274341824
1.74 -0.138174932707746
1.75 -0.136776413657407
1.76 -0.135398333157993
1.77 -0.134040315899904
1.78 -0.132701993585541
1.79 -0.131383006291442
1.8 -0.130083001133926
1.81 -0.128801631903473
1.82 -0.127538561048374
1.83 -0.126293457129188
1.84 -0.125065995322574
1.85 -0.123855858657779
1.86 -0.122662735740739
1.87 -0.121486321809236
1.88 -0.12032631885851
1.89 -0.119182435013878
1.9 -0.118054383781383
1.91 -0.116941885573024
1.92 -0.115844665736905
1.93 -0.114762455550253
1.94 -0.11369499251304
1.95 -0.112642018513912
1.96 -0.111603281045714
1.97 -0.11057853343693
1.98 -0.109567533349036
1.99 -0.108570043262411
2 -0.107585831458826
};
\end{axis}

\end{tikzpicture}
        \caption{{\small numerical $\frac{d}{d \gamma} {\rm MMSE}(\gamma)$}}
    \end{subfigure}
    \hfil
        \begin{subfigure}{.2\textwidth}
        \centering
\begin{tikzpicture}[scale=0.25]

\begin{axis}[
tick align=outside,
tick pos=left,
x grid style={white!69.0196078431373!black},
xlabel={$\gamma$},
xmin=0.4505, xmax=1.5395,
xtick style={color=black},
y grid style={white!69.0196078431373!black},
ymin=0.189261722424671, ymax=0.515473959029323,
ytick style={color=black},
label style={font=\fontsize{40}{42}},
tick label style={font=\fontsize{20}{22}}
]
\addplot [ultra thick, red, dashed]
table {%
0.5 0.500646130092748
0.51 0.494121242889215
0.52 0.487821157512282
0.53 0.481737386981197
0.54 0.475861971107514
0.55 0.47018744912564
0.56 0.464706837290375
0.57 0.45941359080087
0.58 0.454301584425027
0.59 0.449365097776506
0.6 0.444598792577118
0.61 0.439997693760298
0.62 0.435557178333087
0.63 0.431272966164425
0.64 0.427141096144281
0.65 0.423157933347651
0.66 0.419320164562847
0.67 0.415624775339315
0.68 0.412069063579182
0.69 0.408650637201457
0.7 0.40536741479457
0.71 0.402217630653942
0.72 0.399199850233568
0.73 0.396312981340824
0.74 0.393556268952275
0.75 0.390929347707765
0.76 0.388432266746635
0.77 0.386065499354946
0.78 0.383830006486614
0.79 0.381727284800643
0.8 0.379759431502005
0.81 0.377929238539682
0.82 0.376240289698846
0.83 0.374697069242731
0.84 0.373305179725697
0.85 0.372071557918331
0.86 0.371004708016429
0.87 0.370115163661284
0.88 0.369415998797592
0.89 0.368923504672022
0.9 0.368658279728437
0.91 0.368646792966961
0.92 0.36892374950637
0.93 0.369535815439632
0.94 0.370548027600652
0.95 0.372055656667882
0.96 0.374207349191335
0.97 0.377241931408763
0.98 0.38193792032915
0.99 0.3909646034213
1 0.399226006666848
1.01 0.402007383541089
1.02 0.402470927746084
1.03 0.40158045890679
1.04 0.399701986305166
1.05 0.39707840173873
1.06 0.393877797437584
1.07 0.390223283926471
1.08 0.386208732027158
1.09 0.381907820144243
1.1 0.377379626040976
1.11 0.372672286141343
1.12 0.367825509999434
1.13 0.362872373570081
1.14 0.357840633871886
1.15 0.352753719790846
1.16 0.347631496537384
1.17 0.342490864711007
1.18 0.337346236256529
1.19 0.332209919478337
1.2 0.327092432886743
1.21 0.322002764538492
1.22 0.316948587247824
1.23 0.311936437797019
1.24 0.306971871669342
1.25 0.302059591695979
1.26 0.297203557551829
1.27 0.292407082017688
1.28 0.287672911574743
1.29 0.283003298213375
1.3 0.278400060531349
1.31 0.273864635867872
1.32 0.269398129292015
1.33 0.265001353320838
1.34 0.260674862837607
1.35 0.256418988431814
1.36 0.252233863737738
1.37 0.248119449079261
1.38 0.244075554010607
1.39 0.240101856463317
1.4 0.236197918608246
1.41 0.232363204249989
1.42 0.228597090343588
1.43 0.224898878638684
1.44 0.221267808320836
1.45 0.217703063188892
1.46 0.214203780994566
1.47 0.210769061293076
1.48 0.207397971163541
1.49 0.204089551361246
};
\end{axis}

\end{tikzpicture}
        \caption{{\small numerical $\frac{d^2}{d \gamma^2} {\rm MMSE}(\gamma)$}}
    \end{subfigure}
    \hfil
        \begin{subfigure}{.2\textwidth}
        \centering
        \input{Supplementary/Examples/Rademacher_spectrum/dddMSE_Rademach_wCL_numeric}
        \caption{{\small numerical $\frac{d^3}{d \gamma^3} {\rm MMSE}(\gamma)$}}
    \end{subfigure}
    \hfil
    \begin{subfigure}{.2\textwidth}
        \centering
        \input{Supplementary/Examples/Rademacher_spectrum/ddddMSE_Rademach_wCL_numeric}
        \caption{{\small numerical $\frac{d^4}{d \gamma^4} {\rm MMSE}(\gamma)$}}
    \end{subfigure}
    \captionsetup{singlelinecheck = false, justification=justified}
    \caption{{\small Analysis of the MMSE in example \ref{Ex-1}. In plots (a),(b),(c) MMSE and its first and second derivatives are computed using the the expression \eqref{RIE-MSE}. Plots (d), (e), (f) are the numerical differentiation of plots (a),(b), (c) respectively. The fourth derivative of the MMSE is computed from the curve in plot (c) by numerical second order differentiation. The first three plots shows that the MMSE, its first and second derivatives are continuous. But, plots (f), (g) suggests that the $\frac{d^2}{d \gamma^2} {\rm MMSE}(\gamma)$ has a vertical tangent at $\gamma_c = 1$, and MMSE has a phase transition of third order at this point.}}
    \label{fig: MMSE-Rad-analys-app}
\end{figure}

From the expression of $\rho_Y$ in appendix \ref{Deriv-Ex-1} and from Theorem 2 we provide integral representations of the derivatives of the MMSE w.r.t. $\gamma$. These integrals are computed numerically and the result illustrated in figures \ref{fig: MMSE-Rad-analys-app} (a-b-c) for the MMSE and its first and second derivatives. For the third and the fourth derivatives the integral representation become unwieldy, so we only computed it numerically from the second derivative, as shown in figures \ref{fig: MMSE-Rad-analys-app} (f),(g). Based on these plots, we see that the third derivative (of the MMSE) at $\gamma_c = 1$ does not seem to exist, and therefore the free energy \eqref{free-energy-def} (or mutual information) might have a fourth order phase transition at this point. Further numerical analysis in appendix \ref{Deriv-Ex-1} is compatible with a  behavior of the function ${\rm MMSE}(\gamma)$ close to the point $\gamma_c=1$ of the form
\begin{equation}
    {\rm MMSE}(\gamma) \approx {\rm MMSE}(1) + {\rm MMSE}'(1) (\gamma - 1) + \frac{1}{2} {\rm MMSE}''(1) (\gamma - 1)^2  + \alpha (\gamma - 1)^3 \big( \ln |\gamma - 1| + \beta \big) + o\big((\gamma - 1)^3\big)
\end{equation}
with $\alpha \approx -0.06125$, $\beta \approx 1.411$.

\subsection{Signal with Bernoulli spectrum}\label{Ex-2}
\begin{figure}
\captionsetup{singlelinecheck = false, justification=justified}
    \centering
    \input{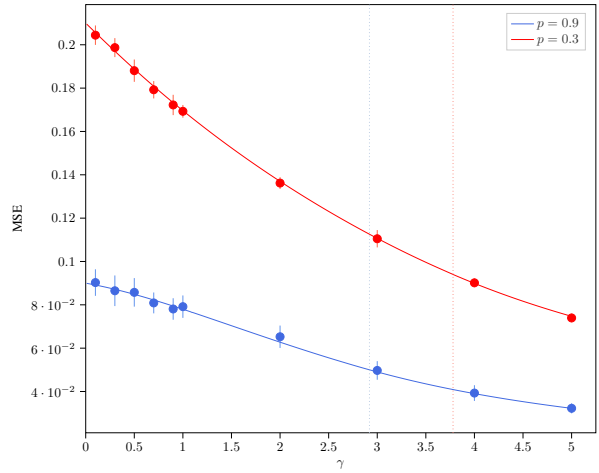}
    \caption{{\small MMSE computed for the signal with Bernoulli spectrum $\rho_S = p \delta_{0} + (1-p) \delta_{+1}$ for $p=0.3$ and $0.9$. The MMSE is continuous w.r.t. $\gamma$. The vertical dashed lines corresponds to the values of $\gamma$, i.e.,  $2.92$ for $p=0.9$ and $3.78$ for $p=0.3$, where the disjoint intervals of the support of $\rho_Y$ merge. We do not observe any phase transition for any low order derivative at these values. The MSE of RIE is computed for $N=1000$, and the results are averaged over 20 runs.}}
    \label{fig:Th-MMSE & RIE-MSE for Bernoulli}
\end{figure}

Let $\rho_S = p \delta_{0} + (1-p) \delta_{+1}$. The corresponding signal matrix is not full-rank but it has a rank linear in $N$. For this prior, the spectrum of $\rho_Y = \rho_{\sqrt{\gamma} S}\boxplus \rho_{\rm sc}$ is computed in appendix \ref{Deriv-Ex-2} using a similar technique as in the previous example. Depending on the SNR parameter $\gamma$ the support of $\rho_Y$ can be a single interval or is composed of two disjoint intervals as shown on figure \ref{fig:limiting spectrum of rho_Y Ex 2}. The MMSE and the MSE of RIE are illustrated in figure \ref{fig:Th-MMSE & RIE-MSE for Bernoulli} for the two values $p=0.9$ and $p=0.3$.

\begin{figure}
\captionsetup{singlelinecheck = false, justification=justified}
    \centering
    \input{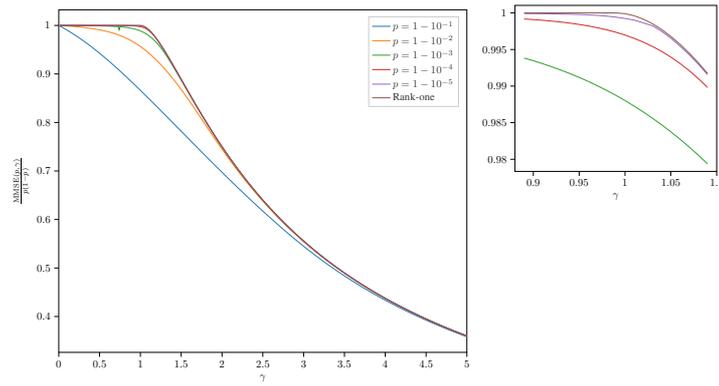}
    \caption{{\small Normalized MMSE  for the signal with spectrum $\rho_S = p \delta_{0} + (1-p) \delta_{+1}$ for $p \to 1$. The MMSE of the rank-one problem is also plotted for comparison.}}
    \label{fig:Th-MMSE & RIE-MSE for Bernoulli - sparse}
\end{figure}

In figure \ref{fig:Th-MMSE & RIE-MSE for Bernoulli - sparse}, the suitably normalized MMSE is plotted for the highly sparse case where $p$ tends to $1$. The MMSE is normalized by dividing by $p(1-p)$. We observe that as $p\to 1$ the MMSE approaches the MMSE of the rank-one symmetric matrix estimation problem, which has a  phase transition at $\gamma_c = 1$.

\subsection{Wishart Matrix}\label{Ex-3}
\begin{figure}
\captionsetup{singlelinecheck = false, justification=justified}
    \centering
    \input{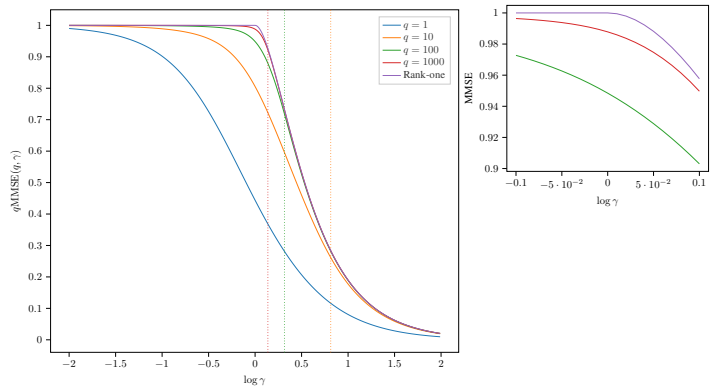}
    \caption{{\small Normalized MMSE for the signal with the Marchenko-Pastur spectral distribution for large $q$'s . The MMSE of the rank-one problem is also plotted for comparison. The vertical dashed lines corresponds to the critical value where the support of $\rho_Y$ splits into two intervals. We do not find any phase transition on low order derivatives}}
    \label{fig:Th-MMSE & RIE-MSE for Wishart - sparse}
\end{figure}

In this example, we consider the signal matrix $\bS$ to be $\frac{1}{N}\bX \bX^\intercal$, where $\bX \in \bR^{N \times M}$ has i.i.d. standard Gaussian entries. We look at the limit of aspect ratio $\frac{N}{M} \to q$. Then, the limiting spectral distribution of $\bS$ is a rescaling of the usual \textit{Marchenko-Pastur} distribution by the factor $\alpha$:
\begin{equation}
    \rho_S(x) = \big( 1 - \frac{1}{q} \big)^+ \delta(x) + \frac{\sqrt{\Big(x - \big( \frac{1}{\sqrt{q}}-1 \big)^2\Big)\Big( \big(\frac{1}{\sqrt{q}} + 1\big)^2 -x \Big) }}{2 \pi  x}.
    \label{scaled-MP}
\end{equation}
The limiting spectral distribution of $\bY$ is computed in appendix \ref{Deriv-Ex-3}. For $q > 1$ the support of $\rho_Y$ is the union of two disjoint intervals if $\gamma > q \big( \sqrt[3]{q} - 1 \big)^{-3}$, and is a single interval otherwise. As in the previous example, we expect that in the high sparse regime, the MMSE behaves like the low-rank case. In figure \ref{fig:Th-MMSE & RIE-MSE for Wishart - sparse} the MMSE is illustrated large $q$'s. 

\subsection{Finite-rank deformation of a Wigner matrix as signal}
First, consider the case where $\bS$ is a standard Wigner matrix, then by independence of $\bS$ and $\bZ$, $\bY$ is also a Wigner matrix with variance $\gamma + 1$, and $\rho_Y$ is a semi-circle law of variance $\gamma + 1$. From \eqref{simp-asymp-mI}, we find that $\frac{1}{N^2} I_N(\bS; \bY)$ converges to $\frac{1}{4} \ln (\gamma + 1)$. This limit could also be obtained using the Gaussianity of entries of the matrices.

Now, let $\bS$ be a finite rank deformation of a Wigner matrix, $\bS = \bA +  \bzt$, where $\bA$ is a {\it finite-rank} symmetric matrix, and $\bzt \in \bR^{N \times N}$ is a symmetric Gaussian matrix with variance $\frac{1}{N}$ for non-diagonal, and $\frac{2}{N}$ for diagonal entries. We have the observation $\bY = \sqrt{\gamma} \bS + \bZ$. Since $\bA$ has finite rank, the limiting spectral measure of $\bS$ is the semicircle law, and average mutual information $\frac{1}{N^2} I_N (\bS; \bY)$ converges to $\frac{1}{4} \ln (\gamma + 1)$, \eqref{simp-asymp-mI}.
Since $\bZ$ and $\bzt$ are independent, the observation matrix $\bY$ has the same distribution as the matrix $\tilde{\bY} = \sqrt{\gamma} \bA + \sqrt{\gamma + 1} \tilde{\bZ}$, where $\tilde{\bZ}$ is a symmetric Gaussian matrix, so $\cH(\bY) = \cH(\tilde{\bY})$. Define the matrix $\bY' = \sqrt{\frac{\gamma }{\gamma + 1}} \bA + \bZ'$. By symmetry of the matrices, we have:
\begin{equation}
    \mathcal{H}_N (\bY') = \mathcal{H}_N (\bY) - \frac{N (N + 1)}{2} \ln \sqrt{\gamma + 1}.
\label{H(Y) & H(Y')}
\end{equation}
Here 
$\bY'$ has the form of the observation matrix in the low-rank matrix estimation, which has been extensively studied in \cite{lelarge2019fundamental, dia2016mutual}, and in particular, it has been shown that under suitable assumptions, the average mutual information $\frac{1}{N} I_N(\bA; \bY')$ converges to a finite value. By definition of mutual information, and \eqref{H(Y) & H(Y')}, we have:
\begin{equation}
    \begin{split}
        \frac{1}{N} I_N(\bA; \bY') &= \frac{1}{N} \mathcal{H}_N (\bY') - \frac{1}{N} \mathcal{H}_N (\bZ') \\
        &= \frac{1}{N} \mathcal{H}_N (\bY) - \frac{1}{N} \mathcal{H}_N (\bZ') - \frac{N + 1}{4} \ln (\gamma +1) \\
        &= \frac{1}{N} I_N(\bS; \bY)  - \frac{N + 1}{4} \ln (\gamma +1). \\
    \end{split}
    \label{MI realtion}
\end{equation}
Dividing by $N$ and taking the limit, we find the asymptotic mutual information 
    \begin{equation}
        \begin{split}
            \lim_{ N \to \infty} \frac{1}{N^2} I_N (\bS ; \bY) &= \lim_{ N \to \infty} \frac{1}{N^2} I_N(\bA; \bY') + \frac{N+1}{4 N}  \ln (\gamma +1) \\
            &= \frac{1}{4} \ln (\gamma +1).
        \end{split}
    \end{equation}
where we used the fact that 
$\frac{1}{N} I_N(\bA; \bY')$ has a finite limit.
Moreover, from \eqref{MI realtion} the finite-size correction term for the asymptotic average mutual information can be derived as 
\begin{equation}
    \lim_{N \to \infty} N \big( \frac{1}{N^2} I_N (\bS ; \bY) -  \frac{1}{4} \ln (\gamma +1) \big) =  \lim_{N \to \infty} \frac{1}{N} I_N(\bA; \bY') + \frac{1}{4}  \ln (\gamma +1).
    \label{finite-correction}
\end{equation}
Therefore when $\bS$ is a finite rank deformation of the Wigner matrix, the finite-size correction term is directly related to the mutual information in the low-rank matrix estimation problem, which may exhibit a phase transition.

Now, let $\bA$ be a {\it rank-one} matrix. $\bS = \frac{\sqrt{\eta}}{N} \bx \bx^\intercal + \bzt $ where $\bx \in \bR^N$ has i.i.d. components distributed according to the standard normal distribution. The matrix $\bY' = \frac{\sqrt{\frac{\eta \gamma}{\gamma + 1}}}{N} \bx \bx^\intercal + \bZ'$ is a rescaling of the rank-one model studied in \cite{lelarge2019fundamental}, and since mutual information is invariant under rescaling, using Theorem 1 in \cite{lelarge2019fundamental}, we have:
\begin{equation}
    \lim_{N \to \infty} \frac{1}{N} I_N(\bx ; \bY') = \begin{cases}
    \frac{1}{4} \frac{\eta \gamma}{\gamma + 1} &{\rm if} \, \frac{\eta \gamma}{\gamma + 1} \leq 1, \\
    \frac{1}{4} \frac{\gamma + 1}{\eta \gamma} + \frac{1}{2} \ln \frac{\eta \gamma}{\gamma + 1} &{\rm else.}
    \end{cases}
\end{equation}
Fix $\eta \leq 1$. For all $\gamma > 0$, we have $\frac{\eta \gamma}{\gamma + 1} \leq 1$ and there is no phase transition in the mutual information. On the other hand, for $\eta > 1$ mutual information has a phase transition at $ \gamma = \frac{1}{\eta - 1}$. From random matrix theory \cite{benaych2011eigenvalues}, we know that For $\eta \leq 1$, in the limit $N \to \infty$, all the eigenvalues of $\bS$ are inside the bulk, whereas for $\eta > 1$ one eigenvalue (which is the largest one) is outside the bulk. Therefore, for this particular signal, we can relate the phase transition of the correction term to the existence of an eigenvalue outside the bulk of $\rho_S$ in the asymptotic limit.

\section{Further details for proof of Theorem 1}\label{proof-thm1-app}
\subsection{Proof of lemma \ref{converg-Expec}}\label{proof of prop1}

To prove lemma \ref{converg-Expec} we first need four preliminary lemmas.

\begin{lemma}
For any $N$, and $N \times N$ symmetric matrices $\bA, \bB$ with spectral radius $r^{(\bA)}, r^{(\bB)}$
\begin{equation*}
    - \frac{1}{2} r^{(\bA)} r^{(\bB)}  \leq \mathcal{J}_N (\bA, \bB) \leq  \frac{1}{2} r^{(\bA)} r^{(\bB)}.
\end{equation*}
\label{bound-on-J}
\end{lemma}
\begin{proof}
Let $\bA = \bU_{\bA} \bm{\Lambda}_{\bA} \bU^\intercal_{\bA}$,  $\bB = \bU_{\bB} \bm{\Lambda}_{\bB} \bU^\intercal_{\bB}$ be the eigendecomposition of $\bA$, $\bB$. We can write
\begin{equation*}
\begin{split}
    \mathcal{I}_N (\bA, \bB) = \int D \bU  e^{\frac{N}{2} \Tr  \bU \bm{\Lambda}_{\bA} \bU^\intercal \bm{\Lambda}_{\bB}} = \int D \bU  e^{\frac{N}{2}  \sum_{i,j}^N \lambda_i^{(\bA)} \lambda_j^{(\bB)} U_{ij}^2 }.
\end{split}
\end{equation*}
For each for all $i,j \in \{1, \hdots, N\}$, $- r^{(\bA)} r^{(\bB)} \leq \lambda_i^{(\bA)} \lambda_j^{(\bB)} \leq  r^{(\bA)} r^{(\bB)}$. Therefore, we get
\begin{equation*}
\begin{split}
     \mathcal{I}_N (\bA, \bB) &= \int D \bU  e^{\frac{N}{2}  \sum_{i,j}^N \lambda_i^{(\bA)} \lambda_j^{(\bB)} U_{ij}^2 }  \\
     & \leq \int D \bU e^{\frac{N}{2} r^{(\bA)} r^{(\bB)} \sum_{i,j}^N  U_{ij}^2 } \\
    &= \int D \bU e^{\frac{N^2}{2}  r^{(\bA)} r^{(\bB)} } \\
    &= e^{\frac{N^2}{2}  r^{(\bA)} r^{(\bB)} }.
\end{split}
\end{equation*}
Similarly we can obtain $\mathcal{I}_N (\bA, \bB) \geq e^{-\frac{N^2}{2} r^{(\bA)} r^{(\bB)}}$. Therefore
\begin{equation*}
    -\frac{1}{2} r^{(\bA)} r^{(\bB)} \leq \mathcal{J}_N(\bA, \bB) \leq \frac{1}{2} r^{(\bA)} r^{(\bB)}.
\end{equation*} 
\end{proof}


\begin{lemma}
Let $r^{(\tilde{\tilde{\bY}})}$ denote the spectral radius of the matrix $\tilde{\tilde{\bY}} = \sqrt{\gamma} \bLambda^0  +  \tilde{\bZ}$. For $k > 2 + \sqrt{\gamma} C $, we have
\begin{equation*}
    \begin{split}
        \bP \{ r^{(\tilde{\tilde{\bY}})} \geq k \} \leq 4 e^{-\frac{N}{4}(k-\sqrt{\gamma}C - 2)^2}.
    \end{split}
\end{equation*}
\label{bound-on-prob-of-spectral-radius}
\end{lemma}
\begin{proof}
Denote the top and bottom eigenvalues of $\tilde{\tilde{\bY}}$ by $\lambda_{\rm max}^{(\tilde{\tilde{\bY}})}$, $\lambda_{\rm min}^{(\tilde{\tilde{\bY}})}$. By Weyls' inequality,
\begin{equation*}
    \begin{split}
        \lambda_{\rm max}^{(\tilde{\tilde{\bY}})} &\leq \sqrt{\gamma} \max_i \lambda^0_i  + \lambda_{\rm max}^{(\tilde{\bZ})} \leq \sqrt{\gamma} C + \lambda_{\rm max}^{(\tilde{\bZ})}\\
        \lambda_{\rm min}^{(\tilde{\tilde{\bY}})} &\geq  \sqrt{\gamma} \min_i \lambda^0_i + \lambda_{\rm min}^{(\tilde{\bZ})} \geq  -\sqrt{\gamma}C + \lambda_{\rm min}^{(\tilde{\bZ})}
    \end{split}
\end{equation*}
where $\lambda_{\rm max}^{(\tilde{\bZ})}$,$\lambda_{\rm min}^{(\tilde{\bZ})}$ are the top and bottom eigenvalues of $\tilde{\bZ}$. Thus, we can write
\begin{equation*}
    \begin{split}
        \bP \{ \lambda_{\rm max}^{(\tilde{\tilde{\bY}})} \geq k \} & \leq \bP \big\{  \lambda_{\rm max}^{(\tilde{\bZ})} + \sqrt{\gamma} C \geq k \big\} \\
        & = \bP \{ \lambda_{\rm max}^{(\tilde{\bZ})} \geq k -  \sqrt{\gamma} C\}.
    \end{split}
\end{equation*}
By \cite{davidson2001local} (Theorem II.11), for $k -  \sqrt{\gamma} C > 2$, we have
\begin{equation*}
    \begin{split}
        \bP \{ \lambda_{\rm max}^{(\tilde{\bZ})} \geq k -  \sqrt{\gamma} C \}  \leq e^{-\frac{N}{4}(k-\sqrt{\gamma}C - 2)^2}
    \end{split}
\end{equation*}
and therefore we get
\begin{equation}
    \begin{split}
        \bP \{ \lambda_{\rm max}^{(\tilde{\tilde{\bY}})} \geq k \}  \leq e^{-\frac{N}{4}(k+\sqrt{\gamma}C - 2)^2}.
    \end{split}
\label{bound-on-top-ev}
\end{equation}
For the bottom eigenvalue we have
\begin{equation}
    \begin{split}
        \bP \{ \lambda_{\rm min}^{(\tilde{\tilde{\bY}})} \leq -k \} & \leq \bP \big\{ -\sqrt{\gamma}C + \lambda_{\rm min}^{(\tilde{\bZ})}  \leq -k \} \\
        &= \bP \big\{ \lambda_{\rm min}^{(\tilde{\bZ})}  \leq \sqrt{\gamma}C -k \} \\
        &\leq e^{-\frac{N}{4}(k-\sqrt{\gamma}C - 2)^2}.
    \end{split}
\label{bound-on-smallest-ev}
\end{equation}
From \eqref{bound-on-top-ev}, \eqref{bound-on-smallest-ev}, we get
\begin{equation*}
    \begin{split}
        \bP \{ r^{(\tilde{\tilde{\bY}})} \geq k \} & \leq \bP \{ |\lambda_{\rm max}^{(\tilde{\tilde{\bY}})} | \geq k \} + \bP \{ |\lambda_{\rm min}^{(\tilde{\tilde{\bY}})}| \geq k \}  \\
        & \leq \bP \{ \lambda_{\rm max}^{(\tilde{\tilde{\bY}})}  \geq k \} + \bP \{ \lambda_{\rm max}^{(\tilde{\tilde{\bY}})}  \leq - k \} + \bP \{ \lambda_{\rm min}^{(\tilde{\tilde{\bY}})} \geq k \} +  \bP \{ \lambda_{\rm min}^{(\tilde{\tilde{\bY}})} \leq -k \}\\
        & \leq 2 \bP \{ \lambda_{\rm max}^{(\tilde{\tilde{\bY}})}  \geq k \} + 2 \bP \{ \lambda_{\rm min}^{(\tilde{\tilde{\bY}})} \leq -k \}\\
        & \leq 4 e^{-\frac{N}{4}(k-\sqrt{\gamma}C - 2)^2}
    \end{split}
\end{equation*}
for $k > 2 + \sqrt{\gamma} C$.
\end{proof}

\begin{lemma}
For any polynomial function $g$, and $k$ a sufficiently large constant, we have that 
$$
\lim_{n \to \infty} \bE \big[ g(r^{(\tilde{\tilde{\bY}})}) \mathbb{I}\{r^{(\tilde{\tilde{\bY}})} \geq k\} \big] = 0 .
$$
\label{expec-poly-r}
\end{lemma}
\begin{proof}
By linearity of expectation, it is enough to consider the case $g(x) = x^i$. Let $X = {r^{(\tilde{\tilde{\bY}})}}^i \mathbb{I}\{r^{(\tilde{\tilde{\bY}})} \geq k\}$ a non-negative random variable. We have
\begin{equation*}
    \begin{split}
        \bE [ X ] & = \int_0^{\infty} \bP(X \geq x) dx \\
        &= i \int_0^{\infty } \bP(X \geq x^i) x^{i-1} dx \\
        &= i \int_0^{\infty } \bP\big({r^{(\tilde{\tilde{\bY}})}}^i \mathbb{I}\{r^{(\tilde{\tilde{\bY}})} \geq k\} \geq x^i\big) x^{i-1} dx \\
        &= i \int_0^{\infty } \bP(r^{(\tilde{\tilde{\bY}})} \geq k, r^{(\tilde{\tilde{\bY}})} \geq x) x^{i-1} dx \\
        &= i \int_0^{k} \bP(r^{(\tilde{\tilde{\bY}})} \geq k) x^{i-1} dx + i \int_{k}^{\infty} \bP(r^{(\tilde{\tilde{\bY}})} \geq x) x^{i-1} dx \\
        & \leq 4 e^{-\frac{N}{4}(k-\sqrt{\gamma}C - 2)^2} k^i + 4 i \int_{k}^{\infty} e^{-\frac{N}{4}(x-\sqrt{\gamma}C - 2)^2}  dx \\
        & \leq 4 e^{-\frac{N}{4}(k-\sqrt{\gamma}C - 2)^2} k^i + 4 i \int_{0}^{\infty} e^{-\frac{N}{4}(x-\sqrt{\gamma}C - 2)^2}x^{i-1} dx.
    \end{split}
\end{equation*}
The first term converges to $0$ as $N \to \infty$. The second term involves moments of a Gaussian with variance $\frac{1}{2N}$, one can see that the second term also converges to $0$. Thus,  $\lim_{N \to \infty} \bE[X] = 0$.
\end{proof}

\begin{lemma}
For the sequence of random matrices $ \tilde{\tilde{\bY}}$, the log spherical integral $\mathcal{J}_N \big( \sqrt{\gamma} \bLam^0,\tilde{\tilde{\bY}} \big)$ converges almost surely to a well-defined limit, denoted by $\mathcal{J}[\rho_{\sqrt{\gamma}S}, \,\rho_{\sqrt{\gamma}S} \boxplus \rho_{\rm sc}]$.
\label{as-conv}
\end{lemma}
\begin{proof}
By assumption 1.A the support of $\hat{\mu}_{\sqrt{\gamma} \bLam^0}^{(N)}$ is included in a compact subset of $\bR$,$[-\sqrt{\gamma}C,\sqrt{\gamma}C]$, for all $N \in \mathbb{N}$. Moreover, by the law of large numbers, $\hat{\mu}_{\sqrt{\gamma} \bLam^0}^{(N)}$  converges weakly towards $\rho_{\sqrt{\gamma} S}$.

Consider the sequence $\sqrt{\gamma} \bLambda^0  +  \tilde{\bZ}$. For each matrix in the sequence, the second moment of the empirical spectral distribution is:
\begin{equation*}
    \begin{split}
        \frac{1}{N} \Tr (\sqrt{\gamma} \bLambda^0  +  \tilde{\bZ})^2 &= \frac{1}{N} \gamma \Tr {\bLambda^0} ^2 + \frac{2}{N} \sqrt{\gamma} \Tr \bLambda^0 \tilde{\bZ} + \frac{1}{N} \Tr \tilde{\bZ}^2 .
    \end{split}
\end{equation*}
The first term is bounded (for all $N$) by the construction of $\blam^0$. The last term is also bounded since the second moment of the sequence of Wigner matrices converges to $1$ almost surely. For the second term, we have
\begin{equation*}
    \begin{split}
        \frac{1}{N} \Tr \bLambda^0 \tilde{\bZ} &\leq \frac{1}{N} \sqrt{ \sum {\gamma^0_i}^2} \sqrt{\Tr \tilde{\bZ}^2} \\
        &\leq C \sqrt{\frac{1}{N} \Tr \tilde{\bZ}^2} 
    \end{split}
\end{equation*}
which is bounded for all $N$, since $\frac{1}{N} \Tr \tilde{\bZ}^2$ is a convergent sequence (a.s.). Therefore, the sequence of matrices $\sqrt{\gamma} \bLambda^0  +  \tilde{\bZ}$ has bounded second moment for all $N$ almost surely. Moreover, according to \cite{anderson2010introduction}, by the independence of $\blam^0$ and $\tilde{\bZ}$, the empirical spectral distribution of this sequence converges weakly, almost surely to the free additive convolution of $\rho_{\sqrt{\gamma}S}$ with the semi-circle law $\rho_{\rm sc}$.

Therefore, the conditions of theorem 1 in \cite{guionnet2002large} hold a.s. for the sequence $\bLambda^0, \sqrt{\gamma} \bLambda^0  +  \tilde{\bZ}$. Hence, $\mathcal{J}_N \big( \sqrt{\gamma} \bLambda^0,\tilde{\tilde{\bY}} \big)$ has a well-defined limit which is a function of $\rho_{\sqrt{\gamma}S}$ and $\rho_{\sqrt{\gamma}S} \boxplus \rho_{\rm sc}$, and is dented by$\mathcal{J}[\rho_{\sqrt{\gamma}S}, \,\rho_{\sqrt{\gamma}S} \boxplus \rho_{\rm sc}]$.
\end{proof}

Now, we are ready to prove lemma \ref{converg-Expec}. 

\begin{proof}{\it of lemma \ref{converg-Expec}.}
For simplicity of notation, we denote $\mathcal{J}_N(\sqrt{\gamma} \bLambda^0,\tilde{\tilde{\bY}})$ by $\mathcal{J}_N$, and $\mathcal{J}[\rho_{\sqrt{\gamma}S}, \,\rho_{\sqrt{\gamma}S} \boxplus \rho_{\rm sc}]$ by $\mathcal{J}$. By Jensen's inequality (note that the expectation is over the matrix $\tilde{\bZ}$), we have
\begin{equation}
\big| \bE [ \mathcal{J}_N ] - \mathcal{J} \big| \leq \bE \big[ | \mathcal{J}_N - \mathcal{J} | \big].
\label{Jensen-expec}
\end{equation}
Let $X_N \equiv \mathcal{J}_N - \mathcal{J}$. For $\epsilon > 0$ We can write
\begin{equation}
\begin{split}
    \bE \big[ |X_N| \big] &= \bE \big[ |X_N| \, \mathbb{I} \{ |X_N| \leq \epsilon \} \big] + \bE \big[ |X_N| \,\mathbb{I} \{ |X_N| > \epsilon \} \big] \\
    &\leq \epsilon + \bE \big[ |X_N| \, \mathbb{I} \{ |X_N| > \epsilon \} \big].
\label{converg-Expec-1}
\end{split}
\end{equation}
By lemma \ref{bound-on-J}, $ | \mathcal{J}_N | \leq \frac{1}{2} \sqrt{\gamma} C r^{(\tilde{\tilde{\bY}})} $, so the second term in \eqref{converg-Expec-1} can be bounded as, 
\begin{equation}
    \bE \big[ |X_N| \, \mathbb{I} \{ |X_N| > \epsilon \} \big] \leq \bE \big[ |W_N| \, \mathbb{I} \{ |X_N| > \epsilon \} \big]
\label{converg-Expec-2}
\end{equation}
where 
$$
W_N = \max \Big \{ \big| \mathcal{J} - \frac{1}{2} \sqrt{\gamma} C r^{(\tilde{\tilde{\bY}})} \big|, \big| \mathcal{J} + \frac{1}{2} \sqrt{\gamma} C r^{(\tilde{\tilde{\bY}})} \big| \Big\} = \frac{1}{2} \sqrt{\gamma} C r^{(\tilde{\tilde{\bY}})} + {\rm sign}(\mathcal{J}) \mathcal{J} .
$$
For any positive constant $t$, we have
\begin{equation}
\begin{split}
    \bE \big[ |W_N| \, \mathbb{I} \{ |X_N| > \epsilon \} \big] &= \bE \big[ |W_N| \, \mathbb{I} \{ |X_N| > \epsilon \} \, \mathbb{I} \{ |W_N| \leq t \} \big] + \bE \big[ |W_N| \, \mathbb{I} \{ |X_N| > \epsilon \} \, \mathbb{I} \{ |W_N| > t \} \big] \\
    & \leq \bE \big[ |W_N| \, \mathbb{I} \{ |X_N| > \epsilon \} \mathbb{I} \{ |W_N| \leq t \} \big] + \bE \big[ |W_N| \, \mathbb{I} \{ |W_N| > t \} \big]
\end{split}
\label{converg-Expec-3}
\end{equation}
For the first term in \eqref{converg-Expec-3} we can write
\begin{equation}
\begin{split}
    \bE \big[ |W_N| \, \mathbb{I} \{ |X_N| > \epsilon \} \mathbb{I} \{ |W_N| \leq t \} \big] &\leq t \bE \big[ \mathbb{I} \{ |X_N| > \epsilon \} \big] \\
    &\leq t \, \bP \big( |X_N| > \epsilon  \big)
\end{split}
\label{converg-Expec-4}
\end{equation}
and the second term in \eqref{converg-Expec-3} can be rewritten as 
\begin{equation}
    \bE \big[ |W_N| \, \mathbb{I} \{ |W_N| > t \} \big] = \bE \bigg[ |W_N| \, \mathbb{I} \Big\{ r^{(\tilde{\tilde{\bY}})} > \frac{2}{\sqrt{\gamma} C} \big(t-{\rm sign}(\mathcal{J}) \mathcal{J}\big) \Big\} \bigg].
\label{converg-Expec-5}
\end{equation}
From \eqref{converg-Expec-2}, \eqref{converg-Expec-3}, \eqref{converg-Expec-4}, we obtain
\begin{equation}
    \bE \big[ |X_N| \, \mathbb{I} \{ |X_N| > \epsilon \} \big] \leq t \, \bP \big( |X_N| > \epsilon  \big) + \bE \bigg[ |W_N| \, \mathbb{I} \Big\{ r^{(\tilde{\tilde{\bY}})} > \frac{2}{\sqrt{\gamma} C} \big(t-{\rm sign}(\mathcal{J}) \mathcal{J}\big) \Big\} \bigg].
\label{converg-Expec-6}
\end{equation}
Notice that $W_N$ is a polynomial function of $r^{(\tilde{\tilde{\bY}})}$, so by lemma \ref{expec-poly-r}, vanishes as $N \to \infty$ for sufficiently large constant $t$. By lemma \ref{as-conv}, $\bP \big( |X_N| > \epsilon  \big) \xrightarrow{N \to \infty} 0$. For a fixed $t>0$, the first term in \eqref{converg-Expec-6} goes to $0$ in the limit $ N \to \infty$. Therefore, taking the limit of both sides in \eqref{converg-Expec-1}, for any $\epsilon >0$, we find:
\begin{equation}
    \lim_{N \to \infty} \bE \big[ |X_N| \big] \leq \epsilon .
\end{equation}
From which, by \eqref{Jensen-expec}, we deduce that $\lim_{ N \to \infty} \bE [\mathcal{J}_N] = \mathcal{J}$.
\end{proof}

\subsection{Proof of proposition \ref{pseudo-lip}}\label{proof of prop2}
Consider two matrices with the same  eigenvectors, $\bS = \bU \bLambda \bU^\intercal$ and $\tilde{\bS} = \bU \tilde{\bLambda} \bU^\intercal$, where $\bU$ is a Haar orthogonal matrix, and $\blam$, $\tilde{\blam}$ are distributed according to $P_N^{(1)}(\blam), P_N^{(2)}(\tilde{\blam})$, respectively. For two such matrices, we write $(\bS, \tilde{\bS}) \sim Q_N(\bU, \blam, \tilde{\blam})$, where  $Q_N(\bU, \blam, \tilde{\blam})$ is the joint p.d.f. of $\bU, \blam, \tilde{\blam}$,
\begin{equation*}
    d Q_N(\bU, \blam, \tilde{\blam}) = d \mu_N(\bU) \,\, P_N^{(1)}(\blam) \, d \blam \,\, P_N^{(2)}(\tilde{\blam}) \, d \tilde{\blam}.
\end{equation*}
For $t \in [0,1]$ an interpolating parameter, consider the following observation model:
\begin{equation}
    \begin{cases}
  \bY_1^{(t)} = \sqrt{\gamma t}\bS + \bZ_1\\
  \bY_2^{(t)} = \sqrt{\gamma (1-t)}\tilde{\bS} + \bZ_2
\end{cases}
\label{int-model}
\end{equation}
where $\bZ_1, \bZ_2$ are Wigner matrices independent of each other, and  $(\bS, \tilde{\bS}) \sim Q_N(\bU, \blam, \tilde{\blam})$. The free energy for this model can be written as :
\begin{equation*}
\begin{split}
        F_N(t) &= -\frac{1}{N^2} \bE_{\bY_1^{(t)}, \bY_2^{(t)}} \bigg[ \ln \int d Q_N(\bU, \blam, \tilde{\blam}) e^{\frac{N}{2} \Tr [\sqrt{\gamma t} \bX \bY_1^{(t)} - \frac{\gamma t}{2} \bX^2 +  \sqrt{\gamma (1-t)} \tilde{\bX} \bY_2^{(t)} - \frac{\gamma (1-t)}{2} \tilde{\bX}^2]} \bigg] \\
        &\hspace{-1pt}= -\frac{1}{N^2} \bE_{\bY_1^{(t)}, \bY_2^{(t)}} \bigg[ \ln \int d Q_N(\bU, \blam, \tilde{\blam}) e^{\frac{N}{2} \Tr [\gamma t \bX \bS + \sqrt{\gamma t} \bX \bZ_1 - \frac{\gamma t}{2} \bX^2 + \gamma (1-t) \tilde{\bX} \tilde{\bS}+  \sqrt{\gamma (1-t)} \tilde{\bX} \bZ_2 - \frac{\gamma (1-t)}{2} \tilde{\bX}^2]} \bigg]
\end{split}
\end{equation*}
where $\bX, \tilde{\bX}$  has the same eigenspace, $\bX = \bU \bLambda \bU^\intercal$, $\tilde{\bX} = \bU \tilde{\bLambda} \bU^\intercal$.
Note that, for $t=0$ the only term depending on $\blam$ (in both the inner and outer expectation) is the pdf $P_N^{(1)}(\blam)$ and we can integrate over $\blam$ in both of the expectations, to get $F_N(0) = F_N^{(2)}(\gamma)$. Similarly, we have $F_N(1) = F_N^{(1)}(\gamma)$.

Taking the derivative w.r.t. $t$, we get:
\begin{equation*}
\begin{split}
    \frac{d}{d t}F_N(t) = -\frac{1}{N} \bE \Big[ \frac{\gamma}{2}  \Tr \langle \bX \bS \rangle_t &+ \frac{1}{4} \sqrt{\frac{\gamma}{t}} \Tr \bZ_1 \langle \bX \rangle_t - \frac{\gamma}{4} \Tr \langle \bX^2 \rangle_t  \\
    &- \frac{\gamma}{2}  \Tr \langle \tilde{\bX} \tilde{\bS} \rangle_t - \frac{1}{4} \sqrt{\frac{\gamma}{1-t}} \Tr \bZ_2 \langle \tilde{\bX} \rangle_t + \frac{\gamma}{4} \Tr \langle \tilde{\bX}^2 \rangle_t \Big] 
\end{split}
\end{equation*}
where $\langle . \rangle_t$ denotes the expectation with respect to the posterior distribution of the model \eqref{int-model}. By integration by parts, we have
\begin{equation*}
    \bE \big[ \Tr \bZ_1 \langle \bX \rangle_t ] =  \sqrt{\gamma t} \bE \Big[ \Tr \langle \bX^2 \rangle_t  - \Tr \langle \bX \rangle_t^2 \Big], \hspace{5mm}  \bE \big[ \Tr \bZ_2 \langle \tilde{\bX} \rangle_t ] = \sqrt{\gamma (1-t)} \bE \Big[ \Tr \langle \tilde{\bX}^2 \rangle_t  - \Tr \langle \tilde{\bX} \rangle_t^2 \Big]
\end{equation*}
Therefore,
\begin{equation}
\begin{split}
        \frac{d}{d t}F_N(t) &= -\frac{1}{N} \frac{\gamma}{4} \bE \Big[ 2 \Tr \langle \bX \bS \rangle_t - \Tr \langle \bX \rangle_t^2 - 2 \Tr \langle \tilde{\bX} \tilde{\bS} \rangle_t + \Tr \langle \tilde{\bX} \rangle_t^2 \Big] \\
        &= \frac{1}{N} \frac{\gamma}{4} \bE \big[ \Tr [ \langle \bX \bS \rangle_t -  \langle \tilde{\bX} \tilde{\bS} \rangle_t ] \big]
\end{split}
\label{free-energy-two-coupl}
\end{equation}
In the second line, we used the Nishimori identity. In our setting, this identity states that for a continuous function $g$, given $(\bX_1, \tilde{\bX}_1)$, $(\bX_2, \tilde{\bX}_2)$ two i.i.d. samples from the posterior distribution (given $\bY_1^{(t)}, \bY_2^{(t)}$), the following relationship holds:
\begin{equation*}
    \bE \, \big\langle g(\bX_1, \tilde{\bX}_1, \bX_2, \tilde{\bX}_2) \big \rangle_t = \bE \, \big\langle g(\bX_1, \tilde{\bX}_1, \bS, \tilde{\bS}) \big \rangle_t
\end{equation*}
This identity is a direct consequence of Bayes' theorem. As an example, using this identity we get:
\begin{equation*}
    \bE  \langle \bX \rangle_t^2 = \bE \langle \bX_1 \rangle_t \langle \bX_2 \rangle_t = \bE \langle \bX_1 \bX_2 \rangle_t = \bE \langle \bX_1 \bS \rangle_t
\end{equation*}
Applying the trace on both sides, we find $\bE  \Tr \langle \bX \rangle_t^2 =  \bE \Tr \langle \bX \bS \rangle_t$.

From \eqref{free-energy-two-coupl}, we have
\begin{equation*}
    \begin{split}
        \frac{4 N }{\gamma} \Big|   \frac{d}{d t}F_N(t) \Big| &= \Bigg| \bE \bigg[ \Big \langle \Tr  \big[ \bS (\bX - \tilde{\bX})  -  (\tilde{\bS} - \bS)\tilde{\bX} \big] \Big\rangle_t  \bigg] \Bigg| \\
        & \leq   \bE \Bigg[ \bigg\langle \Big| \Tr  \big[ \bS (\bX - \tilde{\bX})  -  (\tilde{\bS} - \bS)\tilde{\bX} \big] \Big| \bigg\rangle_t  \Bigg] \hspace{5mm} \text{(By Jensen)} \\
        &  \leq \bE \Bigg[ \bigg\langle \Big| \Tr \bS (\bX - \tilde{\bX}) \Big| \bigg\rangle_t  \Bigg]   + \bE \Bigg[ \bigg\langle \Big| \Tr  (\tilde{\bS} - \bS)\tilde{\bX}] \Big| \bigg\rangle_t  \Bigg] \\
        & \leq    \bE \Big[ \| \bS \|_F  \langle \| \bX - \tilde{\bX} \|_F  \rangle_t  \Big]   + \bE \Big[ \| \bS - \tilde{\bS} \|_F \langle \|\tilde{\bX}\|_F \rangle_t  \Big] \\
        & \leq \sqrt{ \bE \Big[ \| \bS \|^2_F \Big] \bE \Big[ \big \langle \| \bX - \tilde{\bX} \|_F  \big \rangle_t^2  \Big] }   + \sqrt{\bE \Big[ \| \bS - \tilde{\bS} \|_F^2 \Big] \bE \Big[ \big \langle \|\tilde{\bX}\|_F \big \rangle_t^2  \Big] } \hspace{10pt} \text{(By Cauchy–Schwarz)} \\
        & \leq \sqrt{ \bE \big[ \| \bS \|^2_F \big] \bE \Big[ \big \langle \| \bX - \tilde{\bX} \|_F^2  \big \rangle_t  \Big] }   + \sqrt{\bE \big[ \| \bS - \tilde{\bS} \|_F^2 \big] \bE \Big[ \big \langle \|\tilde{\bX}\|_F^2 \big \rangle_t \Big] } \hspace{10pt} \text{(By Cauchy–Schwarz)} \\
        & = \sqrt{ \bE \big[ \| \bS \|^2_F \big] \bE \big[  \| \bS - \tilde{\bS} \|_F^2   \big] }   + \sqrt{\bE \big[ \| \bS - \tilde{\bS} \|_F^2 \big] \bE \big[ \|\tilde{\bS}\|_F^2  \big] } \hspace{10pt} \text{(By Nishimori)} \\
        &= \Big( \sqrt{ \bE \big[ \| \bS \|^2_F \big]} + \sqrt{ \bE \big[ \| \tilde{\bS} \|^2_F \big]} \Big) \sqrt{\bE \big[ \| \bS - \tilde{\bS} \|_F^2    \big]} \\
        &= \Big( \sqrt{ \bE_{\blam} \big[ \| \blam \|^2 \big]} + \sqrt{ \bE_{\tilde{\blam}} \big[ \| \tilde{\blam} \|^2 \big]} \Big) \sqrt{\bE_{\blam, \tilde{\blam}} \big[ \| \blam - \tilde{\blam} \|^2 \big] } .
    \end{split}
\end{equation*}
We obtain the result by integrating over $t$ from $0$ to $1$.$\hfill \square$

\subsection{Proof of technical lemmas}\label{proof of lemma 2}

\begin{lemma}\label{W-perm}
Given two vectors $\bu, \bv \in \bR^N$, denote their empirical distributions by $\mu, \nu$ respectively. We have
\begin{equation*}
    W_2 (\mu, \nu ) = \sqrt{\min_{\pi \in \mathcal{S}_N} \frac{1}{N} \| \bu - \bv_{\pi} \|^2 } 
\end{equation*}
where $\bv_{\pi}$ is a permutation of $\bv$.
\end{lemma}
\begin{proof}
By definition we have
\begin{equation*}
    W_2 (\mu, \nu )^2 = \inf_{\gamma \in \Gamma(\mu, \nu) } \bE_{\gamma(x,y)} \big[ (x - y)^2 \big].
\end{equation*}
Any measure in  $\Gamma(\mu, \nu)$ can be represented by a doubly stochastic $N \times N$ matrix. Thus, we have
\begin{equation*}
    W_2 (\mu, \nu )^2 = \inf_{\bm{P} \in \mathcal{B}_N } \frac{1}{N} \sum_{i,j} P_{ij} (u_i - v_j)^2
\end{equation*}
where $\mathcal{B}_N$ denotes the set of doubly stochastic matrices.
This minimization problem is a linear optimization problem on the bounded convex set $\mathcal{B}_N$. By Choquet's theorem \cite{phelps2001lectures}, the solutions to this problem exist and are the extremal points of $\mathcal{B}_N$, which are permutation matrices (by Birkhoff's theorem \cite{birkhoff1946tres}). Therefore, the minimization can be written on the set of permutation matrices to get:
\begin{equation*}
    W_2 (\mu, \nu )^2 = \min_{\pi \in \mathcal{S}_N } \frac{1}{N} \sum_{i,j}  (u_i - v_{\pi(i)})^2.
\end{equation*}
\end{proof} 

\begin{lemma}\label{E-W-0-l}
Suppose $\blam \in \bR^N$ is distributed according to $P_{S,N}(\blam)$, and $\blam^0$ is generated with i.i.d. elements from $\rho_S$. Let $\hat{\mu}_{\blam}, \hat{\mu}_{\blam^0}$ be their empirical distribution. We have:
\begin{equation*}
    \lim_{N \to \infty} \bE_{\blam} \big[ W_2(\hat{\mu}_{\blam}, \hat{\mu}_{\blam^0})^2 \big] = 0.
\end{equation*}
\end{lemma}
\begin{proof}
By the triangle inequality
\begin{equation}
    W_2(\hat{\mu}_{\blam}, \hat{\mu}_{\blam^0}) \leq W_2(\hat{\mu}_{\blam}, \rho_S) + W_2(\hat{\mu}_{\blam^0}, \rho_S).
\end{equation}
The first term approaches $0$ as $N \to \infty$ almost surely, by remark \ref{W2-conv}. By lemma \ref{W-conv-emp}, the second term also converges $0$ as $N \to \infty$. Therefore, we have $W_2(\hat{\mu}_{\blam}, \hat{\mu}_{\blam^0}) \to 0$ almost surely. Consequently, we have that $W_2(\hat{\mu}_{\blam}, \hat{\mu}_{\blam^0})^2 \to 0$ almost surely.
Denote $W_2(\hat{\mu}_{\blam}, \hat{\mu}_{\blam^0})^2$ by $X_N$ which is a non-negative random variable. We have:
\begin{equation}
\begin{split}
    \bE [ X_N ] &= \bE \big[ X_N \, \mathbb{I} \{ X_N \leq \epsilon \} \big] + \bE \big[ X_N \,\mathbb{I} \{ X_N > \epsilon \} \big] \\
    &\leq \epsilon + \bE \big[ X_N \, \mathbb{I} \{ X_N > \epsilon \} \big].
\label{c-Expec-W-1}
\end{split}
\end{equation}
By definition, one can see that $W_2(\hat{\mu}_{\blam}, \hat{\mu}_{\blam^0})^2 \leq 2 ( m^{(2)}_{\hat{\mu}_{\blam}} + m^{(2)}_{\hat{\mu}_{\blam^0}} )$, where $m^{(2)}_{\hat{\mu}_{\blam}} = \frac{1}{N} \sum \lambda_i^2 $ and $m^{(2)}_{\hat{\mu}_{\blam^0}} = \frac{1}{N} \sum {\lambda^0_i}^2$. For the second term in \eqref{c-Expec-W-1} we have
\begin{equation}
\begin{split}
        \bE \big[ X_N \, \mathbb{I} \{ X_N > \epsilon \} \big] &\leq 2 \bE \big[ ( m^{(2)}_{\hat{\mu}_{\blam}} + m^{(2)}_{\hat{\mu}_{\blam^0}} ) \, \mathbb{I} \{ X_N > \epsilon \} \big] \\
        &= 2 m^{(2)}_{\hat{\mu}_{\blam^0}} \bE \big[  \mathbb{I} \{ X_N > \epsilon \} \big] +  2 \bE \big[ m^{(2)}_{\hat{\mu}_{\blam}} \, \mathbb{I} \{ X_N > \epsilon \} \big] \\
        &= 2 m^{(2)}_{\hat{\mu}_{\blam^0}} \bP [   X_N > \epsilon ] +  2 \bE \big[ m^{(2)}_{\hat{\mu}_{\blam}} \, \mathbb{I} \{ X_N > \epsilon \} \big].
\end{split}
\label{c-Expec-W-2}
\end{equation}
 For the last term in \eqref{c-Expec-W-2} is decomposed as
\begin{equation}
    \begin{split}
        \bE \big[ m^{(2)}_{\hat{\mu}_{\blam}} \, \mathbb{I} \{ X_N > \epsilon \} \big] &= \bE \big[ m^{(2)}_{\hat{\mu}_{\blam}} \, \mathbb{I} \{ X_N > \epsilon \} \, \mathbb{I} \{ m^{(2)}_{\hat{\mu}_{\blam}} \leq t \} \big] + \bE \big[ m^{(2)}_{\hat{\mu}_{\blam}} \, \mathbb{I} \{ X_N > \epsilon \} \, \mathbb{I} \{ m^{(2)}_{\hat{\mu}_{\blam}} > t \} \big] \\
        & \leq t \bP [   X_N > \epsilon ] + \bE \big[ m^{(2)}_{\hat{\mu}_{\blam}} \, \mathbb{I} \{ m^{(2)}_{\hat{\mu}_{\blam}} > t \} \big]
    \end{split}
    \label{c-Expec-W-3}
\end{equation}
where $t$ is a fixed sufficiently large constant such that $t > C$. From \eqref{c-Expec-W-1}, \eqref{c-Expec-W-2}, \eqref{c-Expec-W-3}, we get:
\begin{equation}
    \bE [ X_N ] \leq \epsilon + 2 m^{(2)}_{\hat{\mu}_{\blam^0}} \bP [   X_N > \epsilon ] + 2 t \bP [   X_N > \epsilon ] + 2 \bE \big[ m^{(2)}_{\hat{\mu}_{\blam}} \, \mathbb{I} \{ m^{(2)}_{\hat{\mu}_{\blam}} > t \} \big].
        \label{c-Expec-W-4}
\end{equation}
Since $W_2(\hat{\mu}_{\blam}, \hat{\mu}_{\blam^0})^2 \to 0$ almost surely, $\bP [   X_N > \epsilon ]$ approaches 0 as $N \to \infty$. By construction $m^{(2)}_{\hat{\mu}_{\blam^0}}$ is bounded (by the constant $C$), so the second and the third terms approaches 0 as $N \to \infty$. The last term also converges 0, by lemma \ref{conv-mom-0}. Therefore, for arbitrary $\epsilon>0$ we find
\begin{equation*}
    \lim_{N \to \infty} \bE [ X_N ] \leq \epsilon.
\end{equation*}
\end{proof}

\begin{lemma}\label{W-conv-emp}
Let $X_1, \hdots, X_N$ be i.i.d. random variables distributed according to distribution $\mu$, which has finite support. Let $\mu_N$ denote their empirical distribution. Then
\begin{equation*}
    \lim_{N \to \infty} W_2(\mu_N, \mu ) = 0 \hspace{5mm} \text{almost surely.}
\end{equation*}
\end{lemma}
\begin{proof}
By the law of large numbers, $\mu_N \to \mu$ almost surely. Moreover, since $\mu$ has bounded support, the second moment of $\mu_N$ converges to the one of $\mu$. By Theorem 7.12 in \cite{villani2021topics}, we have the convergence in the Wasserstein-2 metric almost surely.
\end{proof}

\begin{lemma}\label{conv-mom-0}
Under assumption 1.B, for $t$ large enough, we have:
\begin{equation*}
    \lim_{N \to \infty} \bE \big[ m^{(2)}_{\hat{\mu}_{\blam}} \, \mathbb{I} \{ m^{(2)}_{\hat{\mu}_{\blam}} > t \} \big] = 0.
\end{equation*}
\end{lemma}
\begin{proof}
Boundedness and almost sure convergence of  $m^{(2)}_{\hat{\mu}_{\blam}}$ imply that $\lim_{N \to \infty} \bE [m^{(2)}_{\hat{\mu}_{\blam}}] = m^{(2)}_{\rho_S}$, which is bounded since $\rho_S$ has compact support.  Denote $X_N = m^{(2)}_{\hat{\mu}_{\blam}} \, \mathbb{I} \{ m^{(2)}_{\hat{\mu}_{\blam}} \leq t \}$. Then, $X_N \to  m^{(2)}_{\rho_S} $ a.s. and by bounded convergence we also have that $\lim_{N \to \infty} \bE [ X_N ] = m^{(2)}_{\rho_S}$. Therefore, 
\begin{equation*}
\begin{split}
    \lim_{N \to \infty} \bE \big[ m^{(2)}_{\hat{\mu}_{\blam}} \, \mathbb{I} \{ m^{(2)}_{\hat{\mu}_{\blam}} > t \} \big] &= \lim_{N \to \infty} \bE \big[ m^{(2)}_{\hat{\mu}_{\blam}}  - X_N \big] = 0.
\end{split}
\end{equation*}
\end{proof}


\section{Derivation of the limiting spectral distribution for the model \ref{Ex-1}}\label{Deriv-Ex-1}

Suppose we want to find the density $\mu(x)$, which is the free convolution of the density $\rho(x)$ with the semi-circle density $\rho_{\rm sc}(x)$. This density is given in \cite{biane1997free} by
\begin{equation}
    \begin{split}
        \mu \big( \psi(u) \big) &= \frac{v(u)}{\pi} \\
        \psi(u) = u + \int_{\bR} \frac{u-x}{(u-x)^2+v(u)^2} \rho(x) \, dx,& \hspace{10pt} v(u) = \inf \Big\{ w \geq 0 \big| \int_{\bR} \frac{\rho(x)}{(u-x)^2+w^2}  \, dx \leq 1 \Big\} .
    \end{split}
\label{Biane-technique}
\end{equation}
This result will be used repeatedly.

To compute the density $\rho_Y = \rho_{\sqrt{\gamma} S} \boxplus \rho_{\rm sc}$ where  $ \rho_{\sqrt{\gamma} S}(x) = \frac{1}{2} \delta(x+\sqrt{\gamma}) + \frac{1}{2} \delta(x-\sqrt{\gamma})$, we compute functions $v(u)$ and $\psi(u)$
\begin{equation}
\begin{split}
        \text{if } \gamma < 1: \hspace{5pt} v(u) &= \begin{cases}
    \frac{1}{\sqrt{2}} \sqrt{1-2(u^2+\gamma) +\sqrt{1+16 \gamma u^2}} &\text{if }|u| \leq \frac{1}{\sqrt{2}}\sqrt{1+2\gamma + \sqrt{1+8\gamma}} \\
    0 & \text{else.}
    \end{cases} \\
    \psi(u) &= \begin{cases}
    \frac{1+ 8u^2-\sqrt{1+16\gamma u^2}}{4 u} &\text{if }|u| \leq \frac{1}{\sqrt{2}}\sqrt{1+2\gamma + \sqrt{1+8\gamma}} \\
    \frac{u(u^2-\gamma +1)}{u^2 - \gamma} & \text{else.}
    \end{cases}
\end{split}
\end{equation}
\begin{equation}
\begin{split}
        \text{if }& \gamma \geq 1: \\
    \hspace{5pt} v(u) &= \begin{cases}
    \frac{1}{\sqrt{2}} \sqrt{1-2(u^2+\gamma) +\sqrt{1+16 \gamma u^2}} & \text{if }\frac{1}{\sqrt{2}}\sqrt{1+2\gamma - \sqrt{1+8\gamma}} \leq|u| \leq \frac{1}{\sqrt{2}}\sqrt{1+2\gamma + \sqrt{1+8\gamma}} \\
    0 & \text{else.}
    \end{cases} \\
    \psi(u) &= \begin{cases}
    \frac{1+ 8u^2-\sqrt{1+16\gamma u^2}}{4 u} &\text{if }\frac{1}{\sqrt{2}}\sqrt{1+2\gamma - \sqrt{1+8\gamma}} \leq |u| \leq \frac{1}{\sqrt{2}}\sqrt{1+2\gamma + \sqrt{1+8\gamma}} \\
    \frac{u(u^2-\gamma +1)}{u^2 - \gamma} & \text{else.}
    \end{cases}
\end{split}
\end{equation}
Solving the equation $\rho_Y \big( \psi(u) \big) = \frac{v(u)}{\pi}$, we find:
\begin{equation}
\begin{aligned}
        &\text{if } \gamma < 1 : \hspace{5pt} \rho_Y(x) = \begin{cases}
    \frac{1}{\sqrt{2}\pi} \sqrt{ 1 - 2 \big( \gamma + \frac{1}{2304} A^2 \big) + \sqrt{1+\frac{\gamma}{144} A^2 } } &\text{if } |x| \leq U(\gamma) \\
    0 &\text{else.}
    \end{cases} \\
            &\text{if } \gamma \geq 1 : \hspace{5pt} \rho_Y(x) = \begin{cases}
    \frac{1}{\sqrt{2}\pi} \sqrt{ 1 - 2 \big( \gamma + \frac{1}{2304} A^2 \big) + \sqrt{1+\frac{\gamma}{144} A^2 } } &\text{if } L(\gamma) \leq |x| \leq U(\gamma) \\
    0 &\text{else.}
    \end{cases} 
\end{aligned}
\label{rho_Y expr}
\end{equation}
where
\begin{equation}
    A = 16 x + \frac{32 \times 2^{1/3}(-3+3 \gamma+x^2) }{B}+ 2^{2/3} B 
\end{equation}
\begin{equation*}
    B = \sqrt[3]{576 x + 1152 \gamma x -128 x^3 + 64 \sqrt{x^2(9+18\gamma-2x^2)^2 - 4(-3+3\gamma+x^2)^3}}
\end{equation*}
\begin{equation}
    L(\gamma) = \frac{\big(-3+\sqrt{1+8 \gamma} \big) \sqrt{1+2\gamma - \sqrt{1+8\gamma}} }{\sqrt{2} \big(-1+\sqrt{1+8 \gamma} \big)}, \hspace{10pt} U(\gamma) = \frac{\big(3+\sqrt{1+8 \gamma} \big) \sqrt{1+2\gamma + \sqrt{1+8\gamma}} }{\sqrt{2} \big(1+\sqrt{1+8 \gamma} \big)}
\end{equation}
\begin{figure}
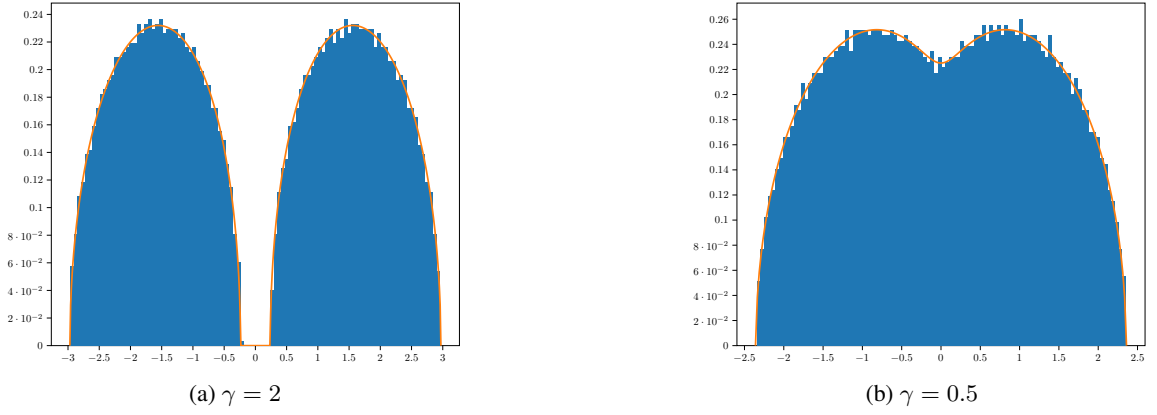
 
    \centering
    \begin{subfigure}{.45\textwidth}
        \centering
        \input{Supplementary/Examples/Rademacher_spectrum/spec-2}
        \caption{$\gamma=2$}
    \end{subfigure}
    \hfill
    \begin{subfigure}{.45\textwidth}
        \centering
        \input{Supplementary/Examples/Rademacher_spectrum/spec-5}
        \caption{$\gamma=0.5$}
    \end{subfigure}
    \captionsetup{singlelinecheck = false, justification=justified}
    \caption{{\small The continuous line is the asymptotic spectral density of the observation matrix $\bY$ in example \ref{Ex-1}. $\rho_Y(x) = \rho_{\sqrt{\gamma} S} \boxplus \rho_{\rm sc}$ where $\rho_{\sqrt{\gamma} S}(x) = \frac{1}{2} \delta(x+\sqrt{\gamma}) + \frac{1}{2} \delta(x-\sqrt{\gamma})$ for $\gamma = 2$ and $\gamma = 0.5$. Compared to the histogram of a realization of size $N=5000$.}}
    \label{fig:limiting spectrum of rho_Y Ex 1}
\end{figure}
From \eqref{rho_Y expr}, one can see that the support of $\rho_Y$ is constituted of two disjoint intervals for $\gamma \geq 1$, and of one single interval for $\gamma < 1$. 

Once we have the expression \eqref{rho_Y expr}, we get from Theorem 2 an explicit integral representation for ${\rm MMSE}(\gamma)$, as well as for the derivatives.  We show the details here for  $\gamma \geq 1$. Since in this example $\rho_Y(x)$ is symmetric we have for $\gamma >1$
\begin{equation}
    {\rm MMSE}(\gamma) = \frac{1}{\gamma} \Big( 1 - \frac{8 \pi^2}{3} \int_{L(\gamma)}^{U(\gamma)} \rho_Y^3(x) \, dx \Big).
    \label{MMSE-Ex1}
\end{equation}
By the Leibniz integral rule (all derivatives are w.r.t $\gamma$) we have using that $\rho_Y$ vanishes at the end-points of the interval $[L(\gamma), U(\gamma)]$
\begin{equation}
\begin{split}
     {\rm MMSE}'(\gamma) &= -\frac{1}{\gamma^2} + \frac{1}{\gamma^2} \frac{8 \pi^2}{3} \int_{L(\gamma)}^{U(\gamma)} \rho_Y^3(x) \, dx \\
     &\hspace{1.5cm}- \frac{1}{\gamma}\frac{8 \pi^2}{3} \Big( \rho_Y^3\big(U(\gamma)\big) U'(\gamma) - \rho_Y^3\big(L(\gamma)\big) L'(\gamma) + \int_{L(\gamma)}^{U(\gamma)} 3 \rho_Y^2(x) \rho_Y'(x) \, dx \Big) \\
     &= -\frac{1}{\gamma^2} + \frac{1}{\gamma^2} \frac{8 \pi^2}{3} \int_{L(\gamma)}^{U(\gamma)} \rho_Y^3(x) \, dx - \frac{8 \pi^2}{\gamma}\int_{L(\gamma)}^{U(\gamma)} \rho_Y^2(x) \rho_Y'(x) \, dx.
\end{split}
\label{dMMSE-Ex1}
\end{equation}
Moreover, $\rho_Y^2(x) \rho_Y'(x)$ can also be checked to vanishes at the end-points of the interval $[L(\gamma), U(\gamma)]$ so
\begin{equation}
\begin{split}
    {\rm MMSE}''(\gamma) &= \frac{2}{\gamma^3} - \frac{1}{\gamma^3} \frac{16 \pi^2}{3} \int_{L(\gamma)}^{U(\gamma)} \rho_Y^3(x) \, dx + \frac{16 \pi^2}{\gamma^2}\int_{L(\gamma)}^{U(\gamma)} \rho_Y^2(x) \rho_Y'(x) \, dx \\
    &\hspace{1cm}- \frac{8 \pi^2}{\gamma}   \int_{L(\gamma)}^{U(\gamma)} \big[ 2\rho_Y(x) \big(\rho_Y'(x))^2 + \rho_Y^2(x) \rho_Y''(x) \big] \, dx.
\end{split}
\label{ddMMSE-Ex1}
\end{equation}
A similar and somewhat simpler calculation also provides integral representations for $\gamma <1$. 

\begin{figure}
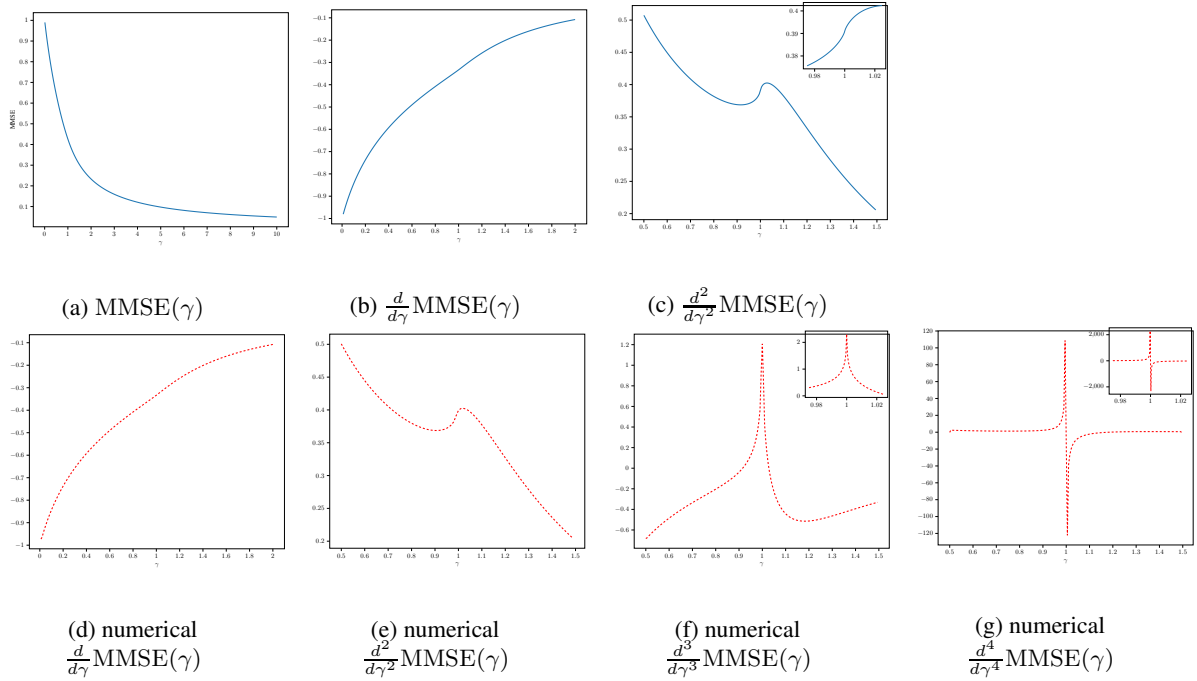
 
    \centering
    \begin{subfigure}{.2\textwidth}
        \centering
        \input{Supplementary/Examples/Rademacher_spectrum/MMSE_Rademach}
        \caption{{\small${\rm MMSE}(\gamma)$}}
    \end{subfigure}
    \hfil
        \begin{subfigure}{.2\textwidth}
        \centering
\begin{tikzpicture}[scale=0.25]

\definecolor{color0}{rgb}{0.12156862745098,0.466666666666667,0.705882352941177}

\begin{axis}[
tick align=outside,
tick pos=left,
x grid style={white!69.0196078431373!black},
xlabel={$\gamma$},
xmin=-0.0895, xmax=2.0995,
xtick style={color=black},
y grid style={white!69.0196078431373!black},
ymin=-1.02422657234277, ymax=-0.0639362725565435,
ytick style={color=black},
label style={font=\fontsize{40}{42}},
tick label style={font=\fontsize{20}{22}}
]
\addplot [ultra thick, color0]
table {%
0.01 -0.980577013261581
0.02 -0.962224004733308
0.03 -0.944829526655763
0.04 -0.928298971625107
0.05 -0.912551102263986
0.06 -0.897515707976817
0.07 -0.883131603898002
0.08 -0.869345161520686
0.09 -0.856109091430111
0.1 -0.843381468366843
0.11 -0.831124942968393
0.12 -0.819306095502232
0.13 -0.807894895150833
0.14 -0.79686425502466
0.15 -0.786189656015376
0.16 -0.775848830293858
0.17 -0.765821492533053
0.18 -0.756089108950357
0.19 -0.746634699664751
0.2 -0.737442670766927
0.21 -0.728498664320263
0.22 -0.719789431866388
0.23 -0.711302721355168
0.24 -0.7030271796138
0.25 -0.694952265447632
0.26 -0.687068174041136
0.27 -0.679365768849003
0.28 -0.67183652197056
0.29 -0.664472460546225
0.3 -0.657266119375222
0.31 -0.650210498107525
0.32 -0.643299022645692
0.33 -0.636525511154056
0.34 -0.629884142550942
0.35 -0.62336942862471
0.36 -0.616976188446309
0.37 -0.610699525471559
0.38 -0.60453480621549
0.39 -0.598477641429888
0.4 -0.592523868424496
0.41 -0.586669535232275
0.42 -0.580910885780934
0.43 -0.575244346685516
0.44 -0.569666514664815
0.45 -0.564174145395255
0.46 -0.558764142851221
0.47 -0.553433549672268
0.48 -0.548179538209861
0.49 -0.542999402209317
0.5 -0.537890549093608
0.51 -0.532850492672337
0.52 -0.52787684663607
0.53 -0.52296731816014
0.54 -0.518119702142741
0.55 -0.513331875627184
0.56 -0.508601792810163
0.57 -0.503927480032418
0.58 -0.49930703127969
0.59 -0.494738603858243
0.6 -0.490220414312868
0.61 -0.485750734428843
0.62 -0.481327887577674
0.63 -0.476950245085949
0.64 -0.472616222820945
0.65 -0.468324277774981
0.66 -0.464072905029667
0.67 -0.459860634293041
0.68 -0.455686026967142
0.69 -0.451547673022319
0.7 -0.447444187904606
0.71 -0.443374209422255
0.72 -0.439336394630809
0.73 -0.435329416607288
0.74 -0.43135196106334
0.75 -0.427402722910533
0.76 -0.423480402817671
0.77 -0.419583702916212
0.78 -0.415711322833699
0.79 -0.411861955048356
0.8 -0.408034279797419
0.81 -0.404226959404358
0.82 -0.400438631857791
0.83 -0.396667903311579
0.84 -0.392913339679952
0.85 -0.389173456745505
0.86 -0.385446707638323
0.87 -0.381731468545628
0.88 -0.378026021031136
0.89 -0.37432852884998
0.9 -0.37063700971249
0.91 -0.366949297852176
0.92 -0.363262993609353
0.93 -0.359575393592293
0.94 -0.355883390789458
0.95 -0.352183325233784
0.96 -0.348470744594035
0.97 -0.344739987340984
0.98 -0.340983366447204
0.99 -0.337189181962098
1 -0.333333333332004
1.01 -0.329372272858879
1.02 -0.325363846032007
1.03 -0.321340404362144
1.04 -0.317319271383051
1.05 -0.313312186996761
1.06 -0.309327762177757
1.07 -0.305372569076412
1.08 -0.301451736228911
1.09 -0.297569315098506
1.1 -0.293728524425683
1.11 -0.289931922493849
1.12 -0.286181533808963
1.13 -0.282478945289661
1.14 -0.278825381129142
1.15 -0.275221762255209
1.16 -0.271668754361758
1.17 -0.268166807215216
1.18 -0.264716187184101
1.19 -0.261317004418437
1.2 -0.257969235752845
1.21 -0.254672744115954
1.22 -0.251427295101638
1.23 -0.248232571155194
1.24 -0.245088183811384
1.25 -0.241993684263658
1.26 -0.238948572504994
1.27 -0.235952305304392
1.28 -0.233004303140768
1.29 -0.230103956256073
1.3 -0.227250629960568
1.31 -0.224443669251763
1.32 -0.221682402891893
1.33 -0.218966146970432
1.34 -0.216294208014061
1.35 -0.21366588574415
1.36 -0.21108047548014
1.37 -0.20853727023958
1.38 -0.206035562589316
1.39 -0.20357464626608
1.4 -0.201153817581655
1.41 -0.198772376659565
1.42 -0.19642962851279
1.43 -0.194124883948154
1.44 -0.19185746040037
1.45 -0.189626682601437
1.46 -0.187431883175695
1.47 -0.185272403149249
1.48 -0.183147592365531
1.49 -0.181056809839477
1.5 -0.178999424052688
1.51 -0.176974813186133
1.52 -0.174982365302126
1.53 -0.173021478494129
1.54 -0.171091560982481
1.55 -0.169192031190222
1.56 -0.16732231777757
1.57 -0.165481859651248
1.58 -0.163670105953317
1.59 -0.161886516033867
1.6 -0.160130559382971
1.61 -0.158401715578728
1.62 -0.156699474198737
1.63 -0.155023334719273
1.64 -0.153372806414886
1.65 -0.151747408243307
1.66 -0.150146668722757
1.67 -0.148570125797039
1.68 -0.147017326705102
1.69 -0.145487827842044
1.7 -0.143981194605547
1.71 -0.142497001260522
1.72 -0.141034830781813
1.73 -0.139594274704106
1.74 -0.138174932973558
1.75 -0.1367764137809
1.76 -0.135398333427616
1.77 -0.134040316152536
1.78 -0.13270199399098
1.79 -0.131383006614417
1.8 -0.130083001180775
1.81 -0.128801632184384
1.82 -0.127538561310128
1.83 -0.126293457270478
1.84 -0.12506599568056
1.85 -0.123855858894505
1.86 -0.122662735873667
1.87 -0.121486322042888
1.88 -0.120326319153458
1.89 -0.119182435142466
1.9 -0.118054384003504
1.91 -0.11694188565478
1.92 -0.11584466580343
1.93 -0.114762455830035
1.94 -0.113694992651001
1.95 -0.11264201860387
1.96 -0.111603281327196
1.97 -0.110578533641026
1.98 -0.109567533429054
1.99 -0.108570043533446
2 -0.107585831637736
};
\end{axis}

\end{tikzpicture}
        \caption{{\small$\frac{d}{d \gamma} {\rm MMSE}(\gamma)$}}
    \end{subfigure}
    \hfil
        \begin{subfigure}{.2\textwidth}
        \centering
        \input{Supplementary/Examples/Rademacher_spectrum/ddMSE_Rademach_wCL}
        \caption{{\small$\frac{d^2}{d \gamma^2} {\rm MMSE}(\gamma)$}}
    \end{subfigure}
    \hfil
    \begin{subfigure}{.2\textwidth}
\begin{tikzpicture}[scale=0.25]

\end{tikzpicture}
    \end{subfigure}\\
    \begin{subfigure}{.2\textwidth}
        \centering
\begin{tikzpicture}[scale=0.25]

\begin{axis}[
tick align=outside,
tick pos=left,
x grid style={white!69.0196078431373!black},
xlabel={$\gamma$},
xmin=-0.0895, xmax=2.0995,
xtick style={color=black},
y grid style={white!69.0196078431373!black},
ymin=-1.01606597707709, ymax=-0.0643248721436706,
ytick style={color=black},
label style={font=\fontsize{40}{42}},
tick label style={font=\fontsize{20}{22}}
]
\addplot [ultra thick, red, dashed]
table {%
0.01 -0.972805017761939
0.02 -0.963796602439484
0.03 -0.944829077960156
0.04 -0.928298602059739
0.05 -0.912550812729216
0.06 -0.897515445099277
0.07 -0.883131394725563
0.08 -0.869344985548928
0.09 -0.856108935918646
0.1 -0.84338133885856
0.11 -0.831124833798976
0.12 -0.819305996897455
0.13 -0.807894807986868
0.14 -0.796864178768975
0.15 -0.786189592565227
0.16 -0.77584877470598
0.17 -0.765821436830029
0.18 -0.756089061359893
0.19 -0.746634659403754
0.2 -0.737442632984997
0.21 -0.728498631513642
0.22 -0.719789401912098
0.23 -0.711302694413804
0.24 -0.703027155583077
0.25 -0.694952244436996
0.26 -0.68706815379362
0.27 -0.679365749163944
0.28 -0.671836504160723
0.29 -0.664472446209162
0.3 -0.657266105215308
0.31 -0.650210485000247
0.32 -0.643299011450108
0.33 -0.63652550001561
0.34 -0.629884132098148
0.35 -0.623369419316369
0.36 -0.616976179067737
0.37 -0.610699516508169
0.38 -0.604534798671778
0.39 -0.59847763479934
0.4 -0.592523862199877
0.41 -0.586669529352317
0.42 -0.580910879853546
0.43 -0.575244341373548
0.44 -0.569666510150765
0.45 -0.564174140345948
0.46 -0.558764138455501
0.47 -0.553433545770218
0.48 -0.548179534821659
0.49 -0.542999398143366
0.5 -0.537890545058457
0.51 -0.53285048971261
0.52 -0.527876843916144
0.53 -0.522967315412857
0.54 -0.518119699841176
0.55 -0.513331873597755
0.56 -0.508601790337911
0.57 -0.50392747763863
0.58 -0.499307029655994
0.59 -0.494738602261784
0.6 -0.49022041217294
0.61 -0.485750732438058
0.62 -0.481327885486461
0.63 -0.476950242694639
0.64 -0.472616222622186
0.65 -0.468324277860778
0.66 -0.464072903308361
0.67 -0.459860633333489
0.68 -0.455686026624618
0.69 -0.451547672482089
0.7 -0.447444187157032
0.71 -0.443374208715323
0.72 -0.43933639243393
0.73 -0.435329413675092
0.74 -0.431351962601636
0.75 -0.42740272538029
0.76 -0.423480402674127
0.77 -0.419583703370118
0.78 -0.415711323489764
0.79 -0.411861955605528
0.8 -0.408034279754384
0.81 -0.404226956799068
0.82 -0.400438629445922
0.83 -0.396667909081783
0.84 -0.392913346082735
0.85 -0.389173458250281
0.86 -0.385446714530843
0.87 -0.381731470807793
0.88 -0.378026016004742
0.89 -0.374328534602682
0.9 -0.370637028322726
0.91 -0.366949315079166
0.92 -0.363262998109453
0.93 -0.359575403181591
0.94 -0.355883426325363
0.95 -0.352183367176801
0.96 -0.348470805690947
0.97 -0.344740116405601
0.98 -0.340983721835821
0.99 -0.337190469446676
1 -0.333327689608505
1.01 -0.329373765750738
1.02 -0.325364286233956
1.03 -0.321340584184515
1.04 -0.317319373562946
1.05 -0.313312254370879
1.06 -0.309327810206802
1.07 -0.305372604467429
1.08 -0.30145176390844
1.09 -0.297569337331663
1.1 -0.293728541912324
1.11 -0.289931936848908
1.12 -0.286181546141903
1.13 -0.282478955478493
1.14 -0.278825389419793
1.15 -0.27522176931084
1.16 -0.271668760517035
1.17 -0.268166812441944
1.18 -0.264716191675462
1.19 -0.261317008285009
1.2 -0.257969239073139
1.21 -0.254672746771199
1.22 -0.25142729724954
1.23 -0.248232573407141
1.24 -0.24508818572537
1.25 -0.241993685661174
1.26 -0.238948573742355
1.27 -0.235952306251458
1.28 -0.233004304009414
1.29 -0.230103956837519
1.3 -0.227250630274488
1.31 -0.224443669818057
1.32 -0.221682403168778
1.33 -0.218966147045802
1.34 -0.216294208394286
1.35 -0.213665886016315
1.36 -0.211080475392875
1.37 -0.208537270099549
1.38 -0.206035562391915
1.39 -0.20357464600958
1.4 -0.201153817682314
1.41 -0.198772376473922
1.42 -0.196429628182557
1.43 -0.194124883874842
1.44 -0.191857460044283
1.45 -0.18962668232107
1.46 -0.187431882978913
1.47 -0.185272402789443
1.48 -0.183147592064166
1.49 -0.181056809633932
1.5 -0.17899942366803
1.51 -0.176974812752985
1.52 -0.174982364898789
1.53 -0.173021478135719
1.54 -0.171091560749258
1.55 -0.169192030875509
1.56 -0.167322317339635
1.57 -0.165481859403815
1.58 -0.163670105560467
1.59 -0.161886515546387
1.6 -0.1601305592609
1.61 -0.158401715348615
1.62 -0.156699473772038
1.63 -0.155023334358529
1.64 -0.153372806194945
1.65 -0.151747407927073
1.66 -0.150146668310823
1.67 -0.148570125556976
1.68 -0.147017326310865
1.69 -0.14548782740932
1.7 -0.143981194363862
1.71 -0.14249700098438
1.72 -0.14103483063348
1.73 -0.139594274341824
1.74 -0.138174932707746
1.75 -0.136776413657407
1.76 -0.135398333157993
1.77 -0.134040315899904
1.78 -0.132701993585541
1.79 -0.131383006291442
1.8 -0.130083001133926
1.81 -0.128801631903473
1.82 -0.127538561048374
1.83 -0.126293457129188
1.84 -0.125065995322574
1.85 -0.123855858657779
1.86 -0.122662735740739
1.87 -0.121486321809236
1.88 -0.12032631885851
1.89 -0.119182435013878
1.9 -0.118054383781383
1.91 -0.116941885573024
1.92 -0.115844665736905
1.93 -0.114762455550253
1.94 -0.11369499251304
1.95 -0.112642018513912
1.96 -0.111603281045714
1.97 -0.11057853343693
1.98 -0.109567533349036
1.99 -0.108570043262411
2 -0.107585831458826
};
\end{axis}

\end{tikzpicture}
        \caption{{\small numerical $\frac{d}{d \gamma} {\rm MMSE}(\gamma)$}}
    \end{subfigure}
    \hfil
        \begin{subfigure}{.2\textwidth}
        \centering
\begin{tikzpicture}[scale=0.25]

\begin{axis}[
tick align=outside,
tick pos=left,
x grid style={white!69.0196078431373!black},
xlabel={$\gamma$},
xmin=0.4505, xmax=1.5395,
xtick style={color=black},
y grid style={white!69.0196078431373!black},
ymin=0.189261722424671, ymax=0.515473959029323,
ytick style={color=black},
label style={font=\fontsize{40}{42}},
tick label style={font=\fontsize{20}{22}}
]
\addplot [ultra thick, red, dashed]
table {%
0.5 0.500646130092748
0.51 0.494121242889215
0.52 0.487821157512282
0.53 0.481737386981197
0.54 0.475861971107514
0.55 0.47018744912564
0.56 0.464706837290375
0.57 0.45941359080087
0.58 0.454301584425027
0.59 0.449365097776506
0.6 0.444598792577118
0.61 0.439997693760298
0.62 0.435557178333087
0.63 0.431272966164425
0.64 0.427141096144281
0.65 0.423157933347651
0.66 0.419320164562847
0.67 0.415624775339315
0.68 0.412069063579182
0.69 0.408650637201457
0.7 0.40536741479457
0.71 0.402217630653942
0.72 0.399199850233568
0.73 0.396312981340824
0.74 0.393556268952275
0.75 0.390929347707765
0.76 0.388432266746635
0.77 0.386065499354946
0.78 0.383830006486614
0.79 0.381727284800643
0.8 0.379759431502005
0.81 0.377929238539682
0.82 0.376240289698846
0.83 0.374697069242731
0.84 0.373305179725697
0.85 0.372071557918331
0.86 0.371004708016429
0.87 0.370115163661284
0.88 0.369415998797592
0.89 0.368923504672022
0.9 0.368658279728437
0.91 0.368646792966961
0.92 0.36892374950637
0.93 0.369535815439632
0.94 0.370548027600652
0.95 0.372055656667882
0.96 0.374207349191335
0.97 0.377241931408763
0.98 0.38193792032915
0.99 0.3909646034213
1 0.399226006666848
1.01 0.402007383541089
1.02 0.402470927746084
1.03 0.40158045890679
1.04 0.399701986305166
1.05 0.39707840173873
1.06 0.393877797437584
1.07 0.390223283926471
1.08 0.386208732027158
1.09 0.381907820144243
1.1 0.377379626040976
1.11 0.372672286141343
1.12 0.367825509999434
1.13 0.362872373570081
1.14 0.357840633871886
1.15 0.352753719790846
1.16 0.347631496537384
1.17 0.342490864711007
1.18 0.337346236256529
1.19 0.332209919478337
1.2 0.327092432886743
1.21 0.322002764538492
1.22 0.316948587247824
1.23 0.311936437797019
1.24 0.306971871669342
1.25 0.302059591695979
1.26 0.297203557551829
1.27 0.292407082017688
1.28 0.287672911574743
1.29 0.283003298213375
1.3 0.278400060531349
1.31 0.273864635867872
1.32 0.269398129292015
1.33 0.265001353320838
1.34 0.260674862837607
1.35 0.256418988431814
1.36 0.252233863737738
1.37 0.248119449079261
1.38 0.244075554010607
1.39 0.240101856463317
1.4 0.236197918608246
1.41 0.232363204249989
1.42 0.228597090343588
1.43 0.224898878638684
1.44 0.221267808320836
1.45 0.217703063188892
1.46 0.214203780994566
1.47 0.210769061293076
1.48 0.207397971163541
1.49 0.204089551361246
};
\end{axis}

\end{tikzpicture}
        \caption{{\small numerical $\frac{d^2}{d \gamma^2} {\rm MMSE}(\gamma)$}}
    \end{subfigure}
    \hfil
        \begin{subfigure}{.2\textwidth}
        \centering
        \input{Supplementary/Examples/Rademacher_spectrum/dddMSE_Rademach_wCL_numeric}
        \caption{{\small numerical $\frac{d^3}{d \gamma^3} {\rm MMSE}(\gamma)$}}
    \end{subfigure}
    \hfil
    \begin{subfigure}{.2\textwidth}
        \centering
        \input{Supplementary/Examples/Rademacher_spectrum/ddddMSE_Rademach_wCL_numeric}
        \caption{{\small numerical $\frac{d^4}{d \gamma^4} {\rm MMSE}(\gamma)$}}
    \end{subfigure}
    \captionsetup{singlelinecheck = false, justification=justified}
    \caption{{\small Analysis of the MMSE in example \ref{Ex-1}. In (a), MMSE is plotted from \eqref{MMSE-Ex1} for $0<\gamma \leq 10$ with step-size $h = 0.01$. The numerical first derivative of the curve in (a) is illustrated in (d), which is computed using five-point stencil \cite{sauer2011numerical} $f'(x) \approx \frac{-f(x+2 h)+8 f(x+h)-8 f(x-h)+f(x-2h)}{12 h}$. In (b), the first derivative is plotted from \eqref{dMMSE-Ex1} with step-size $h = 0.01$, and its numerical first derivative is plotted in (e). The second derivative of MMSE, computed from \eqref{ddMMSE-Ex1}, is depicted in (c) with step-size $h=0.005$. The inset plot is with step-size $h = 0.00025$. The third derivative of MMSE in (f) is obtained from the numerical differentiation of the curve in (c). The fourth derivative is computed using the five-point stencil $f''(x)     \approx \frac{-f(x+2 h)+16 f(x+h)-30 f(x) + 16 f(x-h) - f(x-2h)}{12 h^2}$ from the curve in (c).}}
    \label{fig: dMMSE-app}
\end{figure}

It is not clear how to compute these integrals analytically but precise results can be obtained from numerical integration. Integral represenations of higher derivatives can also be obtained in principle but become unwieldy. In fact the numerical integration of the formula for the second derivative is precise enough to get a good numerical calculation of the third and fourth derivatives. All numerical results  are summearized  in figure \ref{fig: dMMSE-app}.

\begin{figure} 
    \centering
    \begin{subfigure}{0.4\textwidth}
        \centering
        \includegraphics[scale=0.65]{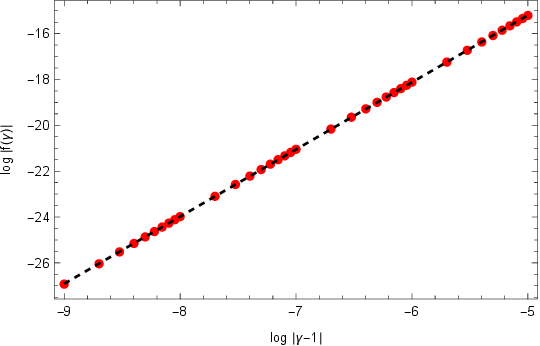}
        \caption{{\small $\log |f(\gamma)|$ for $\gamma \to 1^-$. The fitted line has slope $\alpha \approx 2.929$, and $\log |c| \approx -0.5511 $.}}
    \end{subfigure}
    \hfil
    \begin{subfigure}{0.4\textwidth}
        \centering
        \includegraphics[scale=0.65]{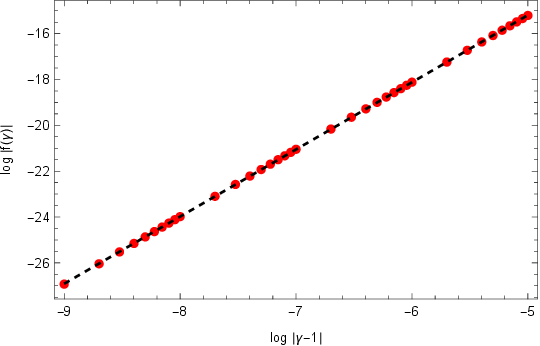}
         \caption{{\small $\log |f(\gamma)|$ for $\gamma \to 1^+$. The fitted line has slope $\alpha \approx 2.929$, and $\log |c| \approx -0.5511 $.}}
    \end{subfigure}
    \caption{{\small $\log |f(\gamma)|$ as a function of $\log |\gamma - 1|$}}
    \label{fig: behavior of MMSE}
\end{figure}

\begin{figure} 
    \centering
    \includegraphics[scale=0.7]{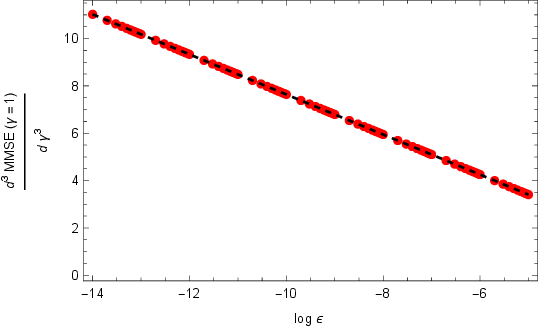}
    \captionsetup{singlelinecheck = false, justification=justified}
    \caption{{\small Numerical third derivative of MMSE at $\gamma = 1$ computed from $\frac{{\rm MMSE}''(1+\epsilon) - {\rm MMSE}''(1-\epsilon)}{2 \epsilon}$} as a function of $\log \epsilon$. A linear function $-0.8463 \log \epsilon - 0.8249$ is fitted to the points.}
    \label{fig:third-derivative MMSE}
\end{figure}

\begin{figure} 
    \centering
    \begin{subfigure}{0.4\textwidth}
        \centering
        \includegraphics[scale=0.65]{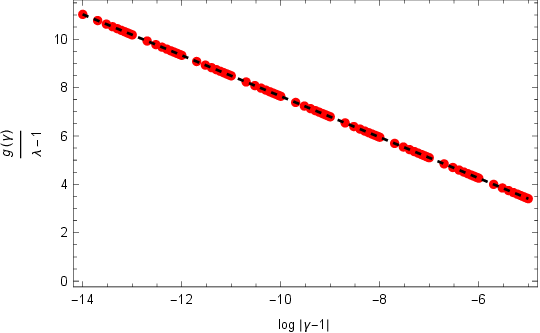}
        \caption{{\small $\frac{g(\gamma)}{\gamma -1}$ for $\gamma \to 1^-$. The fitted line has slope $a \approx -0.8463$, and $ab \approx -0.8249 $.}}
    \end{subfigure}
    \hspace{20 pt}
    \begin{subfigure}{0.4\textwidth}
        \centering
        \includegraphics[scale=0.65]{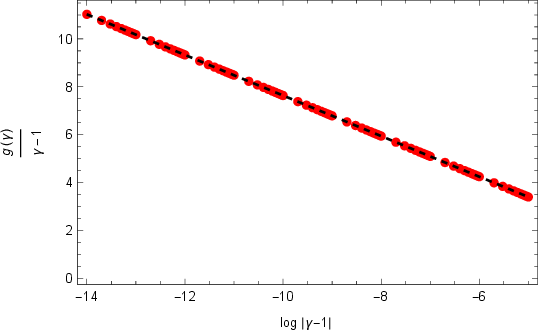}
         \caption{{\small $\frac{g(\gamma)}{\gamma -1}$ for $\gamma \to 1^+$. The fitted line has slope $a \approx -0.8463$, and $ab \approx -0.8249 $.}}
    \end{subfigure}
    \caption{{\small $\frac{g(\gamma)}{\gamma -1}$ as a function of $\log |\gamma - 1|$}}
    \label{fig: behavior of MMSE''}
\end{figure}

Plots (f),(g) in figure \ref{fig: dMMSE-app} suggest the existence of a third-order phase transition at $\gamma_c = 1$. Note that this is the point where the support of $\rho_Y$ transitions from a single to two intervals. Since the singularity seems to appear in teh third derivative we define 
the function 
\begin{equation*}
    f(\gamma) = {\rm MMSE}(\gamma) - {\rm MMSE}(1) + {\rm MMSE}'(1)(\gamma -1) + \frac{1}{2}{\rm MMSE}''(1) (\gamma -1)^2.
\end{equation*}
If one tries the ansatz $f(\gamma)=c \vert \gamma -1\vert^\alpha$ with
$2 < \alpha \leq 3$, or in other words 
$\log |f(\gamma)| \approx \log |c| + \alpha \log |\gamma -1 |$ we find $\alpha\approx 2.929$. This is shown 
in figure \ref{fig: behavior of MMSE} where $f(\gamma)$ is plotted on a log-log scale on both sides of $\gamma_c= 1^{\pm}$. However, the appearance of this exponent is not consistent with the fact that the expression for $\rho_Y$ \eqref{rho_Y expr} is fully algebraic and an excat integration could only give an integer exponent or a logarithmic singularity.
To further investigate the behavior of the MMSE, we study the third numerical derivative obtained from the curve of ${\rm MMSE}''(\gamma)$ using the relation $\frac{d^3}{d \gamma^3}{\rm MMSE}(1) \approx \frac{{\rm MMSE}''(1+\epsilon) - {\rm MMSE}''(1-\epsilon)}{2 \epsilon}$. As plotted in figure \ref{fig:third-derivative MMSE}, $\frac{d^3}{d \gamma^3}{\rm MMSE}(1)$ diverges linearly as $\epsilon$ decays exponentially. This suggests that the behavior of the correction term for the second derivative is of the form $a (\gamma - 1) \big( \log |\gamma - 1| + b \big)$.
Define the function
\begin{equation}
    g(\gamma) = {\rm MMSE}''(\gamma) - {\rm MMSE}''(1)
\end{equation}
for $\gamma$ close to $1$, we have:
\begin{equation}
    \frac{g(\gamma)}{\gamma - 1} \approx a \log |\gamma -1| + ab
\end{equation}
From the plots in figure \ref{fig: behavior of MMSE''}, we deduce that 
$a \approx -0.8463$ and $b \approx 0.9746$.
Therefore, the ${\rm MMSE}''(\gamma)$ can be described by the following expansion close to the point $\gamma =1$
\begin{equation}
    {\rm MMSE}''(\gamma) =  {\rm MMSE}''(1) + a (\gamma - 1) \big( \log |\gamma - 1| + b \big) + o\big((\gamma - 1)\big).
    \label{seond-der-expan}
\end{equation}
From this expansion, we conjecture that the MMSE has a third-order phase transition at $\gamma =1$.
\begin{figure}
    \centering
    \input{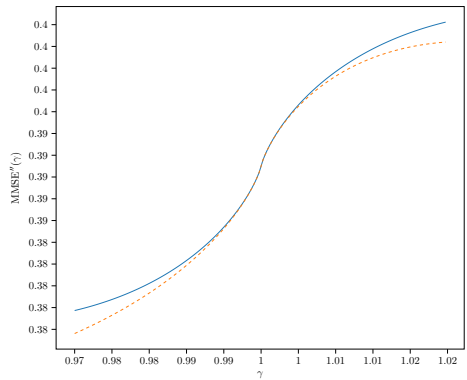}
    \caption{Comparison of the second derivative of the ${\rm MMSE}(\gamma)$ and the expansion \eqref{seond-der-expan} (plotted with dashed line).}
    \label{fig:}
\end{figure}

\section{Derivation of the limiting spectral distribution for the model \ref{Ex-2}}\label{Deriv-Ex-2}
We indicate the main steps to compute the density $\rho_Y = \rho_{\sqrt{\gamma} S} \boxplus \rho_{SC}$ where  $ \rho_{\sqrt{\gamma} S}(x) = p \delta(x) + (1-p) \delta(x-\sqrt{\gamma})$. Following the same procedure as in the previous example, functions $v(u)$ and $\psi(u)$ can be derived as :
\begin{equation*}
v(u) = \begin{cases}
    \sqrt{\frac{1}{2} \bigg[ -2u^2 + 2\sqrt{\gamma} u + \sqrt{\sqrt{\gamma} \big(\sqrt{\gamma} -2u\big) \big(-2\sqrt{\gamma}u + \gamma + 4p -2 \big) + 1 } - \gamma +1 \bigg]}
     &\text{if } u \in {\rm Supp}(v) \\
    0 & \text{else.}
    \end{cases} 
\end{equation*}
\begin{equation*}
    \psi(u) = \begin{cases}
    \frac{-8u^2 + 6\sqrt{\gamma} u + \sqrt{\sqrt{\gamma} \big(\sqrt{\gamma} -2u\big) \big(-2\sqrt{\gamma}u + \gamma + 4p -2 \big) + 1 } - \gamma -1}{2\big(\sqrt{\gamma} -2u\big)}
     &\text{if } u \in {\rm Supp}(v) \\
    u + \frac{p}{u} + \frac{1-p}{u - \sqrt{\gamma}} & \text{else.}
    \end{cases}
\end{equation*}
where
\begin{equation*}
    {\rm Supp}(v) = \Big\{ u \big| g(u) < 0 \Big\}
\end{equation*}
and 
\begin{equation}
g(u) = u^4 - 2\sqrt{\gamma}u^3 + (\gamma - 1)u^2 + 2 p \sqrt{\gamma} u - p \gamma .
\end{equation}
Solving the equation $\rho_Y \big( \psi(u) \big) = \frac{v(u)}{\pi}$, we find the analytical expression for $\rho_Y(x)$ which we omit here.

\begin{figure}
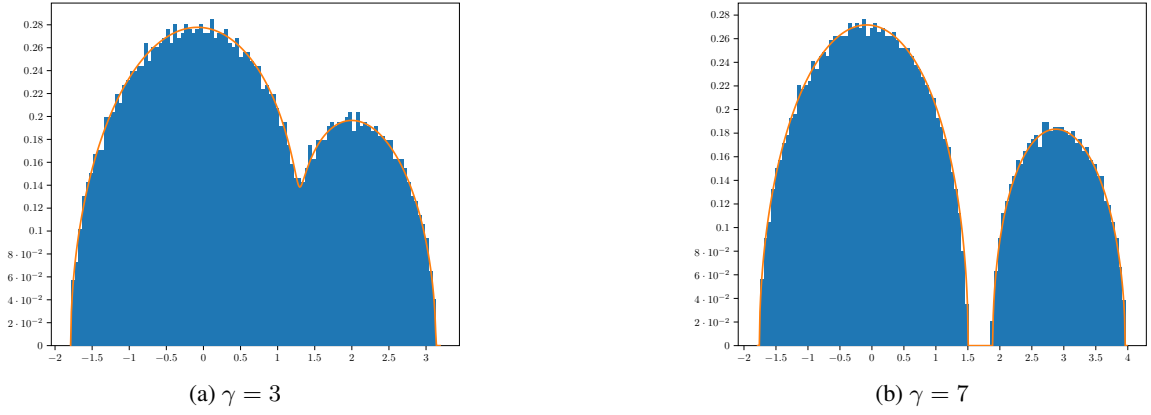

    \centering
    \begin{subfigure}{.45\textwidth}
        \centering
        \input{Supplementary/Examples/Bernoulli_spectrum/spec-3}
        \caption{$\gamma=3$}
    \end{subfigure}
    \hfill
    \begin{subfigure}{.45\textwidth}
        \centering
        \input{Supplementary/Examples/Bernoulli_spectrum/spec-7}
        \caption{$\gamma=7$}
    \end{subfigure}
    \captionsetup{singlelinecheck = false, justification=justified}
    \caption{{\small The continuous line is plotted from the analytical expression for the asymptotic spectral density of the observation matrix $\bY$ in example \ref{Ex-2} for $p=0.7$. $\rho_Y(x) = \rho_{\sqrt{\gamma} S} \boxplus \rho_{\rm SC}$ where $\rho_{\sqrt{\gamma} S}(x) = 0.7 \delta(x) + 0.3 \delta(x-\sqrt{\gamma})$ for $\gamma = 3$ and $\gamma = 7$. The critical value for $p=0.7$ is $\gamma^c \approx 3.78$. This is compared to the histogram of a realization of size $N=5000$.}}
    \label{fig:limiting spectrum of rho_Y Ex 2}
\end{figure}

The set ${\rm Supp}(v)$ determines the support of $\rho_Y$. For a given $0< p <1$ the degree four polynomial $g(u)$ has either two or four real roots, depending on $\gamma$. The former case corresponds to the situation where the support of $\rho_Y$ is a single interval, and the latter corresponds to the case where the support of $\rho_Y$ is a union of two intervals. Using Theorem 3.7 in \cite{janson2007resultant}, a critical value $\gamma_c$ is found such that for $\gamma<\gamma_c$ the polynomial $g(u)$ has two real roots and for $\gamma>\gamma_c$ it has four real roots. We have
\begin{equation}
    \gamma_c = 1 + 3 \sqrt[3]{p^2(1-p)} + 3 \sqrt[3]{p(1-p)^2} .
\end{equation}
An example is illutrated on figure \ref{fig:limiting spectrum of rho_Y Ex 2}. But contrary to the previous example we have not identified any singularity in ${\rm MMSE}(\gamma)$ due to the merging of the two intervals. 

\section{Derivation of the limiting spectral distribution for the model \ref{Ex-3}}\label{Deriv-Ex-3}
In this example, we find the limiting spectral measure $\rho_Y = \rho_{\sqrt{\gamma}S} \boxplus \rho_{\rm sc}$ directly using the free additive convolution formula. 
The limiting spectral measure of $\bS$ is the Marchenko-Pastur law rescaled by factor $q$, which we denote by $\rho_{\rm MP}$. The R-transform is $\cR_{\rho_{\rm MP}} = \frac{1}{q} \frac{1}{1 - z}$. Using the relation $\cR_{a * \mu} (z) = a \cR_{\mu}(a z)$, we have that $R_{\rho_{ \sqrt{\gamma}S}}(z) = \frac{\sqrt{\gamma}}{q} \frac{1}{1 - \sqrt{\gamma} z}$.
The free additive convolution formula is $R_{\rho_{ \sqrt{\gamma}S}}(z) + R_{\rho_{\rm sc}}(z) = R_{\rho_Y}(z)$. Substituting $z$ by the inverse of the Cauchy transform of $\rho_Y$, $G_{\rho_Y}^{-1}$, and using that $R_{\rho_{\rm sc}}(z) =z$ we get:
\begin{equation}
    G_{\rho_Y}(z) +  \frac{\sqrt{\gamma}}{q} \frac{1}{1 -  \sqrt{\gamma} G_{\rho_Y}(z)} + \frac{1}{G_{\rho_Y}(z)} = z .
\end{equation}
Solving this equation for $G_{\rho_Y}(z)$, and using the Stieltjes inversion formula, $\mu(x) = \frac{1}{\pi} \lim_{\epsilon \to 0} \Im \,  \cS_{\mu}(x+i \epsilon)$, we find the density of $\rho_Y$ to be
\begin{equation}
\rho_Y(x) = \begin{cases}
    \frac{ \sqrt[3]{2 A^2} - 2 \big( q \gamma (x^2 - 3) - q \sqrt{\gamma} x + q + 3 \gamma \big) }
    {\pi 2^{\frac{5}{3}} \sqrt{3 q \gamma} \sqrt[3]{A}}
     &\text{if } x \in {\rm Supp}(\rho_Y) \\
    0 & \text{else.}
    \end{cases} 
\end{equation}
where
\begin{equation}
\begin{split}
    A &= q^{\frac{3}{2}} \big(\sqrt{\gamma} x -2 \big) \big( \gamma (2 x^2 - 9) + \sqrt{\gamma} x -1) + 9 \sqrt{q} \gamma \big(\sqrt{\gamma}x +1\big) 
    + \sqrt{f(x)}
\end{split}
\end{equation}
where 
\begin{equation}
f(x) = q \Big( q \big(\sqrt{\gamma} x - 2 \big) \big( 2 \gamma x^2 + \sqrt{\gamma} x - 9 \gamma - 1 \big) + 9 ( \gamma^{\frac{3}{2}} x + \gamma ) \Big)^2 - 4 \Big( q \gamma (x^2 -3) - q \sqrt{\gamma}x + q + 3 \gamma \Big)^3
\end{equation}
and 
\begin{equation}
    {\rm Supp}(\rho_Y) = \Big\{ x \Big| f(x) \geq 0 \Big\} .
\end{equation}

\begin{figure}
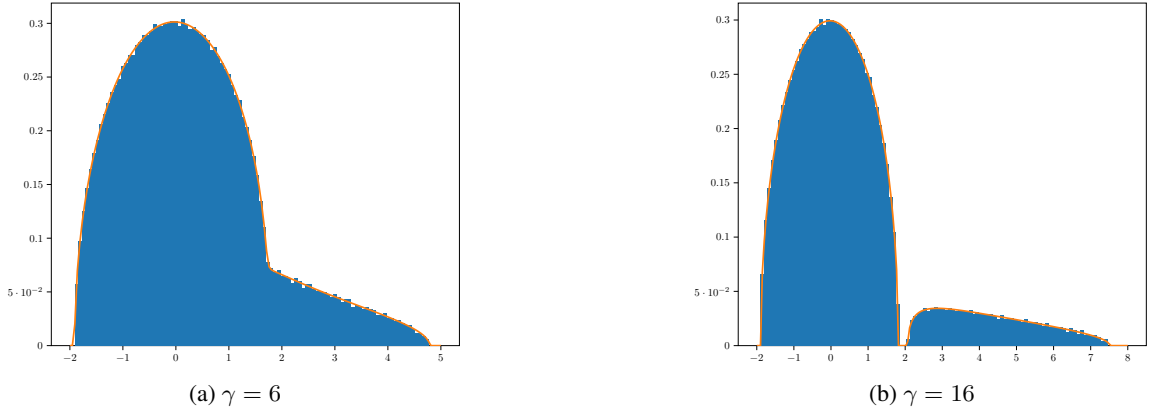

    \centering
    \begin{subfigure}{.45\textwidth}
        \centering
        \input{Supplementary/Examples/Wishart/spec-6,8}
        \caption{$\gamma=6$}
    \end{subfigure}
    \hfill
    \begin{subfigure}{.45\textwidth}
        \centering
        \input{Supplementary/Examples/Wishart/spec-16,8}
        \caption{$\gamma=16$}
    \end{subfigure}
    \captionsetup{singlelinecheck = false, justification=justified}
    \caption{{\small Asymptotic spectral density of the observation matrix $\bY$ in example \ref{Ex-3} for $q= 8$. $\gamma^c = 8$. It is Compared to the empirical density of a realization of size $N=10000$.}}
    \label{fig:limiting spectrum of rho_Y Ex 3}
\end{figure}

If $q \leq 1$, then the support of $\rho_{\sqrt{\gamma}S}$ is only a single interval, and the support of $\rho_Y$ is also an interval. However, for $q>1$, $\rho_{\sqrt{\gamma}S}$ has a delta at zero, and the support of $\rho_Y$ can be a single interval or union of two intervals, depending on $\gamma$.  
For fixed $q > 1$, the intervals merge at the critical value $\gamma_c = \frac{q }{\big( \sqrt[3]{q} -1 \big)^3}$: if $\gamma \leq \gamma_c$, ${\rm Supp}(\rho_Y)$ is a single interval, while if $\gamma > \gamma_c$, ${\rm Supp}(\rho_Y)$ is the union of two intervals. In Fig. \ref{fig:limiting spectrum of rho_Y Ex 3}, the density is plotted for $q = 8$, for which $\gamma_c = 8$.

\end{document}